%
%


%
\expandafter\ifx\csname phyzzx\endcsname\relax\else
 \errhelp{Hit <CR> and go ahead.}
 \errmessage{PHYZZX macros are already loaded or input. }
 \endinput \fi
\catcode`\@=11 
%
%
%
\font\seventeenrm=cmr17
\font\fourteenrm=cmr12 scaled\magstep1
\font\twelverm=cmr12
\font\ninerm=cmr9            \font\sixrm=cmr6
%
\font\fourteenbf=cmbx10 scaled\magstep2
\font\twelvebf=cmbx12
\font\ninebf=cmbx9            \font\sixbf=cmbx6
%
\font\fourteeni=cmmi10 scaled\magstep2      \skewchar\fourteeni='177
\font\twelvei=cmmi12			        \skewchar\twelvei='177
\font\ninei=cmmi9                           \skewchar\ninei='177
\font\sixi=cmmi6                            \skewchar\sixi='177
%
\font\fourteensy=cmsy10 scaled\magstep2     \skewchar\fourteensy='60
\font\twelvesy=cmsy10 scaled\magstep1	    \skewchar\twelvesy='60
\font\ninesy=cmsy9                          \skewchar\ninesy='60
\font\sixsy=cmsy6                           \skewchar\sixsy='60
%
\font\fourteenex=cmex10 scaled\magstep2
\font\twelveex=cmex10 scaled\magstep1
%
\font\fourteensl=cmsl12 scaled\magstep1
\font\twelvesl=cmsl12
\font\ninesl=cmsl9
%
\font\fourteenit=cmti12 scaled\magstep1
\font\twelveit=cmti12
\font\nineit=cmti9
\font\fourteentt=cmtt10 scaled\magstep2
\font\twelvett=cmtt12
\font\fourteencp=cmcsc10 scaled\magstep2
\font\twelvecp=cmcsc10 scaled\magstep1
\font\tencp=cmcsc10
\newfam\cpfam
\newdimen\b@gheight		\b@gheight=12pt
\newcount\f@ntkey		\f@ntkey=0
\def\f@m{\afterassignment\samef@nt\f@ntkey=}
\def\samef@nt{\fam=\f@ntkey \the\textfont\f@ntkey\relax}
\def\rm{\f@m0 }
\def\mit{\f@m1 }         
\def\cal{\f@m2 }
\def\it{\f@m\itfam}
\def\sl{\f@m\slfam}
\def\bf{\f@m\bffam}
\def\tt{\f@m\ttfam}
\def\caps{\f@m\cpfam}
\def\fourteenpoint{\relax
    \textfont0=\fourteenrm          \scriptfont0=\tenrm
      \scriptscriptfont0=\sevenrm
    \textfont1=\fourteeni           \scriptfont1=\teni
      \scriptscriptfont1=\seveni
    \textfont2=\fourteensy          \scriptfont2=\tensy
      \scriptscriptfont2=\sevensy
    \textfont3=\fourteenex          \scriptfont3=\twelveex
      \scriptscriptfont3=\tenex
    \textfont\itfam=\fourteenit     \scriptfont\itfam=\tenit
    \textfont\slfam=\fourteensl     \scriptfont\slfam=\tensl
    \textfont\bffam=\fourteenbf     \scriptfont\bffam=\tenbf
      \scriptscriptfont\bffam=\sevenbf
    \textfont\ttfam=\fourteentt
    \textfont\cpfam=\fourteencp
    \samef@nt
    \b@gheight=14pt
    \setbox\strutbox=\hbox{\vrule height 0.85\b@gheight
				depth 0.35\b@gheight width\z@ }}
\def\twelvepoint{\relax
    \textfont0=\twelverm          \scriptfont0=\ninerm
      \scriptscriptfont0=\sixrm
    \textfont1=\twelvei           \scriptfont1=\ninei
      \scriptscriptfont1=\sixi
    \textfont2=\twelvesy           \scriptfont2=\ninesy
      \scriptscriptfont2=\sixsy
    \textfont3=\twelveex          \scriptfont3=\tenex
      \scriptscriptfont3=\tenex
    \textfont\itfam=\twelveit     \scriptfont\itfam=\nineit
    \textfont\slfam=\twelvesl     \scriptfont\slfam=\ninesl
    \textfont\bffam=\twelvebf     \scriptfont\bffam=\ninebf
      \scriptscriptfont\bffam=\sixbf
    \textfont\ttfam=\twelvett
    \textfont\cpfam=\twelvecp
    \samef@nt
    \b@gheight=12pt
    \setbox\strutbox=\hbox{\vrule height 0.85\b@gheight
				depth 0.35\b@gheight width\z@ }}
\def\tenpoint{\relax
    \textfont0=\tenrm          \scriptfont0=\sevenrm
      \scriptscriptfont0=\fiverm
    \textfont1=\teni           \scriptfont1=\seveni
      \scriptscriptfont1=\fivei
    \textfont2=\tensy          \scriptfont2=\sevensy
      \scriptscriptfont2=\fivesy
    \textfont3=\tenex          \scriptfont3=\tenex
      \scriptscriptfont3=\tenex
    \textfont\itfam=\tenit     \scriptfont\itfam=\seveni
    \textfont\slfam=\tensl     \scriptfont\slfam=\sevenrm
    \textfont\bffam=\tenbf     \scriptfont\bffam=\sevenbf
      \scriptscriptfont\bffam=\fivebf
    \textfont\ttfam=\tentt
    \textfont\cpfam=\tencp
    \samef@nt
    \b@gheight=10pt
    \setbox\strutbox=\hbox{\vrule height 0.85\b@gheight
				depth 0.35\b@gheight width\z@ }}
%
%
%
\normalbaselineskip = 20pt plus 0.2pt minus 0.1pt
\normallineskip = 1.5pt plus 0.1pt minus 0.1pt
\normallineskiplimit = 1.5pt
\newskip\normaldisplayskip
\normaldisplayskip = 20pt plus 5pt minus 10pt
\newskip\normaldispshortskip
\normaldispshortskip = 6pt plus 5pt
\newskip\normalparskip
\normalparskip = 6pt plus 2pt minus 1pt
\newskip\skipregister
\skipregister = 5pt plus 2pt minus 1.5pt
\newif\ifsingl@    \newif\ifdoubl@
\newif\iftwelv@    \twelv@true
\def\singlespace{\singl@true\doubl@false\spaces@t}
\def\doublespace{\singl@false\doubl@true\spaces@t}
\def\normalspace{\singl@false\doubl@false\spaces@t}
\def\Tenpoint{\tenpoint\twelv@false\spaces@t}
\def\Twelvepoint{\twelvepoint\twelv@true\spaces@t}
\def\spaces@t{\relax
      \iftwelv@ \ifsingl@\subspaces@t3:4;\else\subspaces@t1:1;\fi
       \else \ifsingl@\subspaces@t3:5;\else\subspaces@t4:5;\fi \fi
      \ifdoubl@ \multiply\baselineskip by 5
         \divide\baselineskip by 4 \fi }
\def\subspaces@t#1:#2;{
      \baselineskip = \normalbaselineskip
      \multiply\baselineskip by #1 \divide\baselineskip by #2
      \lineskip = \normallineskip
      \multiply\lineskip by #1 \divide\lineskip by #2
      \lineskiplimit = \normallineskiplimit
      \multiply\lineskiplimit by #1 \divide\lineskiplimit by #2
      \parskip = \normalparskip
      \multiply\parskip by #1 \divide\parskip by #2
      \abovedisplayskip = \normaldisplayskip
      \multiply\abovedisplayskip by #1 \divide\abovedisplayskip by #2
      \belowdisplayskip = \abovedisplayskip
      \abovedisplayshortskip = \normaldispshortskip
      \multiply\abovedisplayshortskip by #1
        \divide\abovedisplayshortskip by #2
      \belowdisplayshortskip = \abovedisplayshortskip
      \advance\belowdisplayshortskip by \belowdisplayskip
      \divide\belowdisplayshortskip by 2
      \smallskipamount = \skipregister
      \multiply\smallskipamount by #1 \divide\smallskipamount by #2
      \medskipamount = \smallskipamount \multiply\medskipamount by 2
      \bigskipamount = \smallskipamount \multiply\bigskipamount by 4 }
\def\normalbaselines{ \baselineskip=\normalbaselineskip
   \lineskip=\normallineskip \lineskiplimit=\normallineskip
   \iftwelv@\else \multiply\baselineskip by 4 \divide\baselineskip by 5
     \multiply\lineskiplimit by 4 \divide\lineskiplimit by 5
     \multiply\lineskip by 4 \divide\lineskip by 5 \fi }
\Twelvepoint  
\interlinepenalty=50
\interfootnotelinepenalty=5000
\predisplaypenalty=9000
\postdisplaypenalty=500
\hfuzz=1pt
\vfuzz=0.2pt
\voffset=0pt
\dimen\footins=8 truein
%
%
%
\def\pagecontents{
   \ifvoid\topins\else\unvbox\topins\vskip\skip\topins\fi
   \dimen@ = \dp255 \unvbox255
   \ifvoid\footins\else\vskip\skip\footins\footrule\unvbox\footins\fi
   \ifr@ggedbottom \kern-\dimen@ \vfil \fi }
\def\makeheadline{\vbox to 0pt{ \skip@=\topskip
      \advance\skip@ by -12pt \advance\skip@ by -2\normalbaselineskip
      \vskip\skip@ \line{\vbox to 12pt{}\the\headline} \vss
      }\nointerlineskip}
\def\makefootline{\baselineskip = 1.5\normalbaselineskip
                 \line{\the\footline}}
\newif\iffrontpage
\newif\ifletterstyle
\newif\ifp@genum
\def\nopagenumbers{\p@genumfalse}
\def\pagenumbers{\p@genumtrue}
\pagenumbers
\newtoks\paperheadline
\newtoks\letterheadline
\newtoks\paperfootline
\newtoks\letterfootline
\newtoks\letterinfo
\newtoks\Letterinfo
\newtoks\date
\footline={\ifletterstyle\the\letterfootline\else\the\paperfootline\fi}
\paperfootline={\hss\iffrontpage\else\ifp@genum\tenrm\folio\hss\fi\fi}
\letterfootline={\iffrontpage\LETTERFOOT\else\hfil\fi}
\Letterinfo={\hfil}
\letterinfo={\hfil}
\def\LETTERFOOT{\hfil} 
%
\def\LETTERHEAD{\vtop{\baselineskip=20pt\hbox to
\hsize{\hfil\seventeenrm\strut 
CALIFORNIA INSTITUTE OF TECHNOLOGY \hfil}
\hbox to \hsize{\hfil\ninerm\strut
CHARLES C. LAURITSEN LABORATORY OF HIGH ENERGY PHYSICS \hfil}
\hbox to \hsize{\hfil\ninerm\strut
PASADENA, CALIFORNIA 91125 \hfil}}}
\headline={\ifletterstyle\the\letterheadline\else\the\paperheadline\fi}
\paperheadline={\hfil}
\letterheadline{\iffrontpage \LETTERHEAD\else
    \rm \ifp@genum \hfil \folio\hfil\fi\fi}
\def\monthname{\relax\ifcase\month 0/\or January\or February\or
   March\or April\or May\or June\or July\or August\or September\or
   October\or November\or December\else\number\month/\fi}
\def\today{\monthname\ \number\day, \number\year}
\date={\today}
\countdef\pageno=1      \countdef\pagen@=0
\countdef\pagenumber=1  \pagenumber=1
\def\advancepageno{\global\advance\pagen@ by 1
   \ifnum\pagenumber<0 \global\advance\pagenumber by -1
    \else\global\advance\pagenumber by 1 \fi \global\frontpagefalse }
\def\folio{\ifnum\pagenumber<0 \romannumeral-\pagenumber
           \else \number\pagenumber \fi }
\def\footrule{\dimen@=\prevdepth\nointerlineskip
   \vbox to 0pt{\vskip -0.25\baselineskip \hrule width 0.35\hsize \vss}
   \prevdepth=\dimen@ }
\newtoks\foottokens
\foottokens={}
\newdimen\footindent
\footindent=24pt
\def\vfootnote#1{\insert\footins\bgroup  
   \interlinepenalty=\interfootnotelinepenalty \floatingpenalty=20000
   \singl@true\doubl@false\Tenpoint
   \splittopskip=\ht\strutbox \boxmaxdepth=\dp\strutbox
   \leftskip=\footindent \rightskip=\z@skip
   \parindent=0.5\footindent \parfillskip=0pt plus 1fil
   \spaceskip=\z@skip \xspaceskip=\z@skip
   \the\foottokens
   \Textindent{$ #1 $}\footstrut\futurelet\next\fo@t}
\def\Textindent#1{\noindent\llap{#1\enspace}\ignorespaces}
\def\footnote#1{\attach{#1}\vfootnote{#1}}

\let\footsymbol=\star
\newcount\lastf@@t           \lastf@@t=-1
\newcount\footsymbolcount    \footsymbolcount=0
\newif\ifPhysRev
\def\bumpfootsymbolcount{\relax
   \iffrontpage \bumpfootsymbolNP \else \advance\lastf@@t by 1
     \ifPhysRev \bumpfootsymbolPR \else \bumpfootsymbolNP \fi \fi
   \global\lastf@@t=\pagen@ }
\def\bumpfootsymbolNP{\ifnum\footsymbolcount <0 \global\footsymbolcount =0 \fi
    \ifnum\lastf@@t<\pagen@ \global\footsymbolcount=0
     \else \global\advance\footsymbolcount by 1 \fi }
\def\bumpfootsymbolPR{\ifnum\footsymbolcount >0 \global\footsymbolcount =0 \fi
      \global\advance\footsymbolcount by -1 }
\def\fd@f#1 {\xdef\footsymbol{\mathchar"#1 }}
\def\generatefootsymbol{\ifcase\footsymbolcount \fd@f 13F \or \fd@f 279
	\or \fd@f 27A \or \fd@f 278 \or \fd@f 27B \else
	\ifnum\footsymbolcount <0 \fd@f{023 \number-\footsymbolcount }
	 \else \fd@f 203 {\loop \ifnum\footsymbolcount >5
		\fd@f{203 \footsymbol } \advance\footsymbolcount by -1
		\repeat }\fi \fi }

\def\nonfrenchspacing{\sfcode`\.=3001 \sfcode`\!=3000 \sfcode`\?=3000
	\sfcode`\:=2000 \sfcode`\;=1500 \sfcode`\,=1251 }
\nonfrenchspacing
\newdimen\d@twidth
{\setbox0=\hbox{s.} \global\d@twidth=\wd0 \setbox0=\hbox{s}
	\global\advance\d@twidth by -\wd0 }
\def\removehglue{\loop \unskip \ifdim\lastskip >\z@ \repeat }
\def\roll@ver#1{\removehglue \nobreak \count255 =\spacefactor \dimen@=\z@
	\ifnum\count255 =3001 \dimen@=\d@twidth \fi
	\ifnum\count255 =1251 \dimen@=\d@twidth \fi
    \iftwelv@ \kern-\dimen@ \else \kern-0.83\dimen@ \fi
   #1\spacefactor=\count255 }
\def\step@ver#1{\relax \ifmmode #1\else \ifhmode
	\roll@ver{${}#1$}\else {\setbox0=\hbox{${}#1$}}\fi\fi }
\def\attach#1{\step@ver{\strut^{\mkern 2mu #1} }}
%
%
%
\newcount\chapternumber      \chapternumber=0
\newcount\sectionnumber      \sectionnumber=0
\newcount\equanumber         \equanumber=0
\let\chapterlabel=\relax
\let\sectionlabel=\relax
\newtoks\chapterstyle        \chapterstyle={\Number}
\newtoks\sectionstyle        \sectionstyle={\chapterlabel\Number}
\newskip\chapterskip         \chapterskip=\bigskipamount
\newskip\sectionskip         \sectionskip=\medskipamount
\newskip\headskip            \headskip=8pt plus 3pt minus 3pt
\newdimen\chapterminspace    \chapterminspace=15pc
\newdimen\sectionminspace    \sectionminspace=10pc
\newdimen\referenceminspace  \referenceminspace=25pc
\def\chapterreset{\global\advance\chapternumber by 1
   \ifnum\equanumber<0 \else\global\equanumber=0\fi
   \sectionnumber=0 \makechapterlabel}
\def\makechapterlabel{\let\sectionlabel=\relax
   \xdef\chapterlabel{\the\chapterstyle{\the\chapternumber}.}}
\def\alphabetic#1{\count255='140 \advance\count255 by #1\char\count255}
\def\Alphabetic#1{\count255='100 \advance\count255 by #1\char\count255}
\def\Roman#1{\uppercase\expandafter{\romannumeral #1}}
\def\roman#1{\romannumeral #1}
\def\Number#1{\number #1}
\def\BLANC#1{}
\def\titlestyle#1{\par\begingroup \interlinepenalty=9999
     \leftskip=0.02\hsize plus 0.23\hsize minus 0.02\hsize
     \rightskip=\leftskip \parfillskip=0pt
     \hyphenpenalty=9000 \exhyphenpenalty=9000
     \tolerance=9999 \pretolerance=9000
     \spaceskip=0.333em \xspaceskip=0.5em
     \iftwelv@\fourteenpoint\else\twelvepoint\fi
   \noindent #1\par\endgroup }
\def\spacecheck#1{\dimen@=\pagegoal\advance\dimen@ by -\pagetotal
   \ifdim\dimen@<#1 \ifdim\dimen@>0pt \vfil\break \fi\fi}
\def\TableOfContentEntry#1#2#3{\relax}
\def\chapter#1{\par \penalty-300 \vskip\chapterskip
   \spacecheck\chapterminspace
   \chapterreset \titlestyle{\chapterlabel\ #1}
   \TableOfContentEntry c\chapterlabel{#1}
   \nobreak\vskip\headskip \penalty 30000
   \wlog{\string\chapter\space \chapterlabel} }

\def\section#1{\par \ifnum\the\lastpenalty=30000\else
   \penalty-200\vskip\sectionskip \spacecheck\sectionminspace\fi
   \global\advance\sectionnumber by 1
   \xdef\sectionlabel{\the\sectionstyle\the\sectionnumber}
   \wlog{\string\section\space \sectionlabel}
   \TableOfContentEntry s\sectionlabel{#1}
   \noindent {\caps\enspace\sectionlabel\quad #1}\par
   \nobreak\vskip\headskip \penalty 30000 }
\def\subsection#1{\par
   \ifnum\the\lastpenalty=30000\else \penalty-100\smallskip \fi
   \noindent\undertext{#1}\enspace \vadjust{\penalty5000}}

\def\undertext#1{\vtop{\hbox{#1}\kern 1pt \hrule}}
\def\ack{\par\penalty-100\medskip \spacecheck\sectionminspace
   \line{\fourteenrm\hfil ACKNOWLEDGEMENTS\hfil}\nobreak\vskip\headskip }
\def\APPENDIX#1#2{\par\penalty-300\vskip\chapterskip
   \spacecheck\chapterminspace \chapterreset \xdef\chapterlabel{#1}
   \titlestyle{APPENDIX #2} \nobreak\vskip\headskip \penalty 30000
   \TableOfContentEntry a{#1}{#2}
   \wlog{\string\Appendix\ \chapterlabel} }
\def\Appendix#1{\APPENDIX{#1}{#1}}
\def\appendix{\APPENDIX{A}{}}
\def\unnumberedchapters{\let\makechapterlabel=\relax \let\chapterlabel=\relax
   \sectionstyle={\BLANC}\let\sectionlabel=\relax \sequentialequations }
%
%
%
\def\eqname#1{\relax \ifnum\equanumber<0
     \xdef#1{{\noexpand\rm(\number-\equanumber)}}%
       \global\advance\equanumber by -1
    \else \global\advance\equanumber by 1
      \xdef#1{{\noexpand\rm(\chapterlabel\number\equanumber)}} \fi #1}

\def\eqn{\eqno\eqname}

\def\eqinsert#1{\noalign{\dimen@=\prevdepth \nointerlineskip
   \setbox0=\hbox to\displaywidth{\hfil #1}
   \vbox to 0pt{\kern 0.5\baselineskip\hbox{$\!\box0\!$}\vss}
   \prevdepth=\dimen@}}
%

%
%
\def\GENITEM#1;#2{\par \hangafter=0 \hangindent=#1
    \Textindent{$ #2 $}\ignorespaces}
\outer\def\newitem#1=#2;{\gdef#1{\GENITEM #2;}}
\newdimen\itemsize                \itemsize=30pt
\newitem\item=1\itemsize;
\newitem\sitem=1.75\itemsize;     
\newitem\ssitem=2.5\itemsize;     
\outer\def\newlist#1=#2&#3&#4;{\toks0={#2}\toks1={#3}%
   \count255=\escapechar \escapechar=-1
   \alloc@0\list\countdef\insc@unt\listcount     \listcount=0
   \edef#1{\par
      \countdef\listcount=\the\allocationnumber
      \advance\listcount by 1
      \hangafter=0 \hangindent=#4
      \Textindent{\the\toks0{\listcount}\the\toks1}}
   \expandafter\expandafter\expandafter
    \edef\c@t#1{begin}{\par
      \countdef\listcount=\the\allocationnumber \listcount=1
      \hangafter=0 \hangindent=#4
      \Textindent{\the\toks0{\listcount}\the\toks1}}
   \expandafter\expandafter\expandafter
    \edef\c@t#1{con}{\par \hangafter=0 \hangindent=#4 \noindent}
   \escapechar=\count255}
\def\c@t#1#2{\csname\string#1#2\endcsname}
\newlist\point=\Number&.&1.0\itemsize;
\newlist\subpoint=(\alphabetic&)&1.75\itemsize;
\newlist\subsubpoint=(\roman&)&2.5\itemsize;
%

%
%
%
%
\newcount\referencecount     \referencecount=0
\newcount\lastrefsbegincount \lastrefsbegincount=0
\newif\ifreferenceopen       \newwrite\referencewrite
\newif\ifrw@trailer
\newdimen\refindent     \refindent=30pt
\def\NPrefmark#1{\attach{\scriptscriptstyle [ #1 ] }}
\let\PRrefmark=\attach
\def\refmark#1{\relax\ifPhysRev\PRrefmark{#1}\else\NPrefmark{#1}\fi}
\def\refend@{\refmark{\number\referencecount}}
\def\refend{\refend@{}\space }
\def\refsend{\refmark{\count255=\referencecount
   \advance\count255 by-\lastrefsbegincount
   \ifcase\count255 \number\referencecount
   \or \number\lastrefsbegincount,\number\referencecount
   \else \number\lastrefsbegincount-\number\referencecount \fi}\space }
\def\refitem#1{\par \hangafter=0 \hangindent=\refindent \Textindent{#1}}
\def\Ref{\rw@trailertrue\REF}
\def\ref{\Ref\?}

\def\REF#1{\r@fstart{#1}%
   \rw@begin{\the\referencecount.}\rw@end}
\def\REFS#1{\r@fstart{#1}%
   \lastrefsbegincount=\referencecount
   \rw@begin{\the\referencecount.}\rw@end}
\def\r@fstart#1{\chardef\rw@write=\referencewrite \let\rw@ending=\refend@
   \ifreferenceopen \else \global\referenceopentrue
   \immediate\openout\referencewrite=referenc.txa
   \toks0={\catcode`\^^M=10}\immediate\write\rw@write{\the\toks0} \fi
   \global\advance\referencecount by 1 \xdef#1{\the\referencecount}}
{\catcode`\^^M=\active %
 \gdef\rw@begin#1{\immediate\write\rw@write{\noexpand\refitem{#1}}%
   \begingroup \catcode`\^^M=\active \let^^M=\relax}%
 \gdef\rw@end#1{\rw@@end #1^^M\rw@terminate \endgroup%
   \ifrw@trailer\rw@ending\global\rw@trailerfalse\fi }%
 \gdef\rw@@end#1^^M{\toks0={#1}\immediate\write\rw@write{\the\toks0}%
   \futurelet\n@xt\rw@test}%
 \gdef\rw@test{\ifx\n@xt\rw@terminate \let\n@xt=\relax%
       \else \let\n@xt=\rw@@end \fi \n@xt}%
}
\let\rw@ending=\relax
\let\rw@terminate=\relax
\let\splitout=\relax
\def\Closeout#1{\toks0={\catcode`\^^M=5}\immediate\write#1{\the\toks0}%
   \immediate\closeout#1}
%
%
\newcount\figurecount     \figurecount=0
\newcount\tablecount      \tablecount=0
\newif\iffigureopen       \newwrite\figurewrite
\newif\iftableopen        \newwrite\tablewrite
\def\FIG#1{\f@gstart{#1}%
   \rw@begin{\the\figurecount)}\rw@end}
\let\FIGURE=\FIG
\def\Fig{\rw@trailertrue\def\rw@ending{Fig.~\?}\FIG\?}
\def\fig{\rw@trailertrue\def\rw@ending{fig.~\?}\FIG\?}
\def\TABLE#1{\T@Bstart{#1}%
   \rw@begin{\the\tableecount:}\rw@end}
\def\Table{\rw@trailertrue\def\rw@ending{Table~\?}\TABLE\?}
\def\f@gstart#1{\chardef\rw@write=\figurewrite
   \iffigureopen \else \global\figureopentrue
   \immediate\openout\figurewrite=figures.txa
   \toks0={\catcode`\^^M=10}\immediate\write\rw@write{\the\toks0} \fi
   \global\advance\figurecount by 1 \xdef#1{\the\figurecount}}
\def\T@Bstart#1{\chardef\rw@write=\tablewrite
   \iftableopen \else \global\tableopentrue
   \immediate\openout\tablewrite=tables.txa
   \toks0={\catcode`\^^M=10}\immediate\write\rw@write{\the\toks0} \fi
   \global\advance\tablecount by 1 \xdef#1{\the\tablecount}}
%

%
%
%
\def\getfigure#1{\global\advance\figurecount by 1
   \xdef#1{\the\figurecount}\count255=\escapechar \escapechar=-1
   \edef\n@xt{\noexpand\g@tfigure\csname\string#1Body\endcsname}%
   \escapechar=\count255 \n@xt }
\def\g@tfigure#1#2 {\errhelp=\disabledfigures \let#1=\relax
   \errmessage{\string\getfigure\space disabled}}
\newhelp\disabledfigures{ Empty figure of zero size assumed.}
\def\figinsert#1{\midinsert\Tenpoint\medskip
   \count255=\escapechar \escapechar=-1
   \edef\n@xt{\csname\string#1Body\endcsname}
   \escapechar=\count255 \centerline{\n@xt}
   \bigskip\narrower\narrower
   \noindent{\it Figure}~#1.\quad }
%
%
%
\def\masterreset{\global\pagenumber=1 \global\chapternumber=0
   \global\equanumber=0 \global\sectionnumber=0
   \global\referencecount=0 \global\figurecount=0 \global\tablecount=0 }
\def\FRONTPAGE{\ifvoid255\else\vfill\penalty-20000\fi
      \masterreset\global\frontpagetrue
      \global\lastf@@t=0 \global\footsymbolcount=0}

\def\paperstyle{\letterstylefalse\normalspace\papersize}
\def\letterstyle{\letterstyletrue\singlespace\lettersize}
\def\papersize{\hsize=35 truepc\vsize=50 truepc\hoffset=-2.51688 truepc
               \skip\footins=\bigskipamount}
\def\lettersize{\hsize=5.5 truein\vsize=8.25 truein\hoffset=.4875 truein
	\voffset=.3125 truein
   \skip\footins=\smallskipamount \multiply\skip\footins by 3 }
\paperstyle   
%
%
\def\MEMO{\letterstyle \letterinfo={\hfil } \let\rule=\memorule
	\FRONTPAGE \memohead }
\let\memohead=\relax

\def\memit@m#1{\smallskip \hangafter=0 \hangindent=1in
      \Textindent{\caps #1}}
\def\subject{\memit@m{Subject:}}
\def\topic{\memit@m{Topic:}}
\def\from{\memit@m{From:}}
\def\to{\relax \ifmmode \rightarrow \else \memit@m{To:}\fi }
\def\memorule{\medskip\hrule height 1pt\bigskip}
\newwrite\labelswrite
\newtoks\rw@toks

\def\addressee#1{\null\vskip .5truein\line{
\hskip 0.5\hsize minus 0.5\hsize\the\date\hfil}\bigskip
   \ialign to\hsize{\strut ##\hfil\tabskip 0pt plus \hsize \cr #1\crcr}
   \writelabel{#1}\medskip\par\noindent}
\def\rwl@begin#1\cr{\rw@toks={#1\crcr}\relax
   \immediate\write\labelswrite{\the\rw@toks}\futurelet\n@xt\rwl@next}
\def\rwl@next{\ifx\n@xt\rwl@end \let\n@xt=\relax
      \else \let\n@xt=\rwl@begin \fi \n@xt}
\let\rwl@end=\relax
\def\writelabel#1{\immediate\write\labelswrite{\noexpand\labelbegin}
     \rwl@begin #1\cr\rwl@end
     \immediate\write\labelswrite{\noexpand\labelend}}
\newbox\FromLabelBox

\let\labelend=\relax
\def\labelbegin#1\labelend{\setbox0=\vbox{\ialign{##\hfil\cr #1\crcr}}
     \MakeALabel }
\newtoks\FromAddress
\FromAddress={}
\def\MakeFromBox#1{\global\setbox\FromLabelBox=\vbox{\Tenpoint
     \ialign{##\hfil\cr #1\the\FromAddress\crcr}}}
\newdimen\labelwidth		\labelwidth=6in
\def\MakeALabel{\vskip 1pt \hbox{\vrule \vbox{
	\hsize=\labelwidth \hrule\bigskip
	\leftline{\hskip 1\parindent \copy\FromLabelBox}\bigskip
	\centerline{\hfil \box0 } \bigskip \hrule
	}\vrule } \vskip 1pt plus 1fil }
\newskip\signatureskip       \signatureskip=30pt
\def\signed#1{\par \penalty 9000 \medskip \dt@pfalse
  \everycr={\noalign{\ifdt@p\vskip\signatureskip\global\dt@pfalse\fi}}
  \setbox0=\vbox{\singlespace \ialign{\strut ##\hfil\crcr
   \noalign{\global\dt@ptrue}#1\crcr}}
  \line{\hskip 0.5\hsize minus 0.5\hsize \box0\hfil} \medskip }
\newbox\letterb@x
\def\lettertext{\par\unvcopy\letterb@x\par}
\def\multiletter{\setbox\letterb@x=\vbox\bgroup
      \everypar{\vrule height 1\baselineskip depth 0pt width 0pt }
      \singlespace \topskip=\baselineskip }
\def\letterend{\par\egroup}
%
%
%
\newskip\frontpageskip
\newtoks\Pubnum
\newtoks\pubtype
\newif\ifp@bblock  \p@bblocktrue
\def\PH@SR@V{\doubl@true \baselineskip=24.1pt plus 0.2pt minus 0.1pt
             \parskip= 3pt plus 2pt minus 1pt }
\def\PHYSREV{\paperstyle\PhysRevtrue\PH@SR@V}
\def\titlepage{\FRONTPAGE\paperstyle\ifPhysRev\PH@SR@V\fi
   \ifp@bblock\p@bblock \else\hrule height\z@ \relax \fi }
\def\nopubblock{\p@bblockfalse}

\frontpageskip=12pt plus .5fil minus 2pt
\pubtype={\tensl Preliminary Version}
\Pubnum={}
\def\p@bblock{\begingroup \tabskip=\hsize minus \hsize
   \baselineskip=1.5\ht\strutbox \topspace-2\baselineskip
   \halign to\hsize{\strut ##\hfil\tabskip=0pt\crcr
       \the\Pubnum\crcr\the\date\crcr\the\pubtype\crcr}\endgroup}
\def\title#1{\vskip\frontpageskip \titlestyle{#1} \vskip\headskip }
\def\author#1{\vskip\frontpageskip\titlestyle{\twelvecp #1}\nobreak}

\def\address#1{\par\kern 5pt\titlestyle{\twelvepoint\it #1}}
\def\andaddress{\par\kern 5pt \centerline{\sl and} \address}

\def\abstract{\par\dimen@=\prevdepth \hrule height\z@ \prevdepth=\dimen@
   \vskip\frontpageskip\centerline{\fourteenrm ABSTRACT}\vskip\headskip }

%
%
%

\def\\{\relax \ifmmode \backslash \else {\tt\char`\\}\fi }
\def\sequentialequations{\relax\if\equanumber<0\else\global\equanumber=-1\fi}

\def\journal#1&#2(#3){\unskip, \sl #1\unskip~\bf\ignorespaces #2\rm (19#3),}

\def\topspace{\hrule height 0pt depth 0pt \vskip}

\def\Buildrel#1\under#2{\mathrel{\mathop{#2}\limits_{#1}}}
\def\becomes#1{\mathchoice{\becomes@\scriptstyle{#1}}{\becomes@\scriptstyle
   {#1}}{\becomes@\scriptscriptstyle{#1}}{\becomes@\scriptscriptstyle{#1}}}
\def\becomes@#1#2{\mathrel{\setbox0=\hbox{$\m@th #1{\,#2\,}$}%
	\mathop{\hbox to \wd0 {\rightarrowfill}}\limits_{#2}}}

\let\int=\intop         \let\oint=\ointop
\def\lsim{\mathrel{\mathpalette\@versim<}}
\def\gsim{\mathrel{\mathpalette\@versim>}}
\def\@versim#1#2{\vcenter{\offinterlineskip
	\ialign{$\m@th#1\hfil##\hfil$\crcr#2\crcr\sim\crcr } }}
\def\big#1{{\hbox{$\left#1\vbox to 0.85\b@gheight{}\right.\n@space$}}}
\def\Big#1{{\hbox{$\left#1\vbox to 1.15\b@gheight{}\right.\n@space$}}}
\def\bigg#1{{\hbox{$\left#1\vbox to 1.45\b@gheight{}\right.\n@space$}}}
\def\Bigg#1{{\hbox{$\left#1\vbox to 1.75\b@gheight{}\right.\n@space$}}}
%
%
%
\let\sec@nt=\sec
\def\sec{\relax\ifmmode\let\n@xt=\sec@nt\else\let\n@xt\section\fi\n@xt}
\def\obsolete#1{\message{Macro \string #1 is obsolete.}}
\def\firstsec#1{\obsolete\firstsec \section{#1}}
\def\firstsubsec#1{\obsolete\firstsubsec \subsection{#1}}
\def\thispage#1{\obsolete\thispage \global\pagenumber=#1\frontpagefalse}
\def\thischapter#1{\obsolete\thischapter \global\chapternumber=#1}
\def\REFSCON{\obsolete\REFSCON\REF}
\def\splitout{\obsolete\splitout\relax}
\def\prop{\obsolete\prop \propto }
\def\nextequation#1{\obsolete\nextequation \global\equanumber=#1
   \ifnum\the\equanumber>0 \global\advance\equanumber by 1 \fi}
\def\BOXITEM{\afterassigment\B@XITEM\setbox0=}
\def\B@XITEM{\par\hangindent\wd0 \noindent\box0 }
\def\phyzzx{PHY\setbox0=\hbox{Z}\copy0 \kern-0.5\wd0 \box0 X}
%
%
\everyjob{\xdef\today{\monthname\ \number\day, \number\year}}
        
%


\hoffset=0.2truein
\voffset=0.1truein
\hsize=6truein

\def\CALT#1{\hbox to\hsize{\tenpoint \baselineskip=12pt
	\hfil\vtop{\hbox{\strut CALT-68-#1}
	\hbox{\strut DOE RESEARCH AND}
	\hbox{\strut DEVELOPMENT REPORT}}}}

\def\CALTECH{\smallskip
	\address{California Institute of Technology, Pasadena, CA 91125}}
\def\TITLE#1{\vskip 1in \centerline{\fourteenpoint #1}}
\def\AUTHOR#1{\vskip .5in \centerline{#1}}

\def\ABSTRACT#1{\vskip .5in \vfil \centerline{\twelvepoint \bf Abstract}
	#1 \vfil}

\def\sqr#1#2{{\vcenter{\hrule height.#2pt
      \hbox{\vrule width.#2pt height#1pt \kern#1pt
        \vrule width.#2pt}
      \hrule height.#2pt}}}

\def\section#1#2{
\noindent\hbox{\hbox{\bf #1}\hskip 10pt\vtop{\hsize=5in
\baselineskip=12pt \noindent \bf #2 \hfil}\hfil}
\medskip}

\def\underwig#1{	
	\setbox0=\hbox{\rm \strut}
	\hbox to 0pt{$#1$\hss} \lower \ht0 \hbox{\rm \char'176}}

\def\bunderwig#1{	
	\setbox0=\hbox{\rm \strut}
	\hbox to 1.5pt{$#1$\hss} \lower 12.8pt
	 \hbox{\seventeenrm \char'176}\hbox to 2pt{\hfil}}

\def\MEMO#1#2#3#4#5{
\frontpagetrue
\centerline{\tencp INTEROFFICE MEMORANDUM}
\smallskip
\centerline{\bf CALIFORNIA INSTITUTE OF TECHNOLOGY}
\centerline{\tencp Charles C. Lauritsen Laboratory of High Energy Physics} 
\bigskip
\vtop{\tenpoint \hbox to\hsize{\strut \hbox to .75in{\caps to:\hfil}
\hbox to3in{#1\hfil}
\hbox to .75in{\caps date:\hfil}\quad \the\date\hfil}
\hbox to\hsize{\strut \hbox to.75in{\caps from:\hfil}\hbox to 2in{#2\hfil}
\hbox{{\caps extension:}\quad#3\qquad{\caps mail code:\quad}#4}\hfil}
\hbox{\hbox to.75in{\caps subject:\hfil}\vtop{\parindent=0pt
\hsize=3.5in #5\hfil}}
\hbox{\strut\hfil}}}






\tolerance 10000


\newread\figureread                                     
\def\g@tfigure#1#2 {\openin\figureread #2.fig           
   \ifeof\figureread \errhelp=\disabledfigures          
     \errmessage{No such file: #2.fig}\let#1=\relax \else
    \read\figureread to\y@p \read\figureread to\y@p     
    \read\figureread to\x@p \read\figureread to\y@m     
    \read\figureread to\x@m \closein\figureread         
    \xdef#1{\hbox{\kern-\x@m truein \vbox{\kern-\y@m truein
      \hbox to \x@p truein{\vbox to \y@p truein{        
        \special{pos,inc=#2.fig}\vss }\hss }}}}\fi }    

\def\section#1{\par\ifnum\the\lastpenalty=30000\else
        \penalty-200\vskip\sectionskip\spacecheck\sectionminspace\fi
        \global\advance\sectionnumber by 1 
        \xdef\sectionlabel{\the\sectionstyle\the\sectionnumber}
        \wlog{\string\section\space\sectionlabel}
        \TableOfContentEntry s\sectionlabel{#1}
        \noindent {\caps\enspace\sectionlabel\quad #1}\par
        \nobreak\vskip\headskip\penalty 30000 }


\input epsf
\ifx\epsffile\undefined\message{(FIGURES WILL BE IGNORED)}
\def\insertfig#1#2{}
\else\message{(FIGURES WILL BE INCLUDED)}
\def\insertfig#1#2{{{\baselineskip=4pt
\midinsert\centerline{\epsfxsize=\hsize\epsffile{#2}}{{\centerline{#1}}}\
\medskip\endinsert}}}

\def\insertfigpage#1#2{{{\baselineskip=4pt
\pageinsert\centerline{\epsfysize=6.5in\epsffile{#1}}
 {{\bigskip{#2}}}\
\smallskip\endinsert}}}

\def\insertcomfigpage#1#2{{{\baselineskip=4pt
\pageinsert\centerline{\epsfysize=6.5in\epsffile{#1}}
 {{\smallskip{#2}}}\
\endinsert}}}

\def\insertnarrowfigpage#1#2{{{\baselineskip=4pt
\pageinsert\centerline{\epsfxsize=5in\epsffile{#1}}
 {{\smallskip{#2}}}\
\endinsert}}}

\def\insertccomfigpage#1#2{{{\baselineskip=4pt
\pageinsert\centerline{\epsfysize=6in\epsffile{#1}}
 {{\smallskip{#2}}}\
\endinsert}}}

\def\insertlongfigpage#1#2{{{\baselineskip=4pt
\pageinsert\centerline{\epsfysize=8in\epsffile{#1}}
 {{{#2}}}\
\endinsert}}}

\def\insertwiidefigpage#1#2{{{\baselineskip=1pt
\pageinsert\centerline{\epsfxsize=6.5in\epsffile{#1}}
\vskip-1in
 {{{#2}}}\
\endinsert}}}

\def\insertwiiderfigpage#1#2{{{\baselineskip=1pt
\pageinsert\centerline{\epsfxsize=7.5in\epsffile{#1}}
\vskip-2in
 {{{#2}}}\
\endinsert}}}

\def\insertxfigpage#1#2{{{\baselineskip=4pt
\pageinsert\centerline{\epsfxsize=6in\epsffile{#1}}
 {{\bigskip{#2}}}\
\medskip\endinsert}}}

\def\bcap#1{{\bf Figure {#1}:\/}}

\fi


\TITLE{Evolution of a Non-Abelian Cosmic String Network}
\AUTHOR{Patrick McGraw\footnote{\dag}{E-mail: mcgraw@physics.unc.edu}}
\CALTECH
\andaddress{
Institute of Field Physics, Department of Physics 
and Astronomy\break
University of North Carolina, Chapel Hill, NC  27599
}

\ABSTRACT{We describe a numerical simulation of the evolution of an $S_3$ cosmic
string network which takes fully into
account the non-commutative nature of the cosmic string fluxes and the 
topological obstructions which hinder strings from moving past each other
or intercommuting.  The influence of initial conditions, string tensions, 
and other parameters on the network's evolution is explored.  Contrary to some previous suggestions, we find no strong evidence of the ``freezing'' required for a string-dominated
cosmological scenario.  Instead, the results in a broad 
range of regimes are consistent with the familiar scaling law; i.e., a constant number of strings per horizon
volume.   The size of this number, however, can vary quite a bit, as can other 
overall features.  There is a surprisingly strong dependence on the statistical properties of the initial 
conditions.  We also observe a rich variety of interesting new structures,   such as light string webs
stretched between heavier strings, which are not seen in Abelian networks.}  

PACS Number:  98.80.Cq

\section{Introduction}
A generic feature of many spontaneously broken gauge theories is the existence
of topological solitons, such as strings (or flux tubes), domain walls and monopoles. Many grand
unified models predict the formation of such defects during a cosmological phase transition.\Ref\Kibblecosmic{T.W.B.~Kibble, Phys. Rep. {\bf 67}  (1980), 183;
J. Phys. {\bf A9} (1976), 1387.}  The  defects could be
interesting as a potentially observable signature of the symmetry-breaking
pattern, and could also have important consequences for the evolution of the 
universe.  Domain walls and monopoles, if they are stable and occur without
cosmic strings, generally are not considered phenomenologically viable.
Domain walls tend to produce density perturbations that are much too
 strong,\Ref\nodomainwalls{Ya. B. Zel'dovich, I. Yu. Kobzarev, L.B. Okun,
Zh. Eksp. Teor. Fiz. {\bf 67} (1974), 3; Sov. Phys. JETP {\bf 40} (1975), 1.} 
whereas monopoles, if they exist at all, are predicted to form in great 
abundances which are incompatible with current
 observations.\Ref\nomonopoles{J. Preskill, Phys. Rev. Lett. {\bf 43} (1979), 1363}   Among the possible categories of stable topological defects,
string-like defects, or cosmic strings,
are considered the least disastrous for cosmological models,  and may be  useful from a model-building point of view.\Ref\Cosmology{For a review, see: A.~Vilenkin, Phys. Rep. {\bf 121} (1985), 263;  or more recently 
M.B. Hindmarsh and T.W.B. Kibble, Rep. Prog. Phys. {\bf 58} (1995), 477.}  Among the potential applications,  it has been proposed that the gravitational effects of either infinite cosmic strings or closed loops of string may serve
as sources of density perturbations leading to galaxy or cluster
 formation.\REF\denspert{Ya. B. Zel'dovich, Mon. Not. R. Astr. Soc. 192 (1980);
A. Vilenkin and Q. Shafi, Phys. Rev. Lett. {\bf 51} (1983), 1716.}\REF\Zevarg{A. Vilenkin, Phys. Rev. Lett. 46 (1979) 1169,
1496(E).}\refmark{\denspert,\Zevarg}

A persistent string network could conceivably also have profound effects
on the evolution of the universe due to its bulk energy density, quite apart
from the effects of fluctuations.  If a network of strings becomes
frozen so that strings are fixed in comoving coordinates, then they will be stretched by the expansion of the universe.  If the network is 
thought of as composed of a fixed number of segments, the number of segments
per unit volume will be proportional to $a^{-3}$ while the length of each
segment grows as $a$ due to stretching,  and so the total energy density
will scale as 
$$ a^{-3} \times a^1 \sim a^{-2}  $$
where $a$  is the scale parameter representing the size of the universe.  
The energy density of non-relativistic matter, on the other hand, scales as $a^{-3}$,
and as the universe expands the energy in strings will grow relative to that
of matter until it eventually dominates.  As explained below,  such a
frozen network is {\it not} a typical outcome for the types of strings that have been studied to date.  Instead, there are energy loss mechanisms that allow strings to be progressively destroyed, while those that survive continue to move relativistically, their total energy scaling in the same way as that of
matter.  It has been suggested, however,
that non-Abelian strings might behave differently from Abelian ones,
and might indeed lead to a string-dominated universe. 
A universe dominated by very heavy strings is not likely to be a viable 
model, but a cosmological model with a fairly recent transition to 
a string-dominated phase with comparatively light strings has some desirable
properties.\Ref\StringDom{A. Vilenkin,  Phys. Rev. Lett. {\bf 53} (1984), 1016.}
Strings could serve as an interesting form of dark matter: giving 
density parameter  $\Omega \simeq 1$
as required by inflation, while $\Omega_{matter} <  1 $, and possibly resolving 
the apparent discrepancy between estimates of the age of the universe from 
the expansion rate and from stellar ages. 
  By modifying the equation of state of the 
universe, they could mimic some of the effects of a cosmological constant.
There has been a recent revival of interest
in such a scenario,\Ref\Pen{D. Spergel and U.L. Pen,  preprint astro-ph/9611198}    and testing its consistency is one of the chief motivations for the work described in this paper.

The evolution of Abelian cosmic strings has been studied extensively, and we review
some of the salient points here in order to point out contrasts with the 
non-Abelian case.
The simplest and best understood type of cosmic strings are those which 
occur in the Abelian Higgs model \Ref\NielsenOleson{H.B. Nielsen and P. Oleson, Nucl. Phys. B 61, 45 (1973).} and are classified by their integer winding number (an element of $Z$).  We shall refer to these as Z strings.\footnote{*}{These
are often referred to in the literature as $U(1)$ strings since they result
from the complete breaking of a $U(1)$ gauge group, but it seems more
logical to call them by their topological classification,  as is usual
for other types of strings.}  Assuming that only
 the strings with unit winding number
are stable  (type II strings), they can be formed either as infinite strings
or closed loops.  Monte Carlo simulations\Ref\abelianIC{ T. Vachaspati and A.~Vilenkin, Phys. Rev. D 30, 2036 (1984);  M. Hindmarsh and K. Strobl, Nucl. 
Phys. {\bf B 437} (1995), 471.} indicate that the infinite
 strings
constitute a majority (63-80\%) of the total length of the network initially formed by a 
phase transition, with the remainder
comprising a ``scale-invariant'' distribution of closed loops.
A very important dynamical process of these strings is intercommutation, \REF\intercomShellard{E.P.S. Shellard, in {\it Cosmic Strings: the Current
Status}, eds. F.S. Accetta and L.M. Krauss (Singapore, World Scientific,
1988), pp. 25-31} \REF\intercomNumerical{E.P.S. Shellard, Nucl. Phys. {\bf B283}
(1988),264; K. Moriarty, E. Myers and C. Rebbi, Phys. Lett. {\bf 207B} (1988),411,
and in  {\it Cosmic Strings: the Current
Status}, eds. F.S. Accetta and L.M. Krauss (Singapore, World Scientific,
1988), pp. 11-24;  R. Matzner and J. McCracken, ibid. pp. 32-41} \refmark{\Kibblecosmic, \intercomShellard,
\intercomNumerical}
the process in which two colliding strings reconnect as
in Figure  \FIG\IntercommuOne ~\IntercommuOne \/. 
Two infinite strings intercommuting with each other twice (or a single 
string intercommuting with itself)  can form new closed loops from pieces
of the infinite strings, and this is thought to be the principal mechanism
for the destruction of infinite strings.

\insertnarrowfigpage{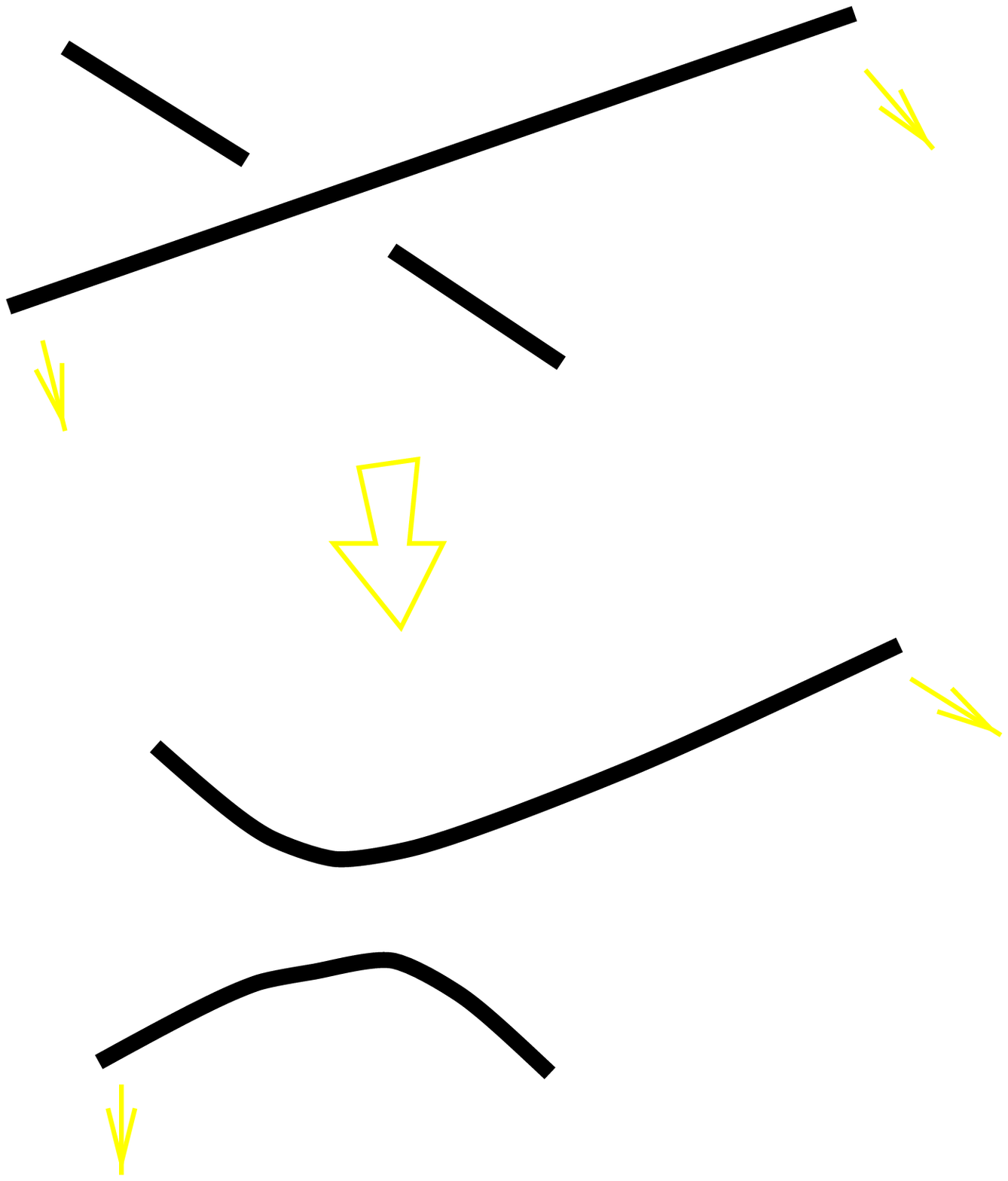}{\bcap\IntercommuOne\/ Intercommutation
of colliding strings.}  

Dimensional analysis and energy conservation arguments \refmark{\Cosmology,  \Zevarg} provide a plausible picture of the strings' evolution: 
   Infinite strings are destroyed by intercommutation
and thus lose energy to the network of loops.  Loops, in turn, shrink as
they lose their energy
to gravitational radiation, and they may also split into smaller
loops by intercommutation.  Any loop will eventually contract to a point 
and be destroyed.  The system reaches a dynamical steady state, or self-similar evolution, characterized by a single evolving length scale which is the 
size of the cosmological horizon.   The rate at which new strings appear
within the growing horizon is balanced by the rate at which they are destroyed.  An approximately constant number (order
unity) of long strings stretch across a Hubble volume at a given time,  and 
intercommutations result in the formation of a similar number of new closed loops per Hubble volume,  which are destroyed within a Hubble time.  
While there is disagreement over details,  numerical simulations 
\Ref\Znumerical{A. Albrecht and N. Turok,  Phys. Rev. Lett. {\bf 54} 1868 (1985);  Phys. Rev. {\bf D40} 973.  D.P. Bennet and F.R. Bouchet,
Phys. Rev. Lett. {\bf 60} 257 (1988), {\bf 63} 2776 (1989);
Phys Rev. {\bf D41} 2408;  B. Allen and A.P.S. Shellard, Phys. Rev. Lett. {\bf 64} 119;  B. Allen and A.P.S. Shellard in {\it The Formation and Evolution of Cosmic Strings} eds. G. Gibbons, S. Hawking and T. Vachaspati, (Cambridge University Press, 1990) pp. 421-448} have 
supported this general picture.  All results are consistent with a scaling 
behavior in which 
 the energy density in strings remains a small, fixed fraction of the 
matter density and there is no string domination.  

We emphasize that intercommutation
is crucial to this picture of the strings' evolution.  Since 
non-Abelian strings in general cannot intercommute, we might expect
different behavior for such strings.

Somewhat less attention has been devoted to the evolution of branched 
networks, in which several strings may join at a vertex.  Branched networks
occur when  a $U(1)$ gauge group is spontaneously broken to $Z_N$ with
$N\geq 3$, or when the unbroken group is a non-Abelian discrete group.
Among the work that has been done is that of 
Vachaspati and Vilenkin \Ref\VV{T.~Vachaspati and A.~Vilenkin, Phys. Rev. D 35, 1131 (1986)} who considered a network of $Z_3$ strings, which have the novel feature that three strings may intersect in a vertex.  $Z_3$ strings tend to form an infinite network of 
vertices connected by string segments, with very few closed loops. 
It had previously been speculated that the nodes in a branched network
could settle to equilibrium positions, thus causing the network to freeze 
as a string-dominated universe requires, but the simulations of [\VV] indicated otherwise.  Instead of reaching an equilibrium, the nodes pull together and
annihilate,  steadily reducing the number of nodes and strings.
The annihilation of vertices leads to a self-similar scaling behavior,
as long as the nodes are able to come close enough to each other to
annihilate.  
The phenomenological consequences of their model are rather similar to 
those of $Z$ strings:  The number of vertices and string segments
per horizon volume remains roughly constant,  and the energy density of the 
network is a small constant fraction of the matter density.  The strings 
never relax to an equilibrium, but continue to move with relativistic 
transverse velocities, following a self-similar evolution pattern much like 
that of $Z$ strings.

The purpose of this paper is to explore the consequences of a network of 
non-Abelian strings.   Such strings are known to exhibit a number of exotic types of interactions\REF\Bais{F. Bais, Nucl. Phys. {\bf B170} [FS1] (1980), 32.}\REF\various{M. Alford, K-M Lee, J. March-Russell and J. Preskill, Nucl. Phys. B 384, 251 (1992),  and references therein.}\refmark{\Bais,\various}.  Particularly significant is the fact that when two non-Abelian strings cross each other, they cannot generally intercommute, nor can they pass through one another without forming new vertices and becoming joined by a new segment of string.\Ref\nocross{G. Toulouse, J. Physique Lett. {\bf 38} (1976),
L-67;  V. Poenaru and G. Toulouse, J. Physique {\bf 38} (1977), 887.}  Linked loops of string cannot usually become unlinked, and vice versa.  One might expect that this would inhibit the decay of a cosmic network by obstructing the removal of string segments.  If new strings are continually being formed through
string collisions,  their energy must come from the already existing 
strings,  and an equilibrium with the strings' transverse velocity
damped out might seem a more likely final state.  Some evidence for slowing down of the network's destruction was reported very recently by Pen and Spergel 
\refmark{\Pen} in a class of models with non-Abelian global strings.   

In this paper, we describe a numerical simulation of a network of $S_3$ strings.
Unlike the authors of reference [\Pen],  we consider gauge strings, which have
no mutual long-range interactions.   Our interest is in understanding the 
qualitative nature of the fate of a non-Abelian string network.  Do collisions
result in a non-diminishing or rapidly increasing number of strings?  
A static equilibrium state which is expected to be conformally stretched 
with the universe's expansion?  Does the network instead decay rapidly into
finite networks and closed loops?  Or does a dynamical self-similar evolution
emerge, as in [\VV]?  
Another question which we hope to illuminate is:  which processes play the most crucial roles in the 
network's evolution?  The importance of intercommutation to the evolution 
of $Z$ strings led to fundamental study of the dynamics of intercommutation;
likewise it is hoped that the results obtained here will suggest which aspects of 
non-Abelian string dynamics are ripe for closer examination.
   
Our method of
simulation is directly inspired by that of reference [\VV]:  we generate initial conditions from a lattice Monte Carlo simulation and then evolve the network according to a highly simplified model of string dynamics which we hope captures the essential features of a string network losing energy.  We find hints of
some quite interesting physics in the interplay between the two types of string
in our model, and a rather surprising dependence of the network's behavior on
the initial conditions.  Concerning the string-dominated cosmological scenario, we reach 
somewhat different conclusions from [\Pen].   For a wide range of conditions, the network's
density follows power laws much like the ones arising in
Abelian networks.  Non-abelian effects can, indeed, slow down the 
network's decay in the sense that they change the coefficients of the power laws, but it appears as if the slowing
down to a stable equilibrium happens only under special circumstances, if at 
all.  

The remainder of this paper is organized as follows:
Section {2} provides a brief summary of the properties of non-Abelian strings
which bear on this simulation.   We discuss some of the subtleties inherent in the description of non-commuting magnetic fluxes, and the necessity of a (gauge fixing) convention to resolve these ambiguities and allow the 
comparison of fluxes of strings. 
Most importantly, we explain why two colliding non-Abelian strings cannot, in 
general, cross or intercommute without forming a new segment of string.  
Section {3} describes the particular $S_3$ model which we have chosen to
simulate.  $S_3$, the permutation group on 3 elements, was chosen as
the gauge group because it is a simple non-Abelian group which  exhibits
all of the important general characteristics
of non-Abelian strings.  Another motivation for this choice
is that $S_3$ contains
$Z_3$ as a subgroup, allowing instructive comparisons with the $Z_3$ network simulations 
of ref. [\VV] .  Section {3} also describes our procedure for simulating the
network's
dynamical evolution, giving enough details
  to allow an understanding of the results.  Additional technicalities
of the procedure are relegated to the appendix. 
Section {4} describes our procedures for generating initial conditions,
and summarizes features of the networks these procedures generate.
We use two different Monte Carlo algorithms which generate initial networks with
somewhat different statistical properties.  The network's evolution turns
out to have a surprisingly strong dependence on the initial conditions.
Section {5} presents results of the dynamical evolution simulation, exploring the 
influence of a number of different variables including the initial conditions
of the network and the ratio of string tensions.  These results are compared with those for a 
$Z_3$ network, which is Abelian. In section {6}, we present our conclusions and suggest 
directions for future work.

An appendix describes
our procedure for keeping track of string fluxes during the simulation
and for establishing them from the lattice Monte Carlo procedure, covering
details not included in section {3}. 
The 
implementation of non-Abelian fluxes in a simulation presents a 
rather difficult problem in its own right.  A careful gauge-fixing 
procedure is required, and some of the subtleties that arise are of 
interest from a field-theoretic point of view.  The algorithm has
been described in greater detail in reference  \REF\mcgraw{P. McGraw, preprint CALT-68-2044, hep-th/9603153;  ``Dynamics of Non-Abelian Aharonov-Bohm Systems,''  Ph.D. thesis, Caltech, 1996. } [\mcgraw].

In this discussion, the strings will be considered as classical objects with
well-defined fluxes (after a gauge has been fixed).  We will not consider quantum-mechanical effects such as
Cheshire charge.\Ref\Cheshire{M. Alford, {\it et al.}, Phys Rev. Lett. 64, 1632 (1990); 65, 668 (E);  Nucl. Phys. B349, 414 (1991) }

\section{Vortices and strings in a non-Abelian discrete gauge theory}
In this section, we review very briefly the definition of
non-Abelian vortices and strings and some of the properties which are important for the current simulation.  The formalism used here was developed for vortices in \REF\Bucher{M. Bucher, Nucl. Phys. {\bf B350} (1991) 163.} [\Bucher] and applied to strings in [\various].

Generically, topological defects of codimension 1 (vortices in two space
dimensions, strings in three)  occur when a gauge symmetry group $G$ is 
spontaneously broken to a subgroup $H$ such that there are non-contractible
closed loops in the vacuum manifold $G/H$.  This is formally expressed by saying
that $\pi_1(G/H)\neq 1$, where $\pi_1$ stands for the first homotopy
group.\REF\homotopy{For more discussion of homotopy groups and their application to topological defects, see:  M. Nakahara, {\it Geometry, Topology, and Physics} (New
York, Adam Hilger, 1990); N.D. Mermin, Rev. Mod. Phys. {\bf 51} (1979), 591.}
\refmark{\Kibblecosmic, \homotopy}  Each element of $\pi_1$ represents a closed path (or, more
precisely, a class of closed paths.)  Paths can be composed by tracing
first one path and then another, giving the group structure of $\pi_1$.
For example, if a $U(1)$ gauge group is broken completely, then $H=1$,
and $\pi_1(G/H)=Z$, the group of the integers.  Closed paths in the 
vacuum manifold are classified by an integer winding number.
 If $G$ is simply connected, then $\pi_1(G/H)\sim \pi_0(H)$, where
$\pi_0$ is the set of disconnected components,  so that strings occur whenever
H is discrete or has components disconnected from the identity.  
We will subsequently assume $G$ is simply connected unless otherwise stated.
A further simplification is when $H$ is discrete,  in which case 
$\pi_0(H)\sim H$.
If $\pi_0(H)$ is 
non-Abelian, then the composition of paths
depends on the order.  Hence the fluxes of strings will be non-commuting 
group elements, and that is what is meant by non-Abelian strings.

The breaking of an underlying $G$ gauge theory to a discrete group $H$ leaves no light propagating gauge fields:  At low energies, in any simply connected region without defects,
the gauge field $A^a_\mu$ is pure gauge.  However, when string defects are present, the
region of true vacuum is not simply connected:  it is 
${\cal R} = {\cal M} - \{D\}$, where
$\{D\}$ is the union of all defect cores (regions of non-vacuum)
 and ${\cal M}$ is the spatial manifold on which the defects exist.  Each string
gives rise to a class of noncontractible closed paths in ${\cal M} - \{D\}$ which encircle the string.  The flux enclosed by any closed loop $\Gamma$ is a group element defined as
a path-ordered exponential of the gauge field:  

$$ {\rm flux} = P \exp ( \oint_\Gamma A\cdot d\ell ).\eqn\pexp$$ 
For any $\Gamma$ within ${\cal R}$  this must be an element of $H$.  This is because the Higgs field is covariantly constant throughout ${\cal R}$ and so the
transformation that results from parallel transport around a loop must leave the
Higgs field invariant.  

The flux of a cosmic string is defined by the above exponential along a path which winds around the string.  In a non-Abelian theory, this definition of the
flux is not gauge invariant, and may depend on the point at which the path 
begins and ends.  However, the flux through any loop
which does not enclose a string is necessarily trivial. A corollary of this fact is that 
two closed loops which share the same beginning and ending point $x_0$, and can be continuously deformed into each other,   have the same
flux.  The relevant structure for the description of the system of defects is
the fundamental group or first homotopy group  $\pi_1({\cal M} - \{D\},x_0)$,
defined with respect to a basepoint $x_0$.  Each string is associated with a generator of the fundamental group.  Once $x_0$ has been (arbitrarily) chosen,  the fluxes of all closed paths (and of all strings) are specified by a homomorphism
from  $\pi_1({\cal M} - \{D\},x_0)$ into $H\sim \pi_1(G/H)$.  The only remaining gauge freedom
is a global one.  However, there is a considerable amount of ambiguity in what we mean by ``the flux'' of one particular string:  an arbitrariness in how exactly
 the set of generators is chosen for the homotopy group
$\pi_1({\cal M} - \{D\},x_0)$.  In figure \FIG\Genambig ~\Genambig \/, 
for example, there are two loops, both beginning and ending at $x_0$, both enclosing the same string without enclosing any others, which are nonetheless
representatives of different homotopy classes (and consequently may be associated with different fluxes).  An intervening string prevents one
path from being continuously deformed to the other.  The fluxes associated with
the two different paths may differ through conjugation by the flux of the other
string.  In an Abelian theory, conjugation is trivial;  not so in a non-Abelian one.  It follows that fluxes cannot meaningfully be compared (say, to determine
if they are the same) if the paths used to define those fluxes pass on opposite
sides of some other string.  Comparisons must be made using ``nearby'' paths.

Since the flux of the same string may described,  depending on convention, by distinct conjugate elements of $H$,
it follows that strings whose fluxes are in the same conjugacy class must 
be degenerate in tension.  It is not true, however, that all fluxes in the same conjugacy class are the same.  
If $a$ and $b$ belong to the same conjugacy class, it is by no means guaranteed
that $a^2$ and $ab$ also are conjugate,  or that $ab=ba$.
Distinctions
among elements within the same conjugacy class can have important consequences
 in any situation where the product or commutator of two 
fluxes is relevant, as is the case when two strings collide.

\insertxfigpage{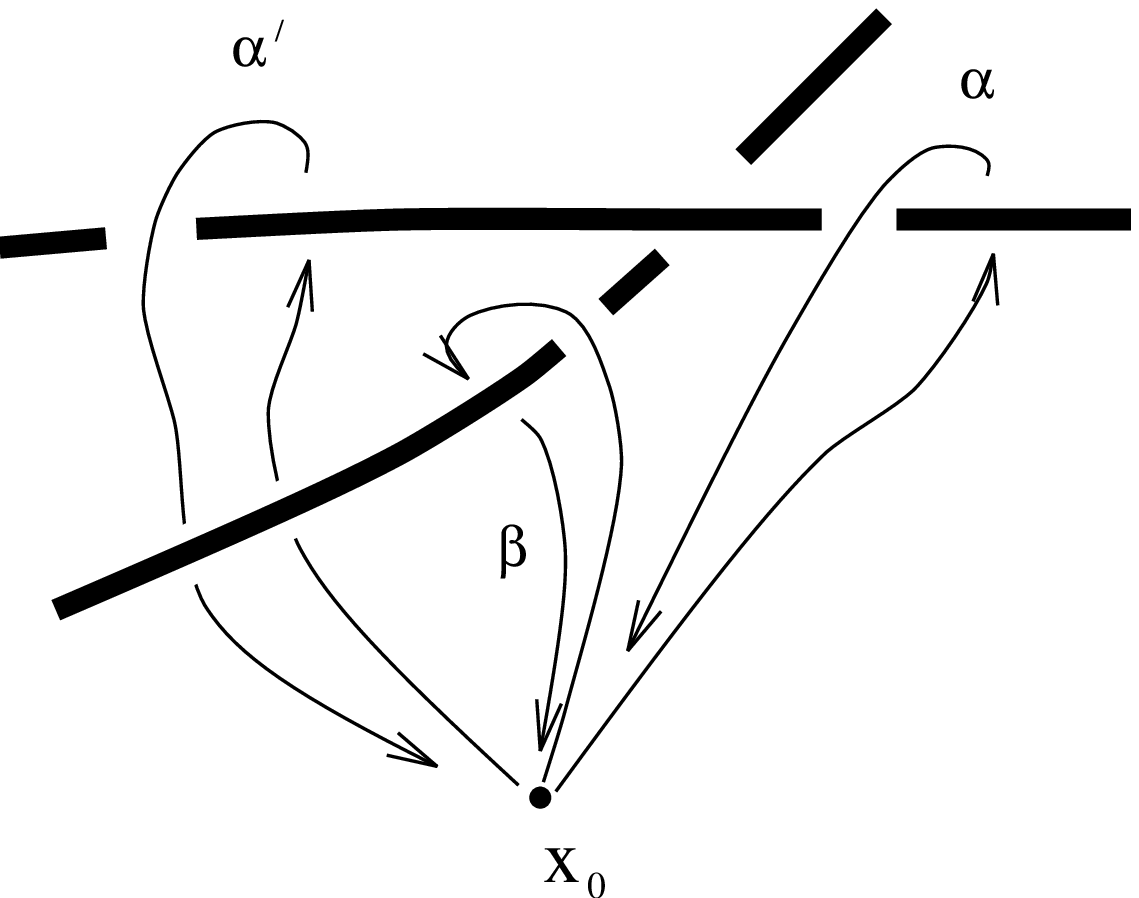}{\bcap\Genambig \/  The paths $\alpha$ and 
$\alpha '$  both enclose the same string and no other strings,  but they cannot
be continuously deformed into each other without crossing another string.  Thus,
they represent different elements of the fundamental group
$\pi_1({\cal M} - \{D\},x_0)$, and so the fluxes associated with them may be 
different.  Specifically, the homotopy classes of $\alpha$ and $\alpha '$ are related through conjugation by another generator: 
$\alpha' \sim \beta\alpha\beta^{-1}$.  (We follow the usual convention of
composing paths from right to left:  $\beta\alpha\beta^{-1}$ means the path formed by traversing first the reverse of $\beta$, then $\alpha$, then $\beta$.
The relation $\sim$ represents homotopy equivalence.)
The associated fluxes are analogously related:
a nontrivial relation if the fluxes don't commute. }

The path dependence of the flux of a string implies an important fact:
two strings with non-commuting fluxes cannot pass through each other without
forming a new segment of string whose flux is the commutator of the fluxes
of the two original strings.    Penetration without the formation of a 
new string would violate flux conservation.\refmark\nocross  This is illustrated in figure
\FIG\generalNCI  ~\generalNCI \/.  Non-commuting strings also cannot 
intercommute.  We will be especially interested in the consequences of
this entanglement process for the evolution of a string network:  it might 
impede the collapse of the network.

\insertfigpage{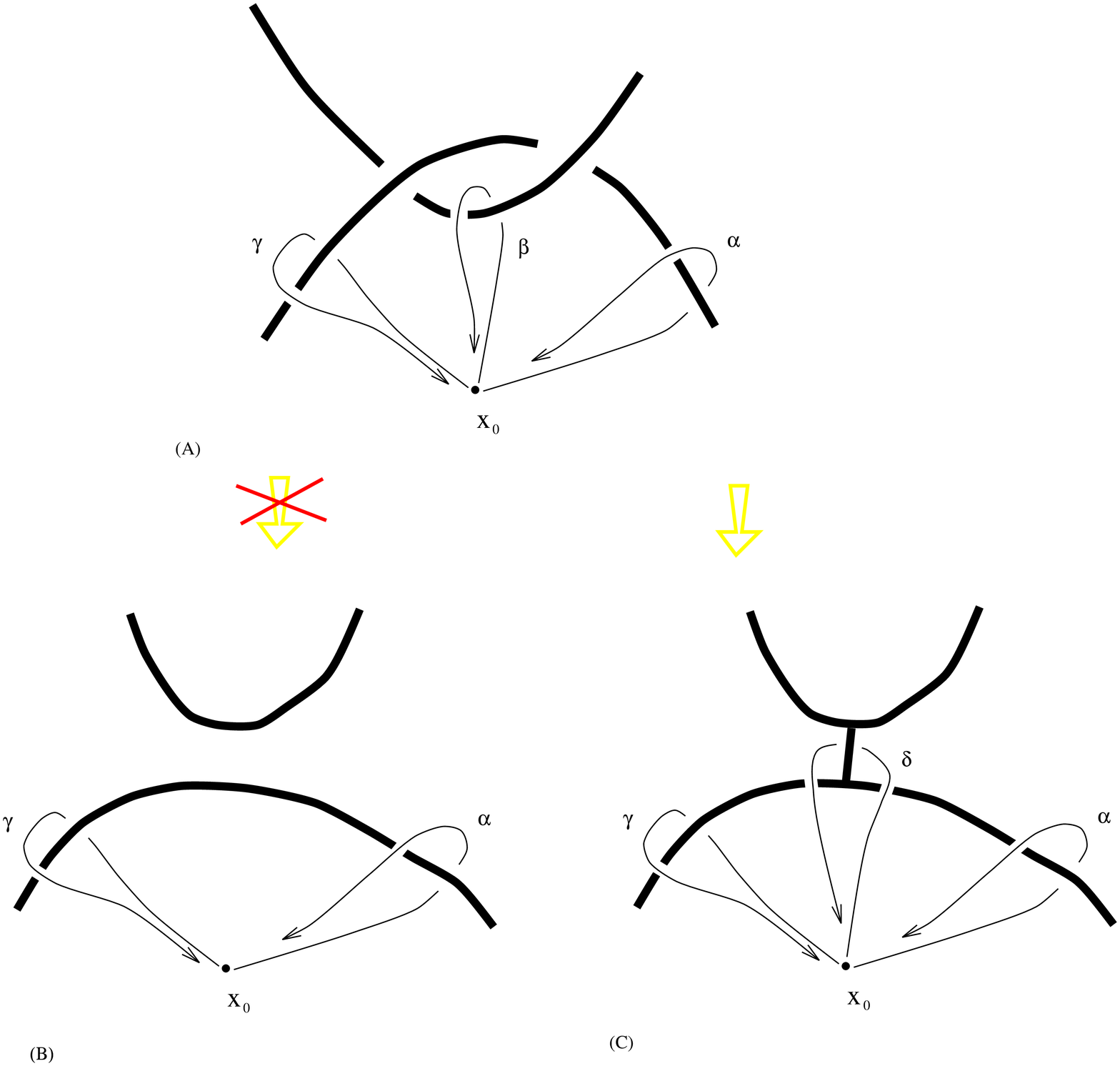}{\bcap\generalNCI \/ Attempt to pass two strings through each other. In (A) the flux of one string may be defined by either of 
the paths $\alpha$ or $\gamma$, and that of the other string by $\beta$.  Let
the  fluxes associated with $\alpha$, $\beta$ and $\gamma$ be $a,b,$ and $c$
respectively.  In this case, $c=bab^{-1}$.  In general, $c\neq a$. Now, if we attempt to pass the strings through each other,  no strings need cross the paths $\alpha$ and $\gamma$,  so the associated fluxes will not change.  But if the strings were to pass through each other freely, as in (B),  $\alpha$ and $\gamma$  would be continuously deformable into each other.  This is impossible
if they have different fluxes.  In order to conserve flux,  the string must branch somewhere and be connected to the other by a new string whose flux
as defined by path $\delta$ in (C) is 
$ca^{-1}=bab^{-1}a^{-1}$.}

\section{Our Model and its Dynamics}
\subsection{$S_3$ strings}

We consider here a model with unbroken gauge group $H=S_3$, the permutation group
on three objects.  The spectrum of this model will include strings  whose fluxes
are elements of $S_3$.  $S_3$ has six elements in all.  The identity $e$ corresponds to the trivial permutation.  There are three odd permutations (two-cycles or transpositions) each leaving one of the three elements invariant and interchanging the other two.   We may denote these, for convenience, by:
$t_1\equiv \{(123)\to (132)\}, t_2\equiv\{(123)\to (321)\}, t_3\equiv\{(123)\to (213)\}$.  In this notation, $t_i$ is the two-cycle which leaves the $i$-th element in the same position.  The two
non-trivial even permutations are the three-cycles, or cyclic permutations,
which we denote here by  $s_+\equiv\{(123)\to (312)\}, s_-\equiv\{(123)\to (231)\}$.  In the more conventional cycle notation,\Ref\Herstein{I.N.~Herstein, {\it Abstract Algebra} (Macmillan, New York 1986).}  we 
have $t_1=(23), t_2=(13), t_3=(12), s_+=(123), s_-=(132)$. 

The 2-cycles form one of the two non-trivial conjugacy classes, and the 3-cycles form another.
Thus our model supports two types of strings, which we shall refer to as odd and even 
strings, or alternately as $t$ and $s$ strings.  The three-cycles generate a $Z_3$
subgroup, so that  our model contains the $Z_3$ model as a subset. Three even strings may meet at a vertex, just as in the $Z_3$ model.  Another type of junction is one where
two 2-cycle (or odd) strings merge to form a 3-cycle (even) string.  
 Figure
\FIG\junctions ~\junctions \/ shows the two types of junctions in our model.

Since each two-cycle is equal to its inverse, oppositely oriented  odd strings are topologically equivalent.  An even string, on the other hand, possesses a natural orientation:  The flux through a path encircling it with one orientation is $s_+$, while it is $s_-$ for the opposite orientation.  In subsequent figures,  even strings will often be denoted by oriented lines, with the string carrying
flux $s_+$ in the direction of the arrow, while odd strings have no arrow, reflecting their lack of orientation. 

\insertfigpage{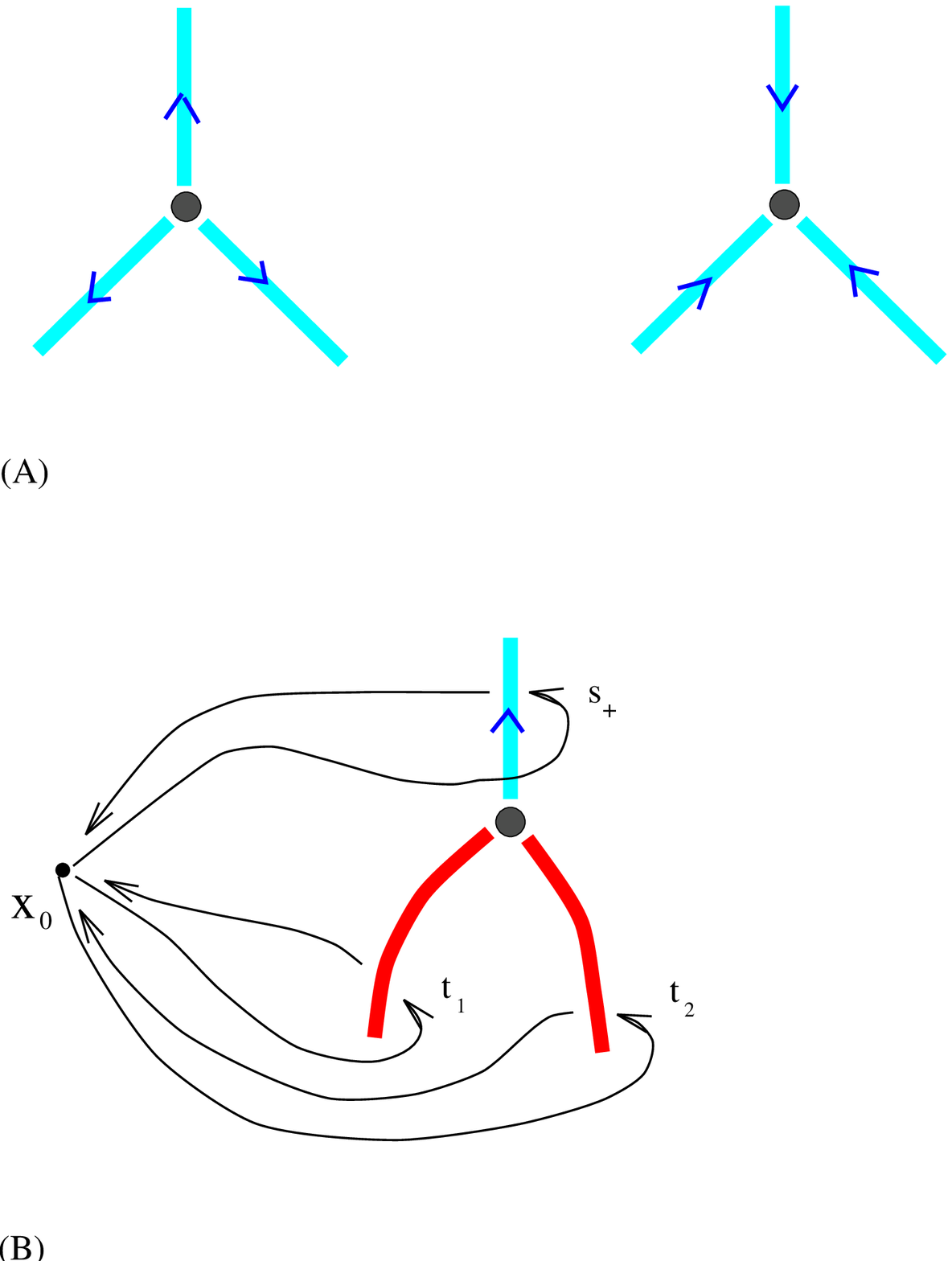}{\bcap\junctions \/ String junctions in the $S_3$ model.  (A) Two possible $sss$ junctions:  three strings with the same flux, $s_+$ or $s_-$, emanate from the node. (Or two $s_+$ strings merge into a single $s_-$, etc.)  (B) One of the class of $stt$ junctions:  Two $t$-strings merge into an $s$-string.  Fluxes are defined with respect to a basepoint 
$x_0$ by the paths shown.  Here, as in many subsequent figures, an $s$-string is drawn as an oriented line.  The string carries flux $s_+$ in the direction of the arrow: i.e., a counterclockwise path around the arrow encloses flux $s_+$}

Note that the parity is conserved at any junction: i.e., an odd string 
entering a junction must be matched by another odd string leaving.  
In this sense,  odd strings can never end, even if they change their flux 
at a junction.   Even strings, on the other hand,  may end at a junction.
It is helpful to view the network as being composed of two interacting
subsystems.  One subsystem consists of infinite or closed t-strings (like
$Z$ strings, they have no free ends).  The 
other comprises a branched Abelian web of s-strings, some of which end on 
t-strings.  The simulation results presented later in this paper exhibit
an interesting interplay between these two subsystems.

\subsection{String Dynamics}

The system we simulate consists of three-way junctions, or nodes, joined by
 $s$ and 
$t$ strings.  The strings are approximated as straight segments between
junctions.  In effect, we are averaging over transverse oscillations of the
strings.  However,  it is possible for a string segment to be interrupted
by a pair of nodes doubly linked to each other as in figure
\FIG\doubleone ~\doubleone.
Doubly linked nodes tend to annihilate fairly quickly, and such configurations
play the role of transient kinks on otherwise straight segments.

\insertnarrowfigpage{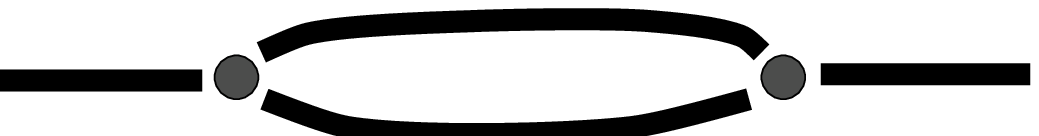}{\bcap\doubleone  \/  Doubly linked nodes.
A pair like this can interrupt an otherwise straight string or act as a 
kink.}

As in [\VV],  the nodes are assumed to undergo damped motion under the
influence of string tensions.  The energy loss leads to the shortening of strings.  As an approximation for the energy loss of a real string 
network,  the model of damped motion of the nodes is most realistic if one supposes that the string junctions are monopoles which carry some unconfined magnetic flux,  as in
a model with the symmetry-breaking pattern
$$G->{{S_3 \times (SU(3)\times SU(2)\times U(1)_{EM}})\over D},$$ 
where 
D is a discrete factor divided out so that the monopoles at string junctions may carry electromagnetic $U(1)$ charge.  Their magnetic charges should then result in radiation damping.  \footnote{*}{A model has been constructed in which
topological $Z_n$ strings become attached to monopoles which also carry other charges.\Ref\coloredmonopoles{M. Hindmarsh and T.W.B. Kibble, Phys. Rev. D 55, 2398 (1985)}  Constructing a model with $S_3$ strings joined at monopoles might
be slightly harder, but it is not our main concern here.  For a hint of how such
a model could arise,  consider
the monopoles that form when an $SU(5)$ group is broken in the familiar way to $SU(3) \times SU(2) \times U(1)/Z_6$.    This transition is
known to yield stable monopoles with $SU(3), SU(2),$ and $U(1)$ flux.\REF\monopoles{M.Daniel, G.Lazarides, and Q. Shafi, Nucl. Phys. B170, 156
(1979)} \REF\SUfivebreak{J. Preskill, Ann. Rev. Nucl. Part. Sci., 34,  461 (1984)}\refmark{\monopoles,\SUfivebreak}  We could imagine a second symmetry-breaking stage in which the 
$SU(3)/Z_3$ factor is broken down to $S_3$ in such a way that the resulting srings also carry nontrivial flux in the $Z_2$ center of $SU(2)$.  Whenever
three such strings join, the resulting net  $Z_2$ flux can unwind through a monopole, which 
has both $SU(2)$ and $U(1)$ flux.}   

Our simulation proceeds in discrete time steps.  During each time step, each node is moved by a displacement proportional to the vector sum of all tensions acting on it.  This type of evolution corresponds to damped motion: ${\rm Force} \prop {\rm Velocity}$. The constant of proportionality is a parameter which may be absorbed into the size of the time step.  Thus, in appropriate units,
$$\Delta {\bf x} = \Delta t \sum_r T_r {\bf n}_r, \eqn\timestep$$ 
where $\Delta t$ is the time step,
 ${\bf n}_r$ is the unit vector along the direction of the $r$-th string connected to the node, and $T_r$ is the magnitude of that string's tension.  $r$ runs from $1$ to $3$  for the three strings that meet
at each node.  
Since our model has two types of string with possibly different tensions,  the ratio
of these two tensions is a variable parameter of the simulation.  When we 
present the results in section 6,  we use units such that the lattice spacing
of the initial-condition Monte Carlo algorithm is 1 and the tensions are
of order unity. (More specifically, the $t$-string tension is normalized to
$1$ while the other tension is varied.)

Each node is moved in turn.  During the motion of a node,  the strings attached
to it may collide with other strings.   The annihilation of two nodes is 
allowed if they approach each other more closely than a certain distance
$d_{ann}$.   The procedures for handling collisions of strings or nodes are 
as follows:\footnote{\dagger}{Some of these procedures differ in minor
details from those described in [\mcgraw].}

i)  {\bf{Intercommutation}:}  If, in the process of moving a node from its initial to
final position, one of its string segments intersects some other segment, then the fluxes of those two segments are compared at the point where the crossing occurs.  If the fluxes commute, then the two segments may either pass through each other unaltered, or intercommute.  The probabilities of these two outcomes may in principle be taken as an adjustable parameter of the simulation, but
we have chosen to let the probability of intercommutation be $1$ in all cases.  It is widely believed that intercommutation is generically the more common outcome whenever two cosmic strings cross.  In the self-dual limit, it can be shown that Nielsen-Oleson strings always intercommute.  Therefore the choice to 
set the intercommutation probability to 1 seems a natural one.

 Intercommutation may occur in two possible situations:  either both strings are 3-cycle ($s$) strings, or both are 2-cycles.  In the latter case, the fluxes of the two strings must in fact be equal.  In an intercommutation, the string ends are rearranged in such a way as to conserve flux.  In the case of two $s$-strings, there is always only one way to rearrange the ends, as shown in figure \FIG\intercom ~\intercom a.  A string end carrying flux $s_+$ to the point of intersection may not be joined to one carrying the inverse flux $s_-$.    When two $t$-strings intercommute, however, there are two possible rearrangements of the ends, owing to the fact that a 2-cycle is equal to its own inverse and 2-cycle or t-strings consequently have no preferred orientation{\footnote{\dagger\dagger}{Strictly speaking, we can only say that there is no 
{\it topological} reason for a t-string to have a preferred orientation.  It is possible that the field equations could have two distinct solutions, corresponding to differently oriented strings, which are topologically equivalent but can be deformed into one another only by surmounting a finite energy barrier.  A situation of this sort occurs in the global vortices of nematic liquid crystals.  This was pointed out to me by J. Preskill.}} (fig. \intercom b.)  In the absence of a reason to prefer one of these rearrangements over the other, the choice is made randomly.  

When an intercommutation
occurs,  we join the segments with a pair of ``kinks''  (doubly linked nodes
as in figure \doubleone)
which will later annihilate and allow the segments to straighten.  By 
delaying the straightening of the rejoined segments, we prevent the highly
non-causal cascades of intercommutations which might otherwise occasionally occur within a single time step and lead to computational infinite loops and
other unpleasant consequences.

\insertfigpage{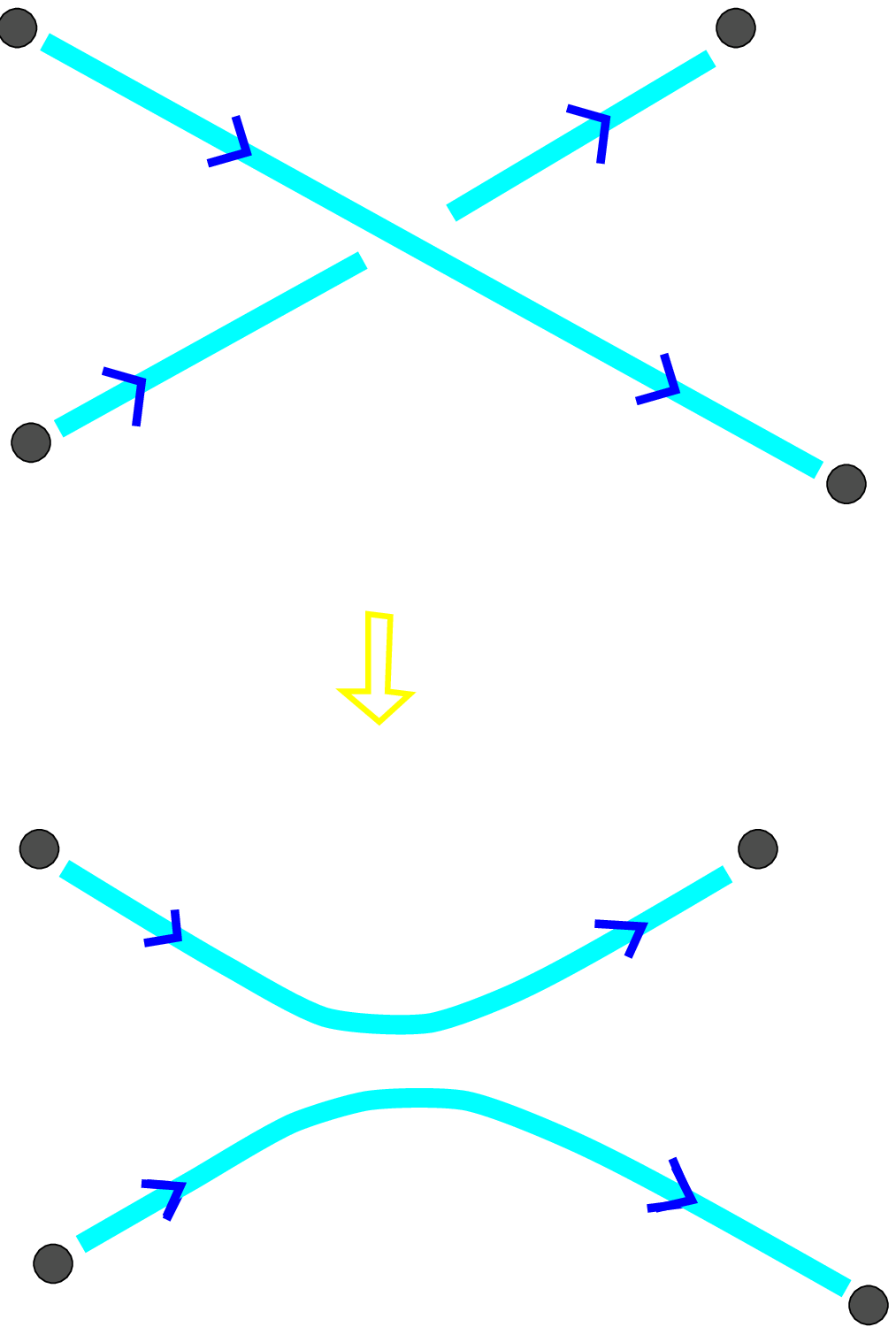}{\bcap{\intercom A}\/  When two three-cycle or $s$-strings intercommute, there is a unique rearrangement which is compatible
with the orientations of the strings.}

\insertxfigpage{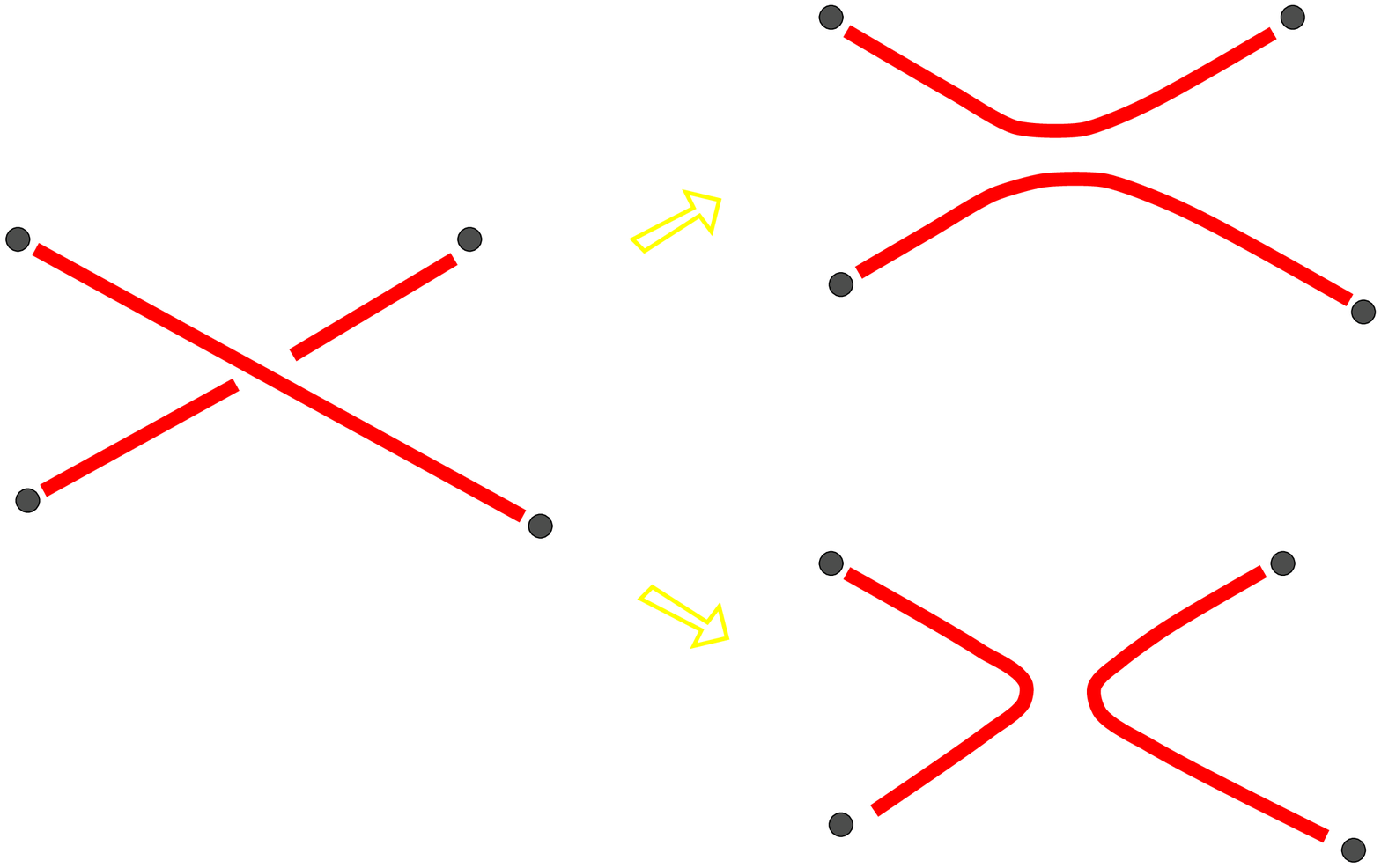}{\bcap{\intercom B}\/  Since $t$-strings (2-cycle
strings) have no orientation, an intercommutation can result in either of two possible
rejoinings of the ends.}

ii)  {\bf Non-commuting collisions (NCC):}  If two non-commuting strings intersect, then it is assumed that they form a new pair of nodes and thus become linked by a new segment.  This may happen in two possible ways.  The two
strings may pass through each other and become linked by a new segment which
stretches between them,  as shown in figure \FIG\NCI ~\NCI a.  We refer to this
outcome as the ``bridge'' configuration.     The flux of the intervening string segment is uniquely determined by the requirement of flux conservation.
(The intervening flux of the bridge must always be a 3-cycle, as the commutator subgroup of $S_3$ is $Z_3$.)  Another possible outcome, consistent with the topology, is that the two colliding segments
stick together in the ``zipper'' configuration shown in figure  ~\NCI b.  The possibility of collisions resulting in zipper formation have recently been mentioned in the 
context of type I Abelian strings\Ref\zipper{L.M.A. Bettencourt, P. Laguna,
and R.A. Matzner, Phys. Rev. Lett. {\bf 78} (1997), 2066;  L.M.A. Bettencourt
and T.W.B. Kibble, Phys. Lett. {\bf B 332} (1994), 297.}.   Which of these two outcomes is more likely is a dynamical and kinematic issue to be
addressed in future work.  It may depend on the relative velocities and
angles of the colliding strings.  In the present simulation, we consider
both possibilities separately and examine their consequences: some runs were
performed with only ``bridge'' NCC's,  and some with only ``zipper'' formation. 
In fact, we find that the choice makes little difference for the network's evolution.

 Whenever 
two new nodes are created by a NCC,  we separate them
immediately by some small initial distance comparable to the time-step  and subesquently allow them to move
normally under the influence of string tensions.  

\insertfigpage{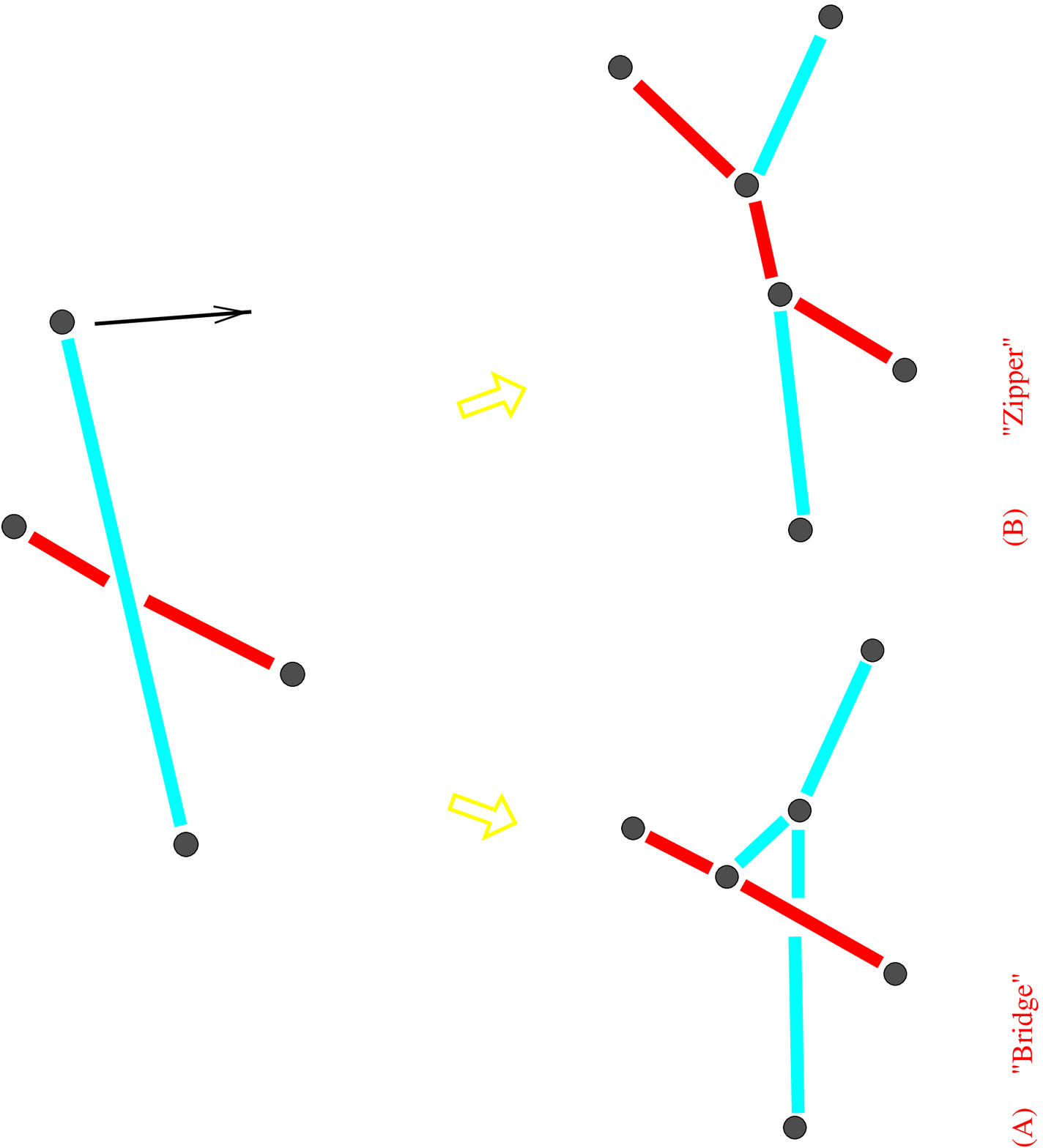}{\bcap\NCI \/ The intersection of two strings whose fluxes
do not commute causes them to become linked by a new segment in the 
``bridge'' configuration (A). Alternatively, they may 
coalesce along part of their length, forming a ``zipper'' (B).}

\FIG\doublelink

\insertnarrowfigpage{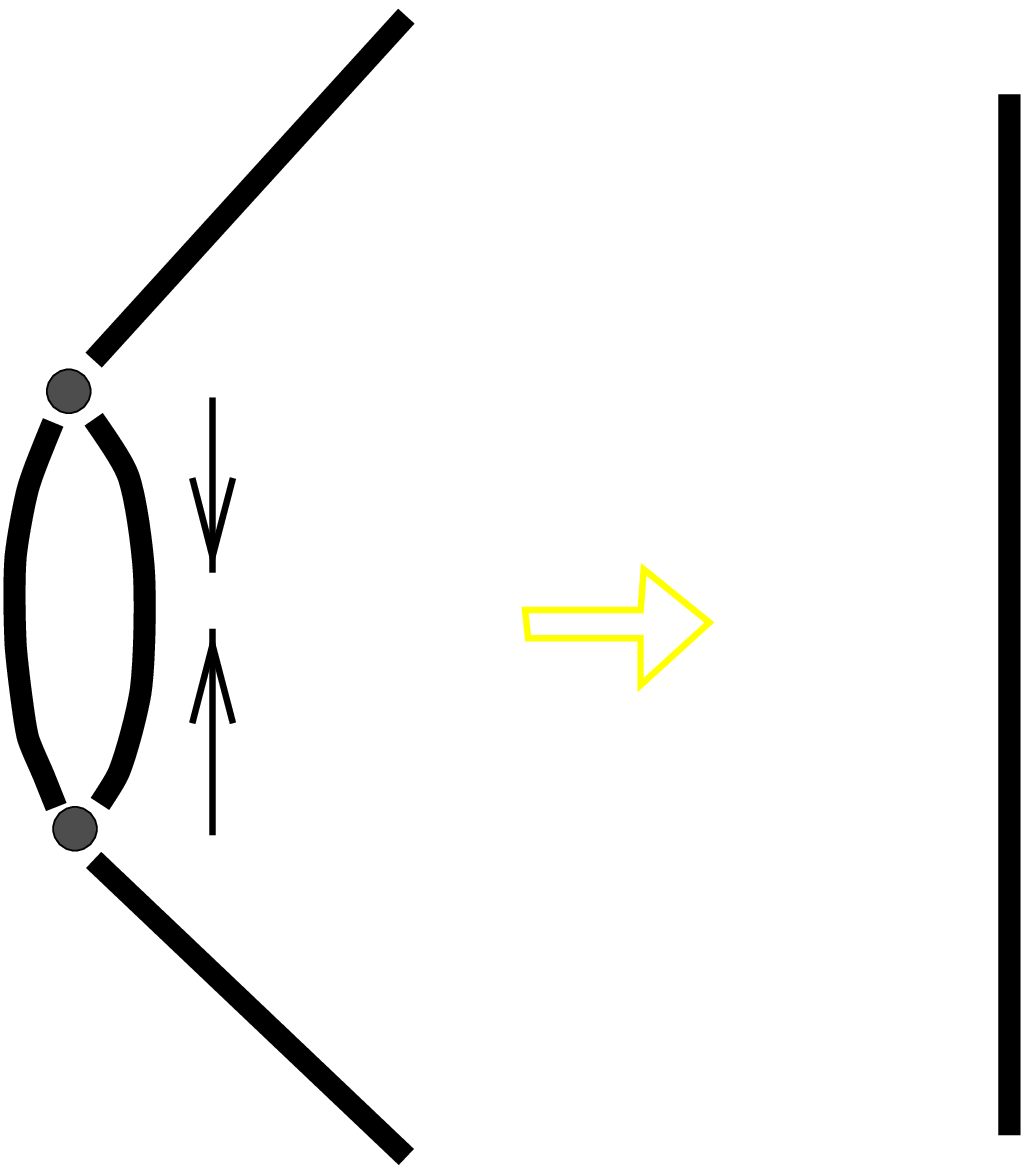}{\bcap\doublelink \/  A pair of nodes linked by
two strings may annihilate, leaving a single string.}

\FIG\sssannih

\insertxfigpage{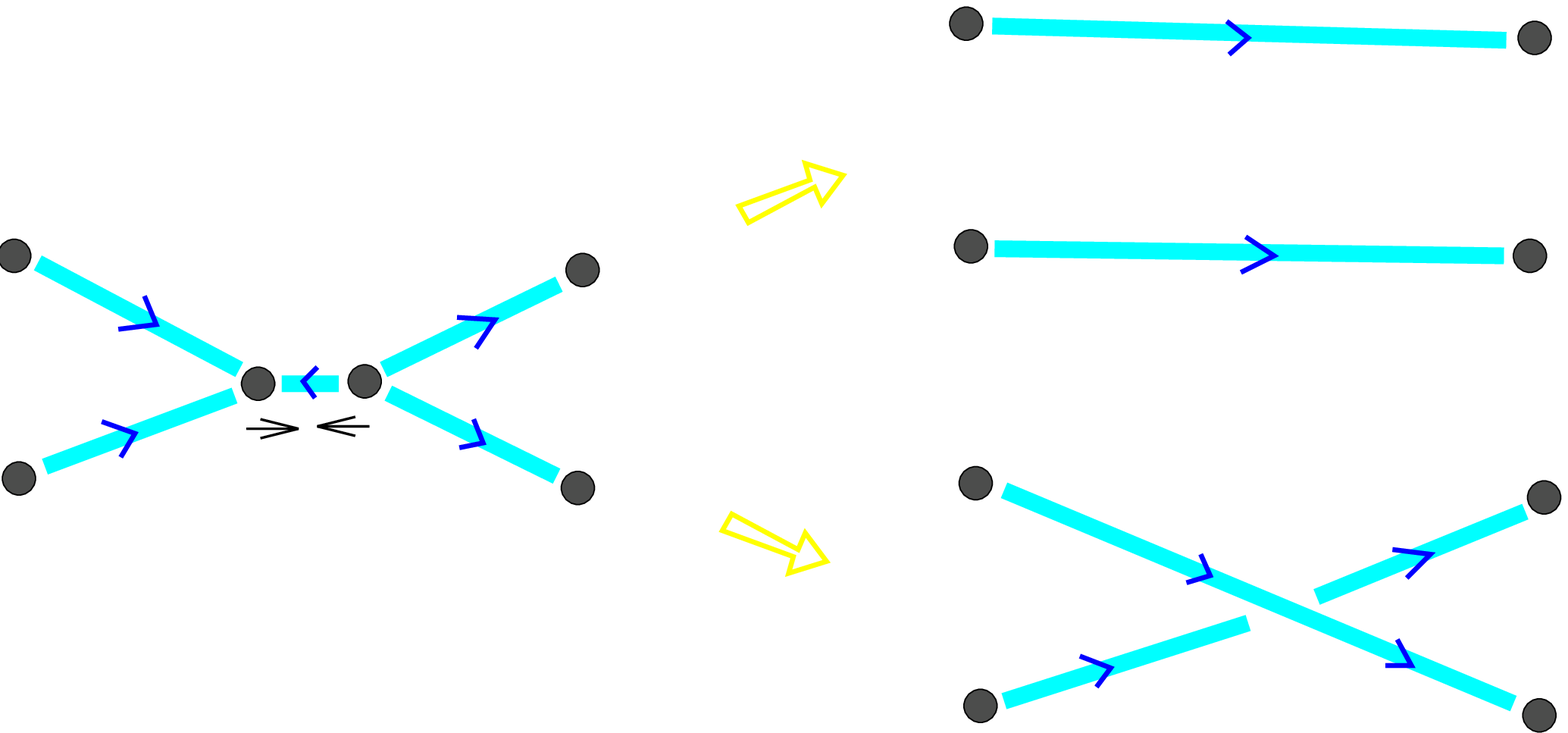}{\bcap\sssannih \/ Annihilation of two $sss$ nodes
joined by a single string.  There are two possible ways to reconnect the strings
consistent with their orientation.}

\FIG\cantannihilate

\insertxfigpage{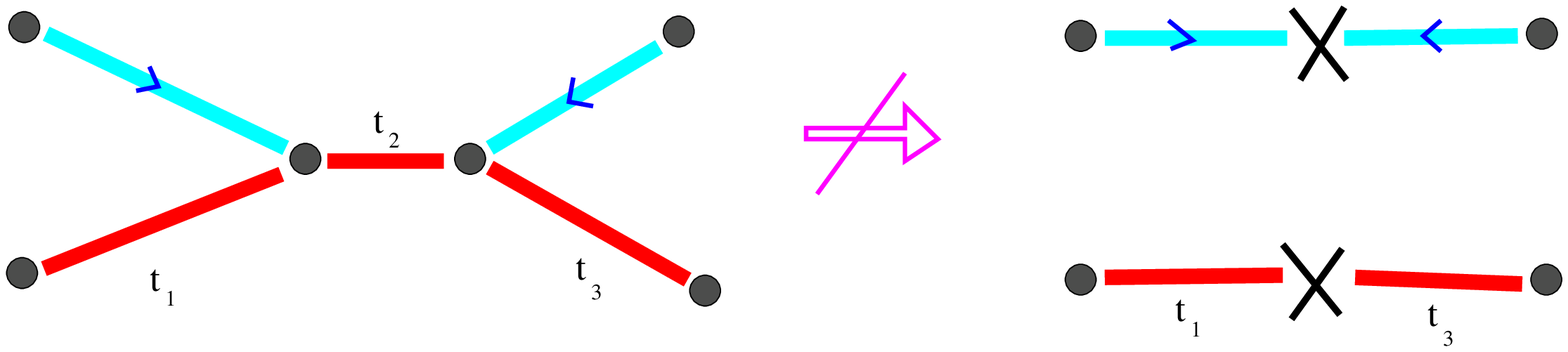}{\bcap\cantannihilate \/ The two nodes shown 
here cannot annihilate, because there is no consistent way to reconnect the 
string ends.}

 \FIG\indannih

iii)  {\bf Annihilation:}  When two nodes approach each other within a distance 
$d_{ann}$ which is a parameter of the simulation,  they are allowed to annihilate.  (In this work, $d_{ann}=0.08$ was chosen.)  
The segment(s) which join the two nodes is eliminated, and the other segments emanating from the two annihilating nodes are joined to each other.   
 
Two nodes are able to annihilate only if there is a consistent way to rearrange the free string ends (i.e., each string is able to find a partner with the same flux).  Annihilation is always possible if the two nodes are doubly 
linked as shown in figure   ~\doublelink.  In the case of double link annihilation,  the segment is straightened (or straightened until an obstruction is encountered, such as a collision with another string.)   Annihilation is also always possible if both junctions are of the sss type, even if they are only singly linked.  In this case, there are two possible rearrangements of the free string ends (figure   ~\sssannih).  One of these two is chosen at random.  When two stt-type junctions approach each other, on the other hand, there can be at most one consistent rearrangement of the free ends,  and it may not be possible for the nodes to annihilate at all.  Annihilation requires that {\it each} of the two segments on one side be matched with one on the other side carrying the {\it same flux}.  Figure   ~\cantannihilate\/ shows an example of a pair of nodes which cannot annihilate
because there is no consistent rejoining of the string ends.

 If two stt-type 
nodes do annihilate, it is easily seen that there can never be more than one consistent rearrangement of string ends.  If two of the outgoing ends are $s$ strings and two are $t$ strings, then there cannot be more than one rearrangement because each string can only be joined with one in the same conjugacy class.  If all outgoing strings are of odd type, then all four cannot have the same flux-- if they did, then the total flux of any pair would be trivial and they would not be connected by a segment.  Nor may any three have the same flux.  It follows that, at best, each string end may reconnect with a unique partner.

When a singly linked pair of nodes annihilates, the segments are first 
rejoined in a kink configuration,  which may straighten later.  

\insertfigpage{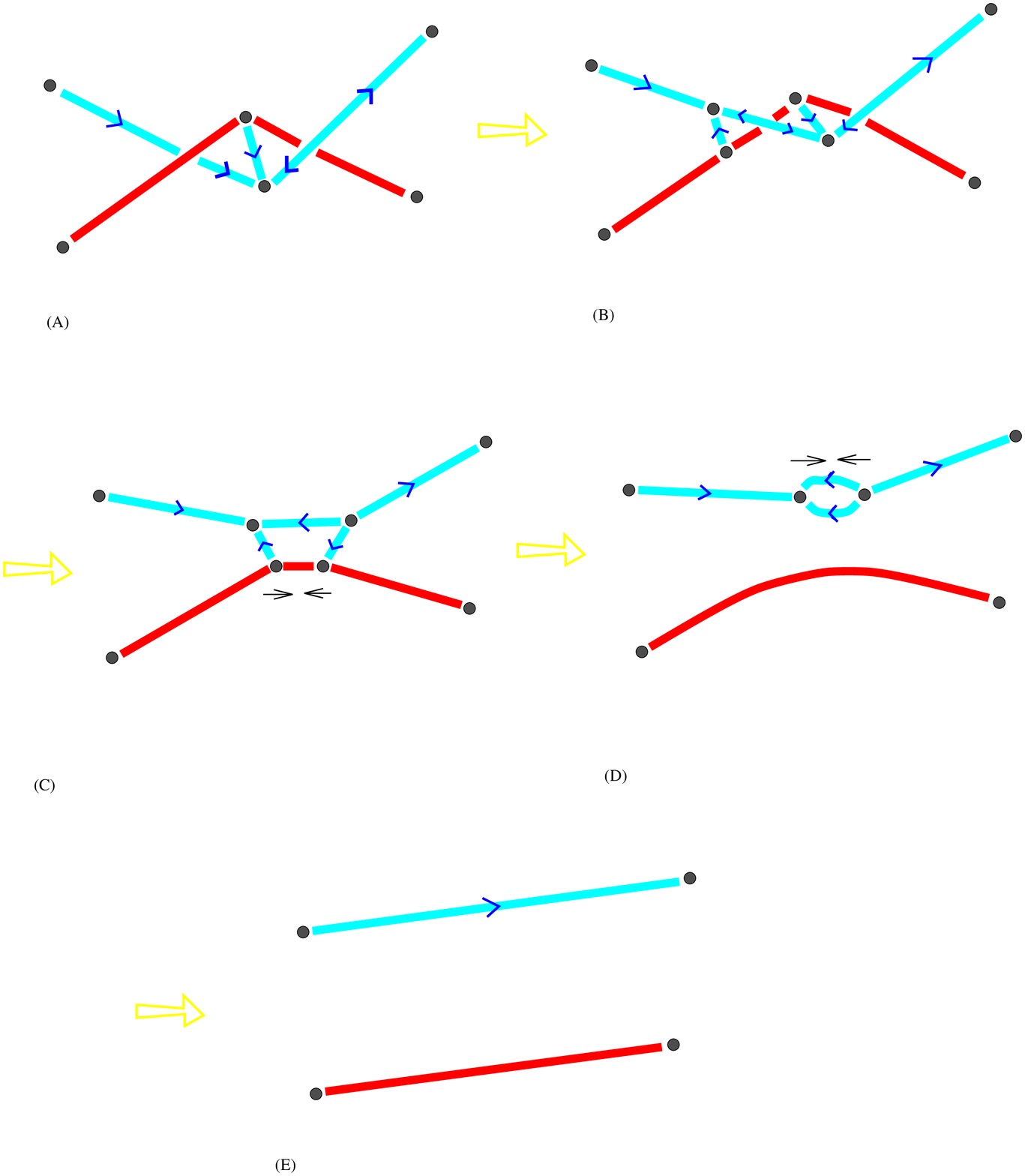}{\bcap\indannih\/ Unlinking of two strings-- the
inverse of the process shown in figure \NCI-- can occur in several steps if
the string tensions pull in the right direction to unlink the strings.  A
linking followed by two annihilations has the net result of removing the 
short intervening segment and unlinking the two longer strings.  In this 
figure, the basepoint is assumed to lie behind the page, so that the definition
of an $s$-string's flux changes when it passes in front of any $t$-string.}
 
Another type of annihilation process, which is the inverse process of bridge 
formation, is not included explicitly in the simulation but may occur through through a multi-step process involving several string intersection and annihilation events
(figure ~\indannih).
We expect that such a process probably {\it will} occur whenever the geometry is appropriate for the unlinking of two strings, so that it is not necessary to perform the unlinking ``by hand'' in a single step within the simulation.

(iv)  {\bf Rearrangement:}
If two nodes approach each other closely
but are prevented from annihilation by flux conservation requirements,  several
outcomes are conceivable.  They may stick together and form a stable junction
of more than three strings.  They may bounce back and move apart again under 
the influence of string tensions.  It is also plausible that the nodes
could rearrange their connections and undergo a sort of quasi-intercommutation,  as
shown in 
 figure \FIG\rearrange  ~\rearrange.  In this simulation, we allow the 
nodes to bounce by introducing a small repulsion at distances shorter 
than $d_{ann}$.   We also allow rearrangement with some probability, 
and examine the consequences of setting that probability either to zero or
to some non-zero value.  The results presented in section {6} indicate that
rearrangement speeds up the network's decay.

\insertxfigpage{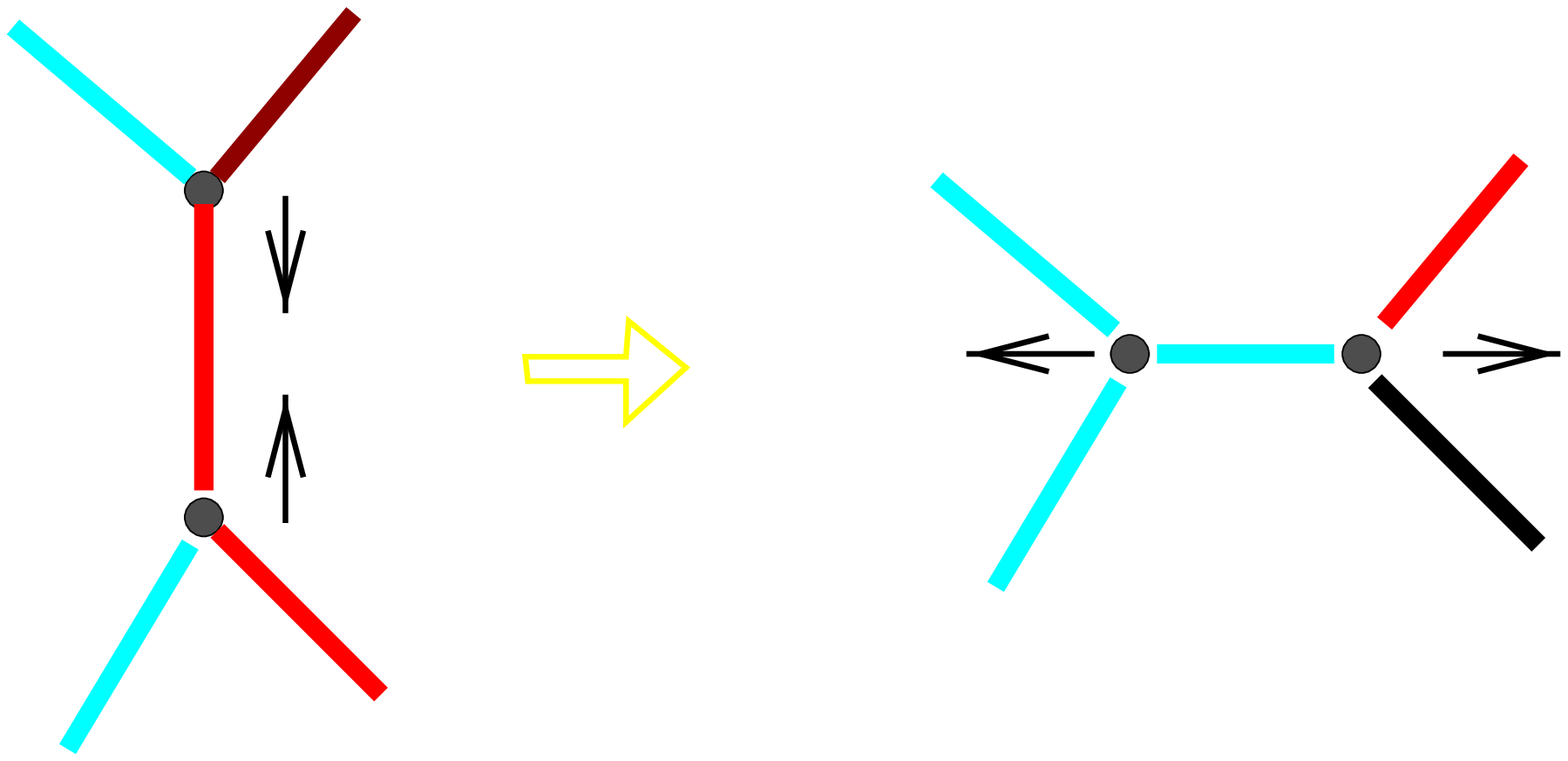}{\bcap\rearrange \/  Rearranging the connections
of two nodes which cannot annihilate.}

We do not include nodes of more than three
strings as fundamental objects,  but  it is quite possible for a pair
to become stuck together very close to each other.  The string 
tension and short distance repulsion allow them to oscillate  at a
short distance, and such a configuration can behave like a single 
junction of four strings.  Such adhesion becomes important 
under certain conditions,  as we will see in section {6}.

\section{Initial Conditions}
In order to perform our dynamical simulation, we must start with an initial configuration.  The generation of initial conditions models
 the symmetry-breaking transition which produces the strings.  As is
frequently done, we use a lattice Monte Carlo procedure to generate an initial string
network.  The lattice spacing is to be identified with the correlation
 length of the Higgs field at the time of string formation.  The Higgs
VEV is thus uncorrelated over distances longer than a lattice spacing, and at each site of the lattice, it takes a random value within the vacuum manifold.  With a suitable interpolation along the length of each link, any plaquette of the lattice is mapped to some closed loop on the vacuum manifold.  If this path is one of the non-contractable loops, then a string must pierce the plaquette.  Each link of the plaquette is associated with an element of $G$ which relates the Higgs values at the two ends of the link.  The product of these elements around a closed plaquette 
must lie within the unbroken group H, and can be taken as the flux of the string which pierces that plaquette. Strings which pierce the faces of a given unit cube must be joined together inside the cube in some way.  If only two faces of the cube have non-trivial flux, then we interpret this as a single string segment passing through the cube.  If three faces are non-trivial, we infer 
that there is a single vertex inside the cube.  Cubes pierced by more than three ends require a more complicated arrangement of nodes and strings inside the cube. There may be more than one consistent way to join the string ends, and
 one
must be chosen arbitrarily. 

Two different lattice Monte Carlo algorithms have been used for the current
simulation to generate different initial distributions of strings.  The 
first,  very simple way to generate a random network of strings is to use an infinite temperature lattice gauge theory:  simply assign a random element of the unbroken group $H=S_3$ to each link of the lattice, and evaluate the product of links on the plaquette to find the flux
through the plaquette.  We refer to this as the lattice gauge Monte Carlo. 
There is no direct reference to a Higgs field in this technique.  

The other method we use is a discrete Higgs simulation analogous to that of  
references \REF\DH{T.~Vachaspati and A.~Vilenkin, Phys. Rev. D 30, 2036 (1984).
T.W.B.~Kibble, Phys. Lett. B 166, 311 (1986)} \REF\Zinitial{M.~Aryal, A.E.~Everett, A.~Vilenkin and T.~Vachaspati, Phys. Rev. D 34, 434 (1986)}
[\DH,\Zinitial].  The essence of this method is that a  discretized vacuum manifold is used. 
 The breaking 
of continuous group $G$ is modelled by using some discrete subgroup
${\cal G} \subset G$ which contains the unbroken group $H$.  Each lattice site is then assigned randomly to one of the discrete cosets in the space ${\cal G}/H$,  corresponding to a choice of vacuum.  With each link of the lattice there is associated an element of ${\cal G}$ which transforms the Higgs field value at one end of the link to the value at the other.  The element relating one coset
to another is not unique;  the possible elements themselves form a coset.  The
convention in this discrete Higgs method is to choose the ``smallest'' possible element for each link variable.  ``Small'' is defined with reference to a metric
on the continuous group $G$:  if all elements are written in the form 
$g=\exp (i \alpha T)$ where $T$ is a normalized element of the Lie algebra of
$G$, then the smallest element is the one with the smallest number $\alpha$.
In this way the Higgs field is effectively interpolated in the smoothest possible way between lattice points.   

A suitable gauge transformation can be performed so that all Higgs field values lie in the same coset, and all link variables lie within $H$, allowing all subsequent computations to be performed 
in terms of only $H$ link variables.  This is a lattice implementation of
unitary gauge.

   For the present 
simulation, we take ${\cal G}$ to be
one of the discrete subgroups of $SU(3)$.  
The simplest choice is the 24-element ``dihedral-like'' subgroup of $SU(3)$ known as $\Delta(24)$. \Ref\SUsub{W.M.~Fairbairn,T.~Fulton, and W.~H.~Klink, J.Math.Phys. 5, 1038 (1964)} This group is generated by the matrices
$$  T_1\equiv        \pmatrix{-1& 0& 0\cr
                               0& 0&-1\cr
                               0&-1& 0},
    S_+ \equiv       \pmatrix{ 0& 1& 0\cr
                               0& 0& 1\cr
                               1& 0& 0},$$
and 
$$  A(1,0)\equiv  {\rm diag}(1,-1,-1).  $$
$\Delta(24)$ 
is the smallest subgroup of $SU(3)$ that contains $S_3$, and in fact it is
isomorphic to $S_4$, the permutation group on 4 elements: an isomorphism 
may be defined which maps $$T_1\leftrightarrow t_1=(2 3), \/\quad\/ S_+ 
\leftrightarrow s_+=(1 2 3)$$  these generate the subgroup $H_0\simeq S_3$ of permutations
on elements 1-3.   $H_0$ may be viewed as the little group of a ``Higgs'' 
vev which has the form $h_0 \equiv (0,0,0,1)$:  permutations of the first
three
positions leave $h_0$ invariant.  
The elements $A(1,0)={\rm diag}(1,-1,-1),  A(0,1)={\rm diag}(-1,-1,1)$ and
$A(1,1)={\rm diag}(-1,1,-1)$  may be mapped to $S_4$ by
$$A(1,0)\leftrightarrow (1 4)(2 3),  A(0,1)\leftrightarrow (2 4)(1 3),
  A(1,1)\leftrightarrow (3 4)(1 2).$$
These act nontrivially on $h_0$, and together with the identity they generate
the 
four distinct left cosets 
$H_0, A(1,0)H_0, A(0,1)H_0,$ and $A(1,1)H_0$.  Each of the three 
nontrivial cosets consists of the set of elements which transform $h_0$ 
to one of three other possible vev's.  For example,  elements of 
$A(1,0)H_0$  take (0,0,0,1) to (1,0,0,0).  

To generate a network of strings,  the discretized Higgs vev is randomly 
assigned to one of its four values at each lattice site.  For a smooth 
interpolation, the link 
between two neighboring sites is chosen to be the smallest within the 
appropriate coset.  In the coset $A(1,0)H_0$, for example,  the two smallest 
elements are $A(1,0)T_2\leftrightarrow (1 4)(2 3)(1 3)$  and 
$A(1,0)T_3\leftrightarrow (1 4)(2 3)(1 2)$.  Both of these have equal
measure;  a random choice may be made between them.
In the identity coset, of course, the identity element is the smallest.  
Each other coset has two smallest elements of the form $A T$, where 
$A\in \{ A(1,0),A(0,1),A(1,1)\}$  and T is one of the two transpositions in 
$S_3$ that fail to commute with $A$.     

After assigning vacuua and group elements, one can then transform to the 
unitary gauge in which the Higgs vev is the same at each site, and all 
flux information is encoded in $S_3$ variables on the links, just as it is
in the lattice gauge Monte Carlo.   

\subsection{Properties of the initial network}

Both of the Monte Carlo algorithms described above create infinite 
branched networks.  The lengths of string segments between branching
are distributed exponentially, reflecting a constant probability of branching per unit length.    In this respect, the 
two methods are similar,  but the resulting networks differ in 
other statistical properties.  In table $1$, 
we summarize some of these features.  For comparison, we also include the corresponding
information for the $Z_3$ system (including both a $Z_3$ lattice-gauge method
and the tetrahedral discrete-Higgs simulation of [\Zinitial].  The $Z_3$ discrete-Higgs numbers are from
reference [\Zinitial]  and the $Z_3$ lattice gauge data are based
on the author's simulations (see also \REF\SURF{The 
$Z_3$ lattice gauge algorithm and the statistics of its initial network were 
also described in the work of L. Nayvelt and J.E. Woods, ``Initial
Configuration and Evolution of $Z_3$ Cosmic Strings,'' unpublished report (Caltech, 1988).} [\SURF]).  For
each method, the fraction of plaquettes pierced by strings of each type is reported.  (In the $Z_3$ case, of course, there is only one type.)
Note that in the lattice gauge method, each group element is weighted equally; 
therefore $2/3$ of all plaquettes are pierced by strings in the $Z_3$ lattice gauge case, and $5/6$ for $S_3$.   Below this are the fractions of cubic lattice cells with 0,2,3,4,5 and 6 of their faces pierced by strings, the number densities of the different types 
of nodes per unit volume, the average length between junctions and the branching length (obtained from 
the exponential decay of the length distribution) for each string species.\footnote{*}{Initially, string lengths are naturally clustered near
integer multiples of the lattice spacing.  The distribution looks
smooth and exponential only when string lengths are placed in bins of at least one
lattice spacing.  For this reason, the decay length is not necessarily 
identical to the average segment length.}     

Pictures of typical initial string networks
(figures
\FIG\ICPictures  ~\ICPictures) illustrate qualitatively the comparison 
between the different $S_3$ initial conditions.  
 It is evident that for both $Z_3$ and $S_3$
systems, the lattice gauge method produces a denser network:  more cube faces
are pierced by strings and more cells have high numbers of strings emerging
through their faces.  Correspondingly,  in the lattice gauge method,  
fewer strings continue through more than one lattice cell without branching. 

Regardless of the initial conditions,  $stt$ junctions outnumber $sss$
junctions in the $S_3$  network,  and we can infer that a majority of
the $s$ strings end on a $t$ string at at least one end.  

Although we have not extracted detailed statistics on the presence of
disconnected loops and finite networks in the initial distributions,
they do not 
appear to form a significant part of the system.  In this respect the
$S_3$ networks are like the $Z_3$ ones.

The results presented in the next section show that in the $S_3$ model, unlike the $Z_3$, differences in the 
statistical properties of the initial networks can have a very pronounced
effect on the network's evolution.


{\pageinsert
\bigskip\bigskip

\vbox{\offinterlineskip
\def\tablerule{\noalign{\hrule}}
\halign{\strut#&\vrule#\tabskip=1em plus2em&
  \hfil#& \vrule#& 
  \hfil#\hfil& \vrule#& 
  \hfil#& \vrule#& 
  \hfil#& \vrule#& 
  \hfil#&\vrule#\tabskip=0pt\cr\tablerule
&&\omit\hidewidth Method\hidewidth&&
  \omit\hidewidth {$Z_3$ L.G.}\hidewidth&&
  \omit\hidewidth {$Z_3$ Higgs}\hidewidth&&
  \omit\hidewidth {$S_3$ L.G.}\hidewidth&&
  \omit\hidewidth {$S_3$ Higgs}\hidewidth&\cr\tablerule
&&Faces with&& && &&  && &\cr
&&{$s$ string}&&0.67&&0.52&&0.33&&0.14&\cr
&&{$t$ string}&&{\--}&&{\--}&&0.50&&0.37&\cr\tablerule
&&Cubes with&& && && && &\cr
&&0 ends && 0.01 && 0.04 && 0.00 && 0.06 &\cr
&&2 ends && 0.12 && 0.34 && 0.01 && 0.32 &\cr
&&3 ends && 0.17 && 0.20 && 0.05 && 0.21 &\cr
&&4 ends && 0.38 && 0.32 && 0.21 && 0.31 &\cr
&&5 ends && 0.24 && 0.09 && 0.20 && 0.02 &\cr
&&6 ends && 0.09 && 0.02 && 0.55 && 0.08 &\cr
\tablerule
&&Density of&& && &&  && &\cr
&&{$sss$ nodes}&&{0.56}&&{0.28}&&0.22&&0.03&\cr
&&{$stt$ nodes}&&{\--}&&{\--}&&1.36&&0.67&\cr
\tablerule
&&Av. length&& && &&  && &\cr
&&bet. junctions&& && &&  && &\cr
&&{$s$ string}&&{2.38}&&{3.71}&&1.02&&1.27&\cr
&&{$t$ string}&&{\--}&&{\--}&&1.05&&1.59&\cr
\tablerule
&&Branching length&& && &&  && &\cr
&&{$s$ string}&&{1.65}&&{3.33}&&0.69&&0.83&\cr
&&{$t$ string}&&{\--}&&{\--}&&0.78&&1.53&\cr
\tablerule}}

\medskip
{\bf Table 1:}\/ Statistics of Initial Conditions.   
\medskip
\bigskip
\endinsert}

\insertlongfigpage{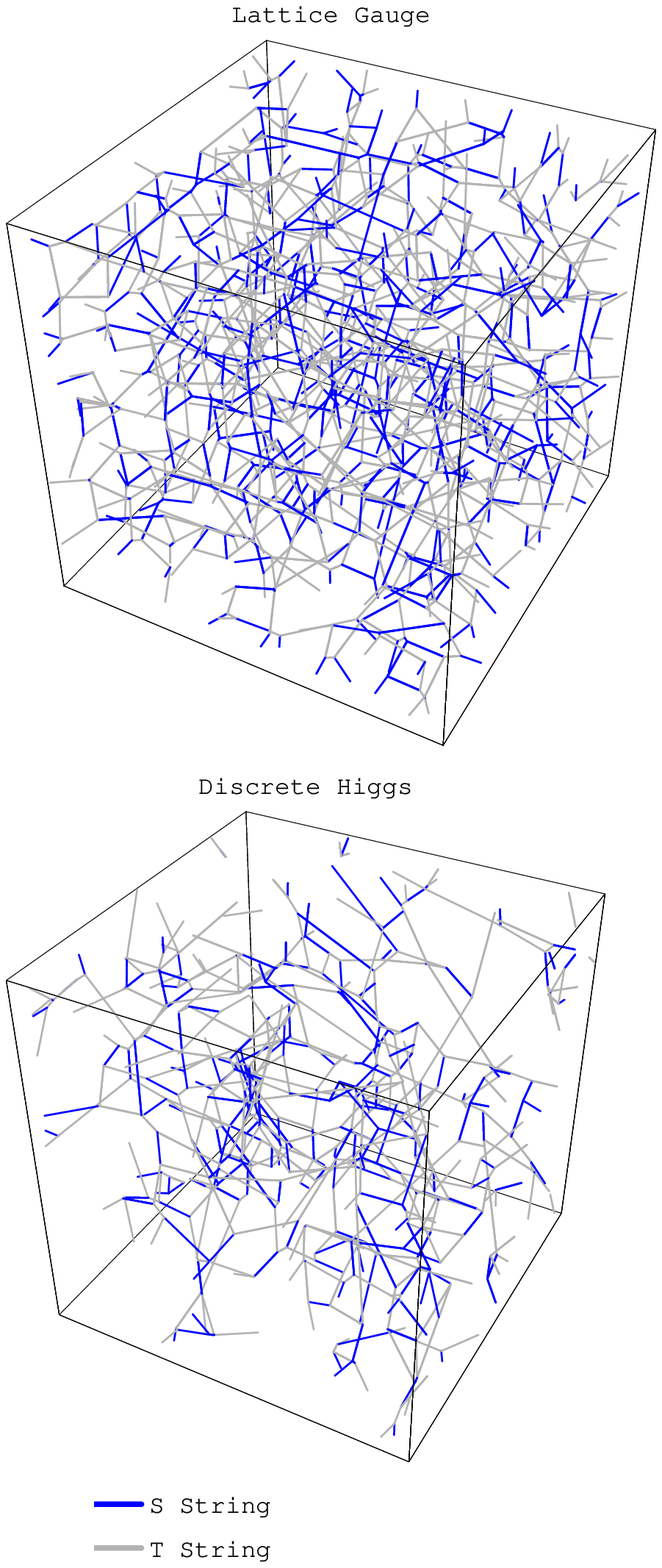}{\bcap\ICPictures\/  Sample Monte Carlo
initial configurations (actually shown after a single time-step of 
dynamical evolution).  The even, or 3-cycle, strings are shown in the darker
color.  Volume shown is $8^3$ in lattice units.}

\section{Dynamical Results}

In this section, we present results from the dynamical simulation.  The 
aim is to give a qualitative picture of the evolution and to determine 
which factors are most important in determining the fate of the network. The 
qualitative nature of the evolution depends on the ratio of the different 
string tensions.  Two contrasting regimes are of interest:  one in 
which the even strings are light, and another in which they are heavy.  
We also examine the influences of other factors.  The results show
a surprisingly marked dependence on statistical properties of the 
initial network.   


\FIG\evthirty
\FIG\densthirty 
\FIGURE\lightintercom
\FIG\lightSpics
\FIG\heavySpics
\FIG\heavyHSpics
\FIG\Llpics
\FIG\Zpics

\FIG\heavySplots
\FIG\lightSPlots
\FIG\equalSPlots

\FIG\histograms
\FIG\histogramt
\FIG\histogramz


This section is organized as 
follows:  First, we review some features of the behavior of Abelian, $Z_3$
strings, for comparison with the $S_3$ results.  The $Z_3$ results are taken in part from [\VV] and in part from simulations by the author.  
The self-similar scaling evolution and the associated power-law behavior 
are demonstrated.  After this come the $S_3$ results.  We describe the 
dependence of these results on parameters of the simulation, focusing especially
on two constrasting regimes of string tensions, and then discussing other
important factors.  Finally,  we make a few comments about fluctuations and the
effect of the finite simulation volume.

\subsection{The $Z_3$ Network and Power-Law Evolution}

Since $S_3$ contains $Z_3$ as a subgroup,  our program can easily be
modified to 
simulate a $Z_3$ network by generating only $s$-strings in the initial conditions.   Figure  \evthirty \/ shows the total
string length and total number of nodes as a function of time for a typical
run on a $30^3$ simulation volume.  The initial conditions were generated with a $Z_3$ lattice gauge method, rather than the discrete Higgs method of [\VV ]. 
(In fact, these data were obtained in a simulation without any intercommutations --- strings were allowed to pass through each other.  There is
no topological obstruction to prevent this in an Abelian network,  and the
inclusion of intercommutations makes only a small difference in the results.)
  The units used for this plot are the ones in which the
initial lattice spacing is $1$   and the string tension is $T=1$.

The time variable plotted on the 
$x$-axis in the figure is $i \Delta t$, where $i$ is the number of elapsed time steps.  (This will be our convention for all remaining plots.)  All distances
and lengths are measured in units of the original lattice spacing.   

A transformation of the data shows more clearly the ``scaling'' behavior of the network.  The scaling hypothesis is that the gross properties
of the network are described by a single length scale, $\ell$, which grows with time
as the network relaxes.  If $\ell$ is the typical distance between nodes, then
the number of nodes per unit volume is $n\sim \ell^{-3}$.  In a scaling solution,  the average length of a string segment between nodes is also
 $\sim \ell$,  while the number of segments per unit volume is proportional
to the node density, $\ell^{-3}$.  Hence the total string length per unit
volume (or the energy stored in strings) scales as $\ell^{-2}$.
In figure \densthirty,  we plot two scale variables with 
the dimensions of length: 
The inverse cube root of the number of nodes per unit volume, which we denote $D_n$,  and 
the inverse square root of the string length per unit volume, $D_l$.  
From figure \FIG\densthirty \/ \densthirty , we can see that both of these
length scales are approximately equal and grow linearly with time.
  The slope of approximately $0.3$ is close to that observed 
in reference [\VV ].  Although the results plotted in fig. \densthirty\/ and those
of [\VV] were obtained from different initial conditions,  the results
agree very closely.  Evidently,  all noticeable differences disappear after 
just a few time steps.  As with $Z$ strings,  the late-time evolution 
is essentially independent of the initial conditions.  

Evolution of the 
$Z_3$ network is self-similar in the sense that network looks statistically
the same at all times except for the increase in scale.  The distribution
of string segment lengths, for example, is exponential at all times,  with
only the scale changing.  This can be seen in the semilogarithmic histograms
of figure \histogramz.  Qualitatively,  a portrait of part of the network
at a given time is indistinguishable from a suitably magnified portrait of
an earlier time.

\insertfigpage{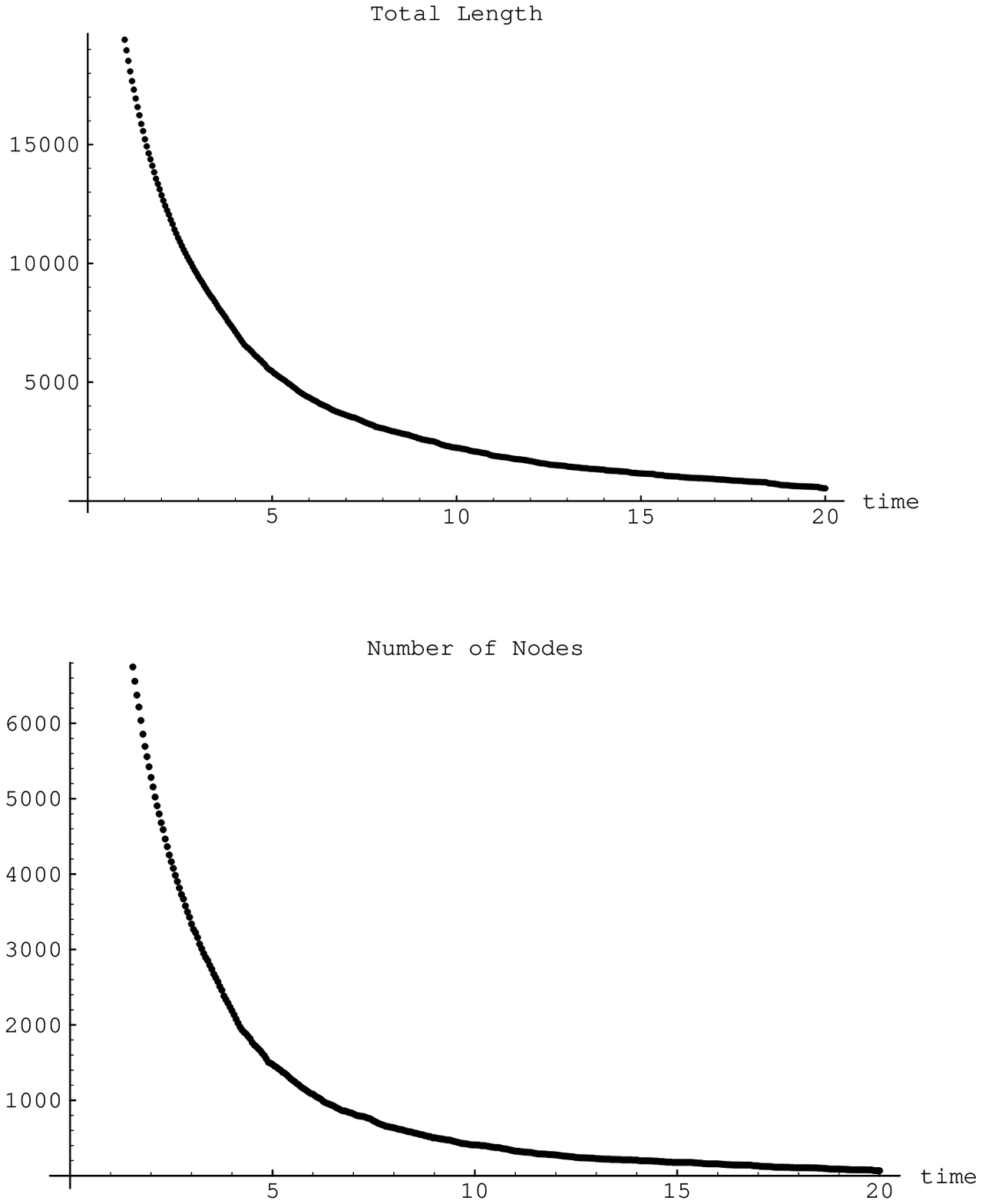}{\bcap\evthirty\/ Number of nodes and total string
length as a function of time for $Z_3$ strings on a $30^3$ volume. } 

\insertfigpage{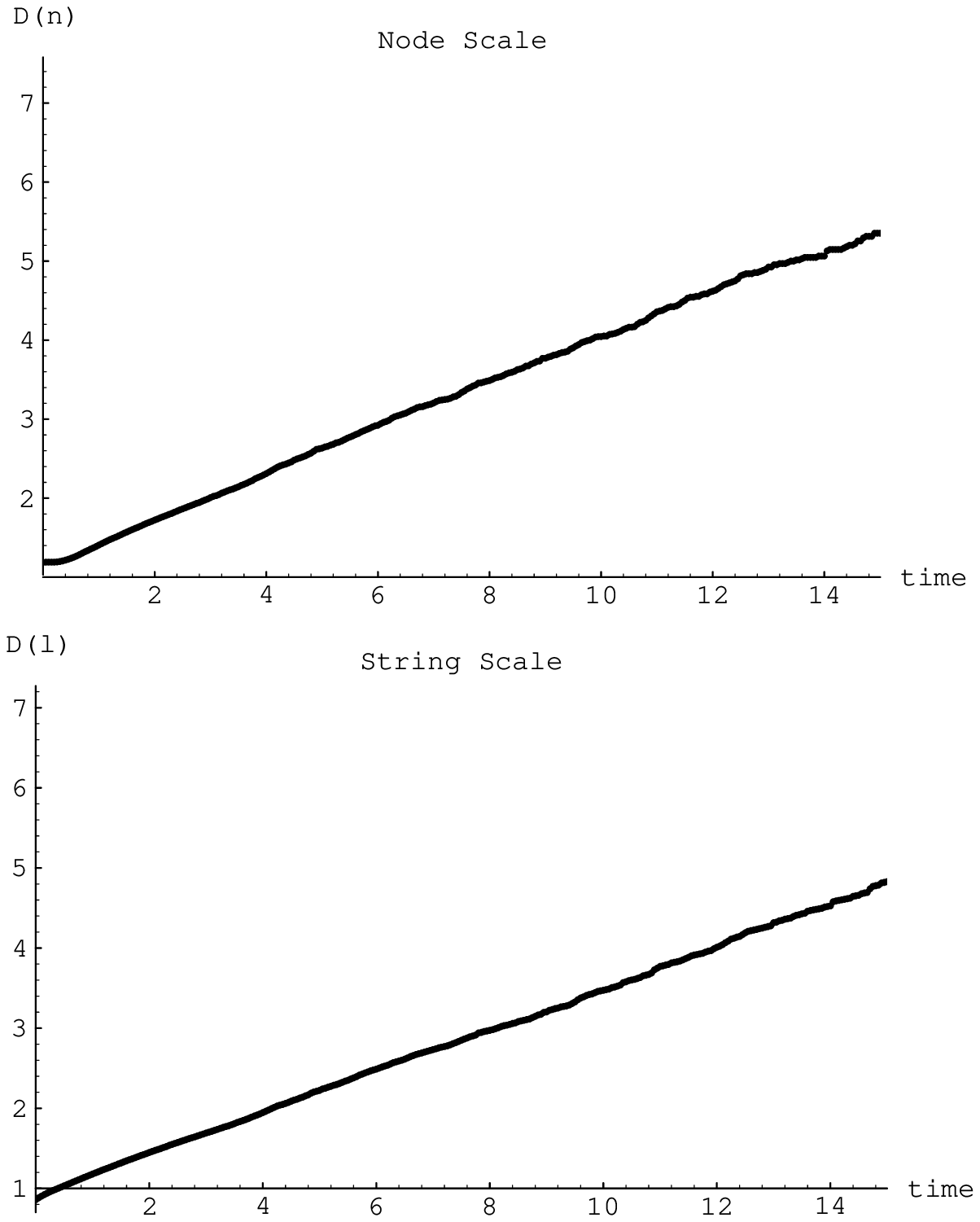}{\bcap\densthirty\/ Scaling behavior of $Z_3$ strings
on a $30^3$ volume.  The typical distance between nodes, $D_n$, grows linearly with
time, and so does the inverse square root of the string energy density, $D_l$.}

\subsection{$S_3$ network:  general comments}

In the remainder of this section we describe results of the $S_3$ simulation,  which
was run using a 
variety of different combinations of initial conditions, string tensions,
and other parameter choices.  Networks generated by lattice-gauge initial
conditions were run on an $8^3$ simulation volume, while the less dense
discrete-Higgs-generated networks were evolved on a volume of $12^3$.
Statistical variables were obtained from averages over several runs.
Computation time constraints made it unfeasible to run many times on larger
volumes,  but a few runs were performed on both larger and smaller volumes in order 
to examine the effects of finite size.    

Key results are displayed graphically in figures \lightSpics \- - \histogramz 
\- and table 2.  Some results for
an Abelian, $Z_3$ network are also shown for comparison.
Figures \lightSpics \- - \Zpics \- show three-dimensional pictures of simulated
cosmic string networks during their evolution.  \Ref\website{Animations are available through the
World Wide Web at http://physics.unc.edu/$\sim$mcgraw/stringmovies.html.} 
The pictures show several of the different patterns of evolution that can 
occur under different conditions.  Pictures of a $Z_3$ network are provided as well.
 The plots in figures \heavySplots \- -
\equalSPlots \- show the evolution of
some length-scale variables as functions of time.  These variables include
  $D_n$, the inverse cube root of
the density of nodes,  the average segment length between junctions for each type
of string,  and the two quantities $D_s$ and $D_t$.  The last two are analogous 
to $D_l$ defined above;  the inverse square root of the string length per
unit volume, computed separately for each type of string.  (Some of these
variables may be redundant.)
With some exceptions discussed below,  these length scale variables
exhibit (after some transients at early times) the linear increase characteristic of self-similar evolution.
Such plots were made for simulations run under a wide variety of conditions
giving a survey of the simulation's parameter space.
A few representative plots are shown here in order to show their typical
shapes.  The remainder are summarized
in Table 2, which gives their slopes.  The last set of figures in this section,
figures \histograms \- - \histogramz , consists of a series of histograms which
show how the distribution of string segment lengths evolves with time.
Distributions are shown for two contrasting cases discussed below,  and
corresponding data is also provided for the $Z_3$ network for the sake of
comparison.

The parameter space surveyed in Table 2 includes two different initial
condition simulations, several different ratios of the string tensions,
and two other binary choices affecting string collisions 
and close encounters between nodes.  The choice between ``bridge'' and
``zipper'' configurations for colliding non-Abelian strings is one
choice (abbreviated B and Z in Table 2.)  The other choice determines what 
happens when two neighboring nodes are within distance $d_{ann}$ but are
topologically unable to annihilate.   In the cases labelled ``R,'' such
nodes undergo a rearrangement of connections (see previous section) with
a probability of $0.2$ per time step.  In the cases labelled ``N,''  no
rearrangements are allowed, and the nodes simply bounce.

As a rough measure of the importance of intercommutations and NCC's,  Table
2 also gives, in the last two columns, the ratio of the total number of NCC
events to the number of intercommutations, and the ratio of the number
of NCC's to the cumulative net number of nodes annihilated.  (For the $Z_3$
network, there are no NCC's and the number of intercommutations is shown 
instead.)

Unless otherwise specified, all numerical quantities are given in units
such that the initial lattice spacing is $1$, the displacement of
each node during a time step is given by 
$\Delta x = \sum_r T_r \Delta t$, and the magnitude of the tension $T_t$ of the 
{\it odd } strings is normalized to $1$.  (This normalization was chosen 
because the special status of odd strings: they are the ones which cannot end,  and in none of the
cases simulated did they show a tendency to become extinct.  Even strings, on 
the other hand, disappear almost completely under certain conditions.) The tension
$T_s$ of
the even strings was varied relative to this fixed value.

One salient feature of the results is hardly unexpected: the ratio of the tensions of 
different string species has a strong effect on the behavior of the network.
The case of heavy $s$ strings, in particular, is an exceptional one which will
be discussed further below.
Much more surprising is the strong difference in behavior between networks
with different initial conditions.  Evidently, different initial conditions
lead to quite different trajectories which appear self-similar.

We will address the issues in the following order:  First, we discuss the 
different evolution patterns that occur with different choices of string
tension, focusing on the contrasting limits of heavy and light even strings.
Then we comment on other effects, including the effects of different initial
conditions.  Finally, at the very end of the section, we will make a few 
remarks about uncertainties and finite-size effects.

\subsection{$S_3$ network with light even strings}

Consider a network in which the even, or 3-cycle, strings have a much lower
tension than the odd strings.  In this case,  the odd strings may pass through
each other with comparative ease by forming new segments of the lighter even string.  The odd strings may shrink with comparatively little energy cost
in the creation of new strings.  Furthermore,  a zipper-type collision of 
two odd strings may have a result which, from the point of view of the 
odd string subsystem,  resembles an intercommutation  (see figure  \lightintercom).
The newly formed light string offers comparatively little resistance to the 
straightening of the rejoined heavy strings.  We might expect that in this limit, 
the odd strings might behave as a network of $Z_2$-like (i.e., unbranched
and unoriented) strings moving 
through a viscous medium formed by the branched $Z_3$ network of even strings.  The odd strings can shorten 
and cut themselves into decaying loops while transferring part of their
energy to the even string network.  The even network, in turn, can dissipate
its energy in much the same way as a $Z_3$ network.

\insertfigpage{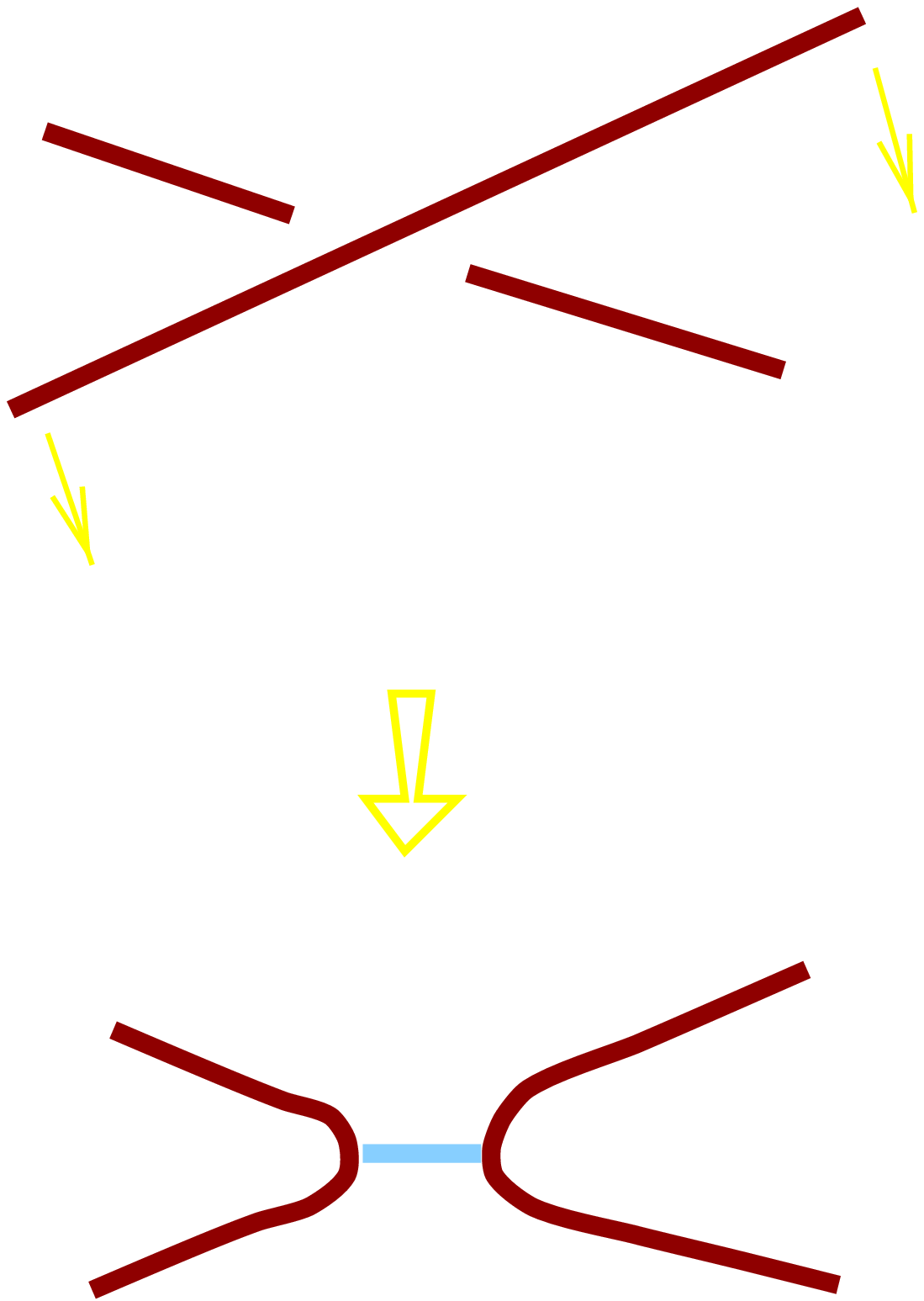}{\bcap\lightintercom\/ Two colliding 
$t$ strings form an $s$-string zipper.  If the $s$ string is ignored,
this event looks very much like an intercommutation.} 

Simulation results seem consistent with the above picture.
Figure
 \lightSpics\/ shows a series of portraits of the 
evolution of a network with tensions $T_s=0.5$,  $T_t=1.0$, beginning 
from lattice gauge initial conditions.  The configuration at time $t=0$
is in fact the one shown shown in fig. \ICPictures.
In the early stages of evolution,  we see a large number of highly crumpled,
apparently Brownian $t$ strings with thick webs of $s$ string stretched between
them.  As time progresses, the odd strings begin to straighten and some closed
loops shrink away and vanish.  However,  even though the odd strings are
shortening at the expense of stretching the even ones,  the population of even strings is 
at the same time
being reduced through the annihilation of nodes, with the result that neither
species of string becomes entirely dominant over the other.  Instead, the
evolution appears self-similar, with an approximately constant ratio between
the amounts of the two different string species.  This constant ratio is evident
in figure \lightSPlots\- from the linear evolution of both length scale variables $D_s$ and $D_t$.   Apparently, after some transient behavior at
very early times, a sort of dynamical equilibrium is established, with energy
being transferred in a steady cascade from the $t$ network to $s$ network
and then lost to damping.

Interestingly,  decreasing the even string tension still further
to $0.25$ does not allow the odd strings to contract more quickly 
(see Table 2) -- evidently
they are impeded by the larger population of even strings.

\subsection{$S_3$ network with heavy even strings}

A situation which contrasts with the case of light $s$ strings is one in which
the $s$ strings have a large tension.  In this subsection,  we discuss 
simulation results for the case with $T_s=2, \/ T_t=1$.
A glance at figures \heavySpics\- and \heavySplots\- reveals that this leads to
quite different results. 
  The $s$, or even,
strings are in this case only marginally stable against decay into pairs of
$t$ strings.  If a heavy $s$ string ends on a $t$ string,
then the two $t$ segments pulling against the $s$ string at the junction cannot
prevent the $s$ segment from shrinking unless the angle between the $t$ segments
is sufficiently small;  in this marginal case of $T_s=2$ they must be collinear in order for the tensions to balance.     In either lattice gauge or discrete Higgs initial networks,
the majority of $s$ strings end on $t$ strings, and so there is very little
to prevent these heavy strings from shrinking rapidly. 
  
The shrinking of an individual $s$ segment frequently brings together 
a pair of nodes which are topologically unable to annihilate.  The nodes stick together because to separate them again would require stretching the $s$ 
segment.  Similar adhesion happens whenever two non-commuting 
$t$-strings collide.  Soon the network consists predominantly of
odd strings stretched between small tangled clusters,  and no further annihilation can occur until separate clusters 
meet,  or the geometry changes sufficiently to allow 
some strings to pull free.   Figure \heavySplots \/
shows the scaling variables as functions of time.  The rapid disappearance
of  $s$ string is evident from the extremely fast increase in the variable 
$D_s$ in the upper plot, and from the lower plot which shows that the 
average length of an $s$ segment quickly drops to near zero.  This is an
exceptional case: in all other
cases besides $T_s=2$, the average segment length for both string types increases with time.  

Even though the $s$ strings do not obey the usual standard scaling behavior,
it appears that the average $t$ segment length does increase 
linearly with time,  as do $D_n$ and $D_t$.  Evidently, the overall system
obeys scaling once most of the $s$ strings have shrunk.

The prevalence of clusters connected by short $s$ segments is apparent in the
snapshots of figure  \heavySpics\- and is illustrated in a different manner
by figure  \histograms, which shows semilogarithmic histograms of the 
distribution of segment lengths at a sequence of times.  The length distributions for the two string types evolve in different ways.  For
lengths larger than one lattice spacing,  the $t$ segments follow an exponential
distribution (which appears linear on semilogarithmic axes) with a 
steadily increasing decay length.  While the few long $s$ segments that remain
are also exponentially distributed, there is no clear tendency for the
decay length of this distribution to increase with time, and the number of
long segments quickly shrinks into insignifigance compared with the 
large population of very short segments in the lowest bin of the histogram.
The presence of so many short segments is associated with  the adhesion of 
many pairs of nodes.  

Compare fig. \histograms\- with the length distributions in the $T_s=0.5$ case, shown in
figure \histogramt.  In the latter, we see that both distributions develop
a peak at short lengths (indicating that some clusters do occur), but the
peaks are less pronounced and the number of such short segments decreases
at a rate commensurate with the remaining population of strings.  The 
exponential decay lengths for both distributions increase with time, as 
expected in a self-similar evolution.   Evidently, the pile-up of clusters
is less significant in the light-$s$-string case.  
Length distributions
for the Abelian network (figure \histogramz) show no excess at all at short 
distances.  

\subsection{Other Effects}

The non-Abelian simulation has been run under a variety of conditions.  For the
most part,  the results are consistent with some type of self similar evolution.  Variables having the dimensions of length
all increase linearly with time.  The exception is for networks with heavy 
$s$ strings, in which the average $s$ segment length decreases.  This
is the only case for which one type of string dominates completely
over the other. In 
all other cases, the amounts of the two string types approach a constant
ratio.   

The evolution of the network has some interesting dependence on parameters
other than the string tensions.  

Most surprising is the very strong dependence on the
statistics of the initial conditions.   When the Higgs initial conditions are
used,  the network decays much faster,  especially in the case of heavy $s$
strings.  Not only the absolute sizes, but the ratios of coefficients
can have quite different values depending on which
initial conditions are used.  This dependence is somewhat mysterious.  
The lattice gauge initial conditions are much more dense than the Higgs, but
the overall density of nodes decreases during the evolution,  soon reaching 
a density comparable to that of the initial Higgs-generated network, yet 
the two continue evolving at very different rates.  Evidently, some statistical
property other than the total density is important, such as, for example,
the distribution of voids,  or even correlations of the fluxes.  A better 
understanding of this phenomenon should be the goal of future work.

{\baselineskip=4pt
{\pageinsert
\bigskip 

\vbox{\offinterlineskip
\def\tablerule{\noalign{\hrule}}
\halign{\strut#&\vrule#\tabskip=1em plus2em&
  \hfil#& \vrule#& 
  \hfil#\hfil& \vrule#& 
  \hfil#& \vrule#& 
  \hfil#& \vrule#& 
  \hfil#& \vrule#& 
  \hfil#& \vrule#& 
  \hfil#& \vrule#& 
  \hfil#&\vrule#\tabskip=0pt\cr\tablerule
&&\omit\hidewidth {Conditions}  \hidewidth&&
  \omit\hidewidth {${d D_s/{d t}}$ }\hidewidth&&
  \omit\hidewidth {$dD_t/dt$  }\hidewidth&&
  \omit\hidewidth {$dD_n/dt$  }\hidewidth&&
  \omit\hidewidth {$d(Avg. S)\over{dt}$  }\hidewidth&&
  \omit\hidewidth {$d(Avg. T)\over{dt}$ }\hidewidth&&
  \omit\hidewidth {$NCC\over{IC}$ }\hidewidth&&
  \omit\hidewidth {$NCC\over{node}$ }\hidewidth&\cr\tablerule
&&{$T_s=2.0$, LG} && && &&  && && && && &\cr
&&{B \/ N}&&0.9&&0.10&&0.11&&$<0$&&0.15 &&5.9&&0.83&\cr
&&{Z \/ N}&&0.7&&0.10&&0.11&&$<0$&&0.15 &&8.6&&0.93&\cr
&&{B \/ R}&&1.4&&0.25&&0.40&&$<0$&&0.37 &&5.6&&0.64&\cr
&&{Z \/ R}&&1.5&&0.29&&0.35&&$<0$&&0.63 &&8.1&&0.91&\cr\tablerule
&&{$T_s=2.0$, H} && && &&  && && && && &\cr
&&{B \/ N}&&2.7&&0.55&&0.69&&$<0$&&1.2&&4.2&&0.23&\cr
&&{Z \/ N}&&2.8&&0.46&&0.59&&$<0$&&1.1 &&4.5&&0.25&\cr
&&{B \/ R}&&3.0&&0.71&&0.91&&$<0$&&2.0 && 2.9&& 0.17&\cr
&&{Z \/ R}&&3.7&&0.82&&1.1&&$<0$&&1.9 && 4.0&&0.22&\cr\tablerule
&&{$T_s=1.0$, LG} && && &&  && && && && &\cr
&&{B \/ N}&&0.15&&0.15&&0.11&&0.08&&0.09 && 5.3&& 0.33&\cr
&&{Z \/ N}&&0.16&&0.12&&0.10&&0.05&&0.08 && 5.8&& 0.23&\cr
&&{B \/ R}&&0.30&&0.20&&0.18&&0.11&&0.17 && 6.3&&0.06 &\cr
&&{Z \/ R}&&0.34&&0.22&&0.17&&0.09&&0.11 && 3.8&&0.06 &\cr\tablerule
&&{$T_s=1.0$, H} && && &&  && && && && &\cr
&&{B \/ N}&&1.0&&0.56&&0.50&&0.19&&0.41 && 3.9&& 0.17&\cr
&&{Z \/ N}&&1.0&&0.53&&0.43&&0.12&&0.33 && 3.0&& 0.10&\cr
&&{B \/ R}&&1.1&&0.65&&0.54&&0.18&&0.35 && 2.8&& 0.05&\cr
&&{Z \/ R}&&1.2&&0.72&&0.61&&0.21&&0.33&& 3.1&& 0.04&\cr\tablerule
&&{$T_s=0.5$, LG} && && &&  && && && && &\cr
&&{B \/ N}&&0.10&&0.10&&0.07&&0.05&&2.9 && 2.9&& 0.37&\cr
&&{Z \/ N}&&0.09&&0.13&&0.07&&0.03&&0.03 && 2.4&& 0.23 &\cr
&&{B \/ R}&&0.17&&0.20&&0.13&&0.07&&0.07 && 2.5&& 0.09&\cr
&&{Z \/ R}&&0.17&&0.26&&0.13&&0.07&&0.07&& 2.3&& 0.07&\cr\tablerule
&&{$T_s=0.5$, H} && && &&  && && && && &\cr
&&{B \/ N}&&0.19&&0.15&&0.10&&0.04&&0.05 && 2.1&& 0.17&\cr
&&{Z \/ N}&&0.20&&0.20&&0.13&&0.07&&0.11 && 6.3&& 0.02&\cr
&&{B \/ R}&&0.47&&0.43&&0.30&&0.18&&0.20 && 2.5&& 0.06&\cr
&&{Z \/ R}&&0.44&&0.39&&0.29&&0.16&&0.17&& 2.0&& 0.05&\cr\tablerule
&&{$T_s=0.25$, $LG$} && && &&  && && && && &\cr
&&{Z \/ R}&&0.09&&0.17&&0.08&&0.04&&0.06 && 1.9 && 0.13&\cr\tablerule
&&{$T_s=1$, $Z_3$} && && &&  && && && && &\cr
&&(with intercommutations)&&0.37&&{\--}&&0.38&&0.25&&{\--} &&{\--} && 0.12&\cr\tablerule}}

\bigskip
{\baselineskip=4pt
{\bf Table 2:}\/ Coefficients for network's evolution:  rates of change for
$D_s$,$D_t$,$D_n$,
and average segment length of each type,  ratio of NCC's to intercommutation events, and cumulative
number of NCC's per node, reported for different conditions.  The final row gives $Z_3$ results.
Abbreviations:  LG=lattice gauge initial conditions, H=Higgs initial conditions,
B=bridge NCC, Z=zipper NCC, R=rearrangements allowed, N=no rearrangements, 
IC=intercommutation, NCC=non-commutative collision.}
\medskip
\endinsert}}

\insertwiidefigpage{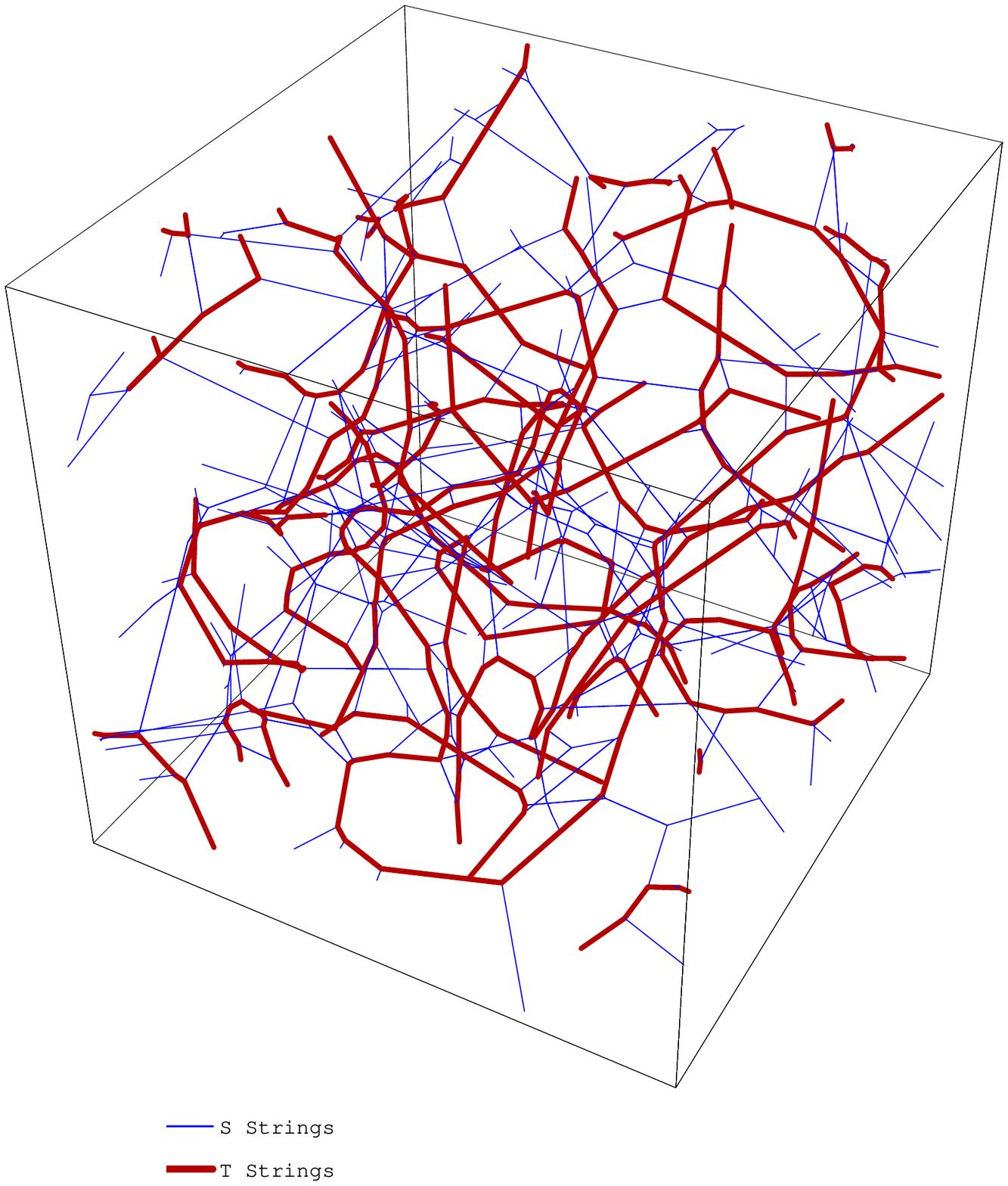}{\bcap{\lightSpics a}\/  A series of snapshots 
showing the evolution of an $S_3$ network with light strings.  The heavier,
odd strings are shown in darker color.  We can observe long strings straightening 
and loops of $t$ string contracting.  The initial network
was generated by the lattice gauge Monte Carlo algorithm:  it is the one
shown in figure \ICPictures.   The full $8^3$ simulation volume is shown.
In the first frame, at $t=2$, the odd strings are quite crumpled and are connected by a 
dense web of even strings.}

\insertwiidefigpage{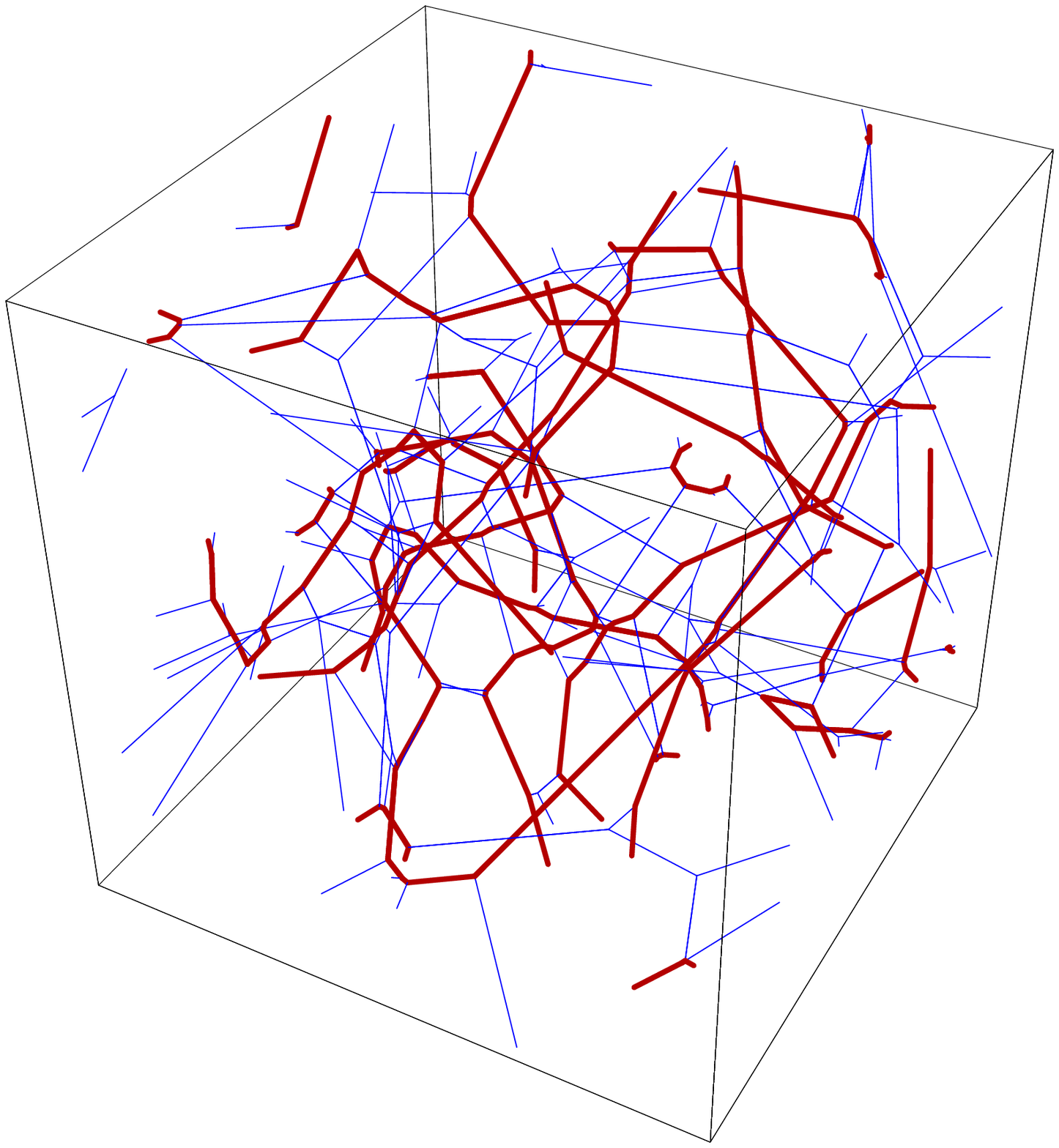}{\bcap{\lightSpics b}\/  Light $s$ strings, $t=4$}  
\insertwiidefigpage{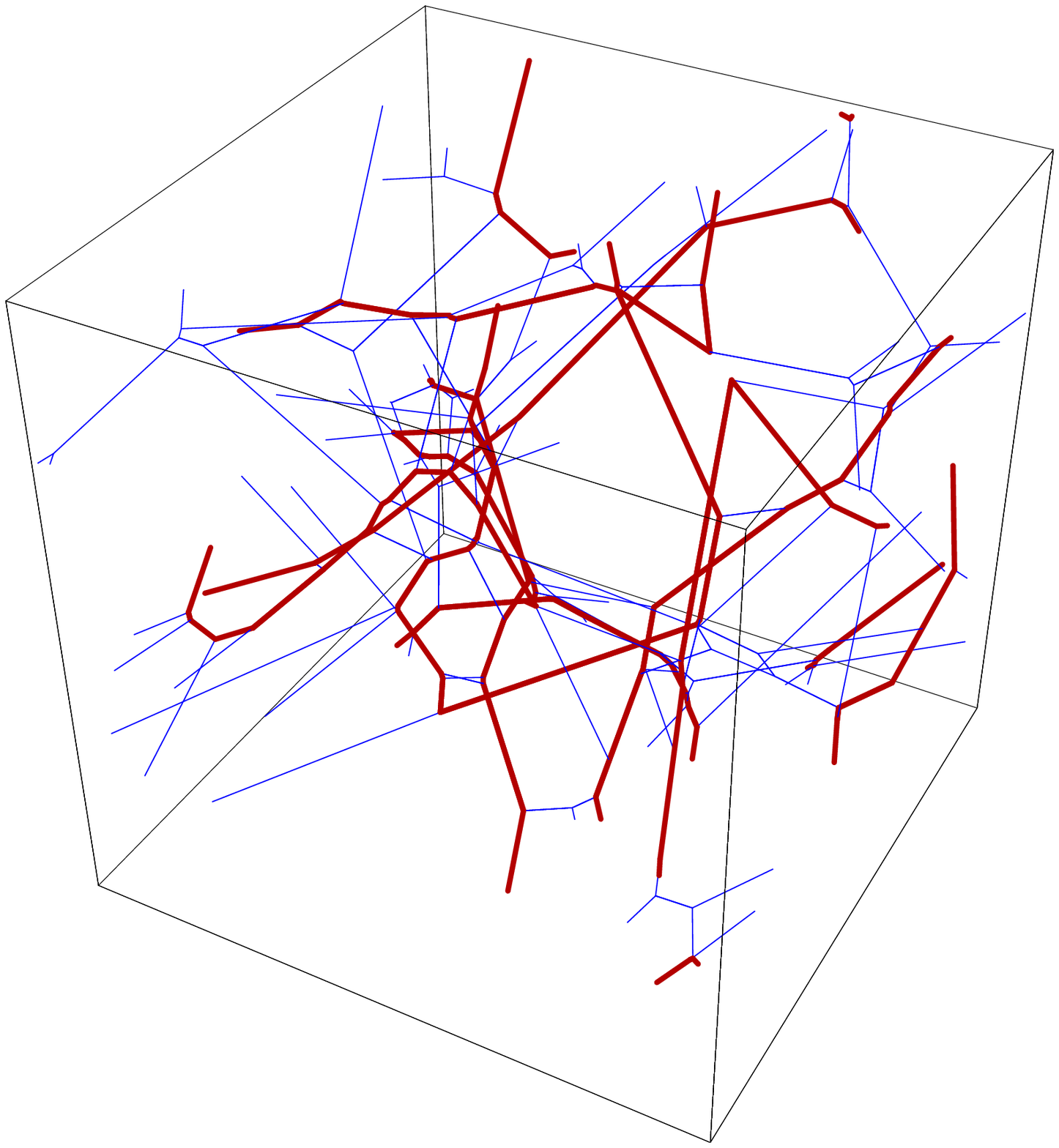}{\bcap{\lightSpics c}\/ Light $s$ strings,  $t=6$}  
\insertwiidefigpage{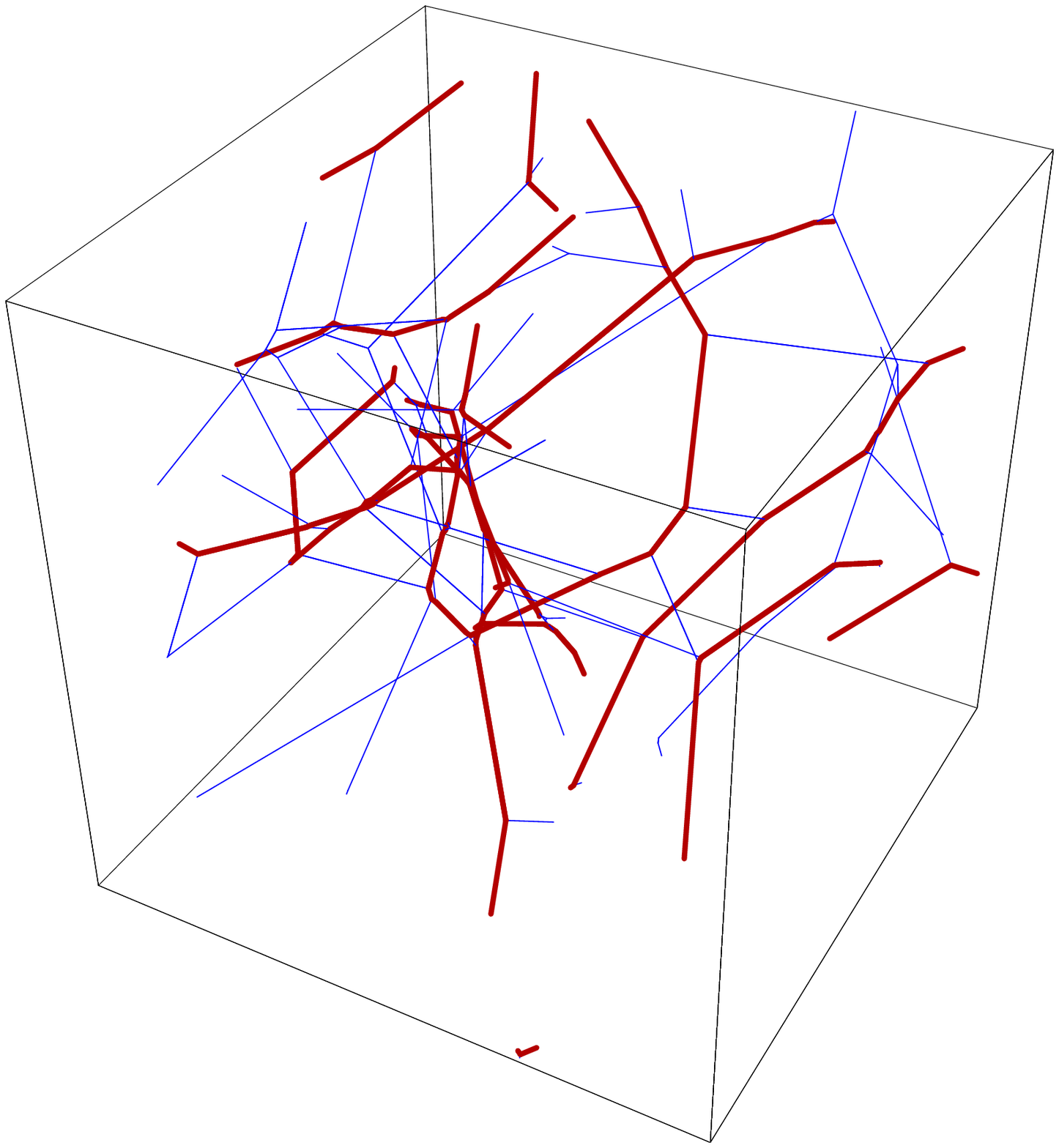}{\bcap{\lightSpics d}\/ Light $s$ strings, $t=8$}

\insertwiidefigpage{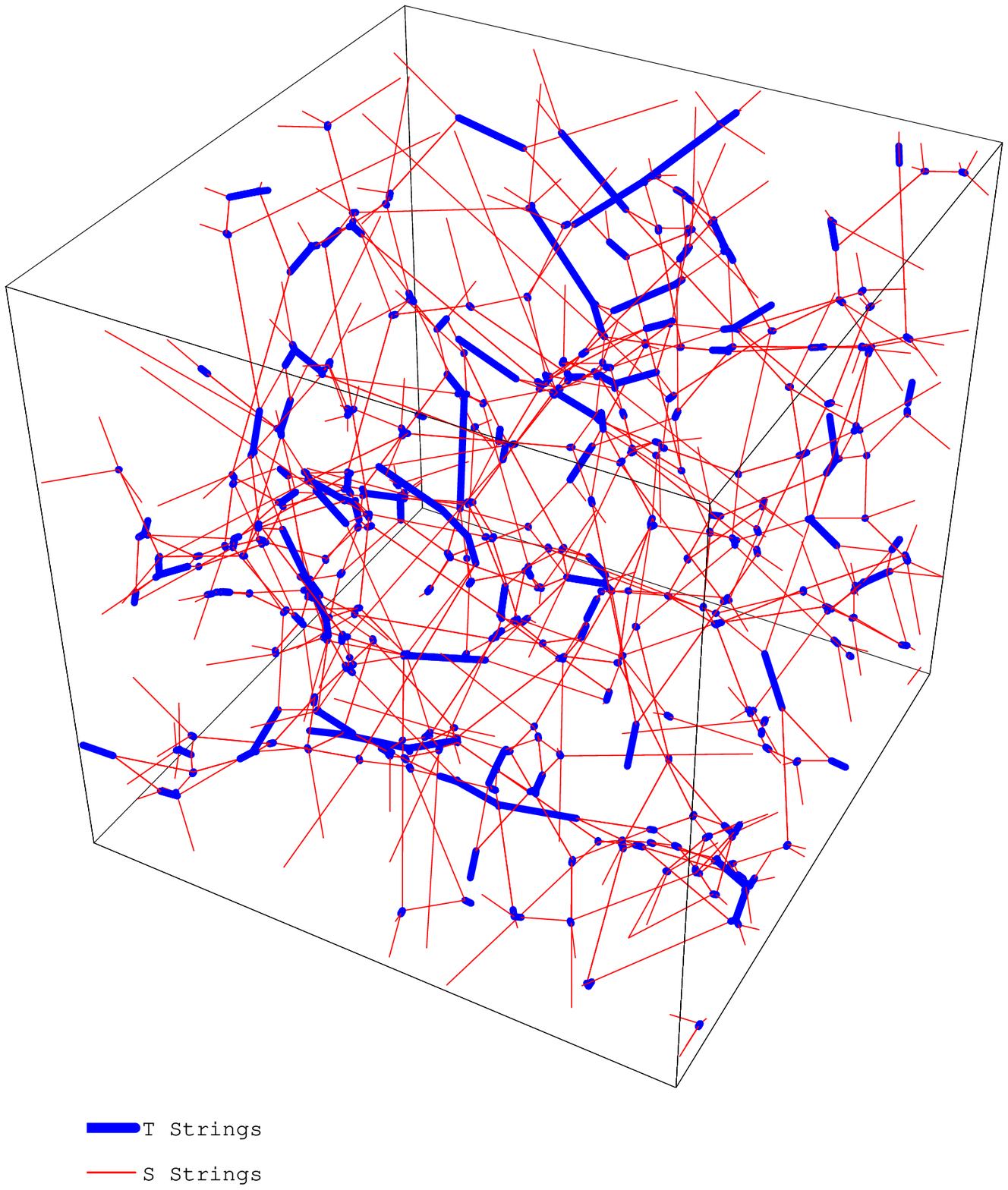}{\bcap{\heavySpics a}\/ Snapshots 
from the evolution of an $S_3$ network with heavy $S$ strings, from lattice 
gauge initial conditions.  The $S$ strings,
or even strings, are shown in thicker lines.  The full $10^3$ simulation 
volume is shown.  Notice the rapid shrinking of even segments, which leads to 
the formation of clusters that are slow to untangle. The first frame shown here
is at $t=1.5$.}

\insertwiidefigpage{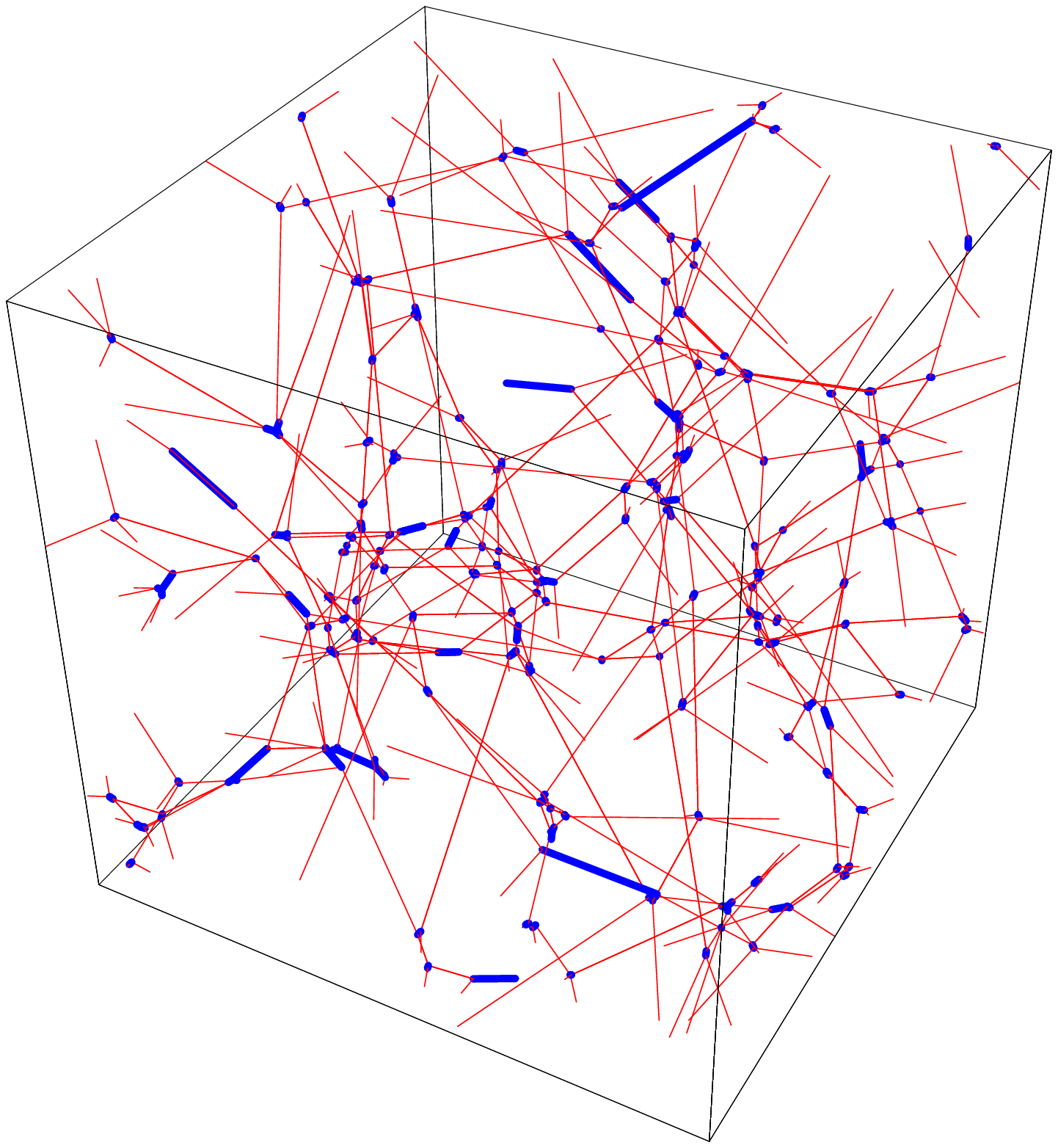}{\bcap{\heavySpics b}\/ Heavy $s$ strings, $t=3.0$}  

\insertwiidefigpage{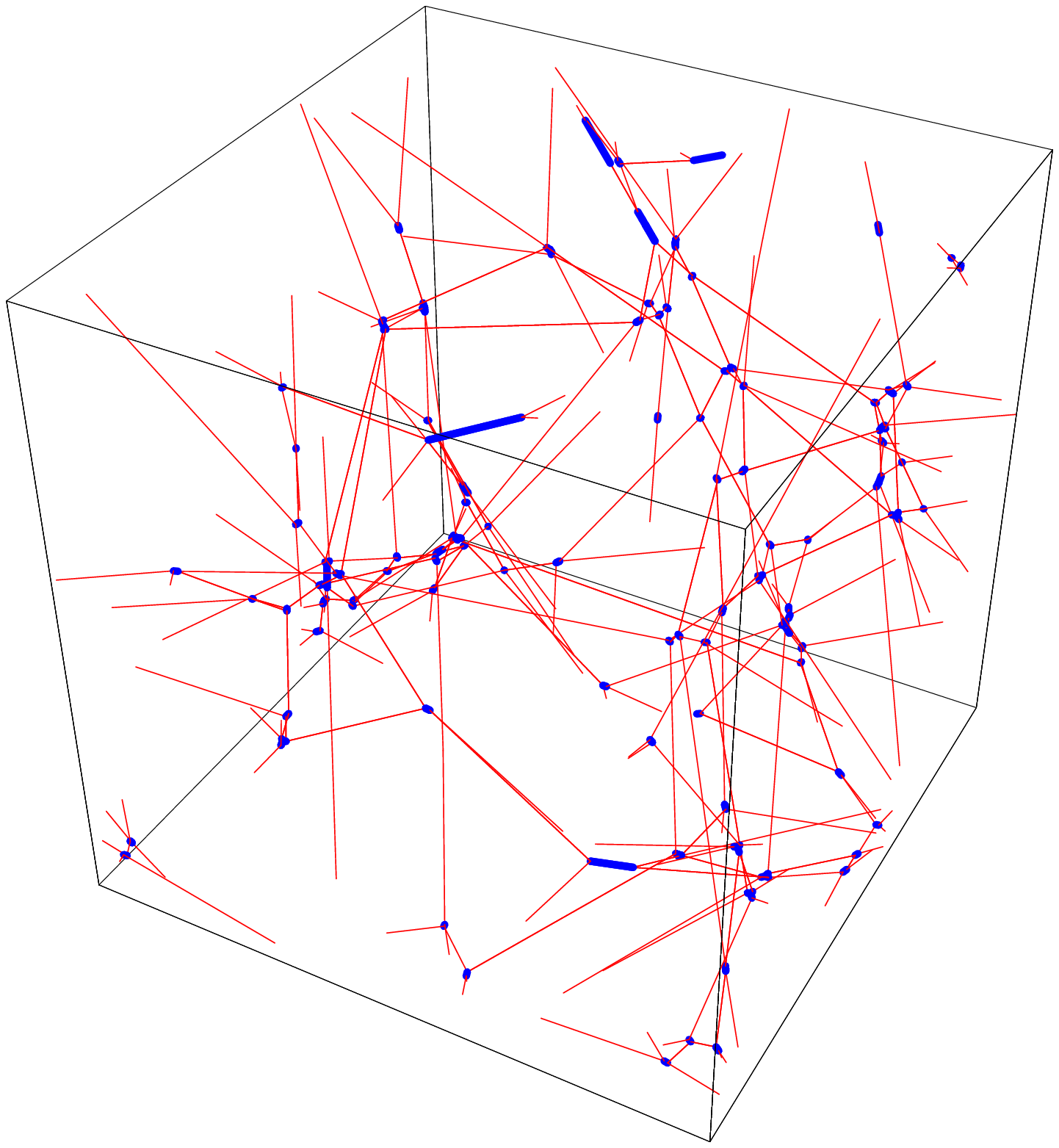}{\bcap{\heavySpics c}\/ Heavy $s$ strings,  $t=4.5$}  

\insertwiidefigpage{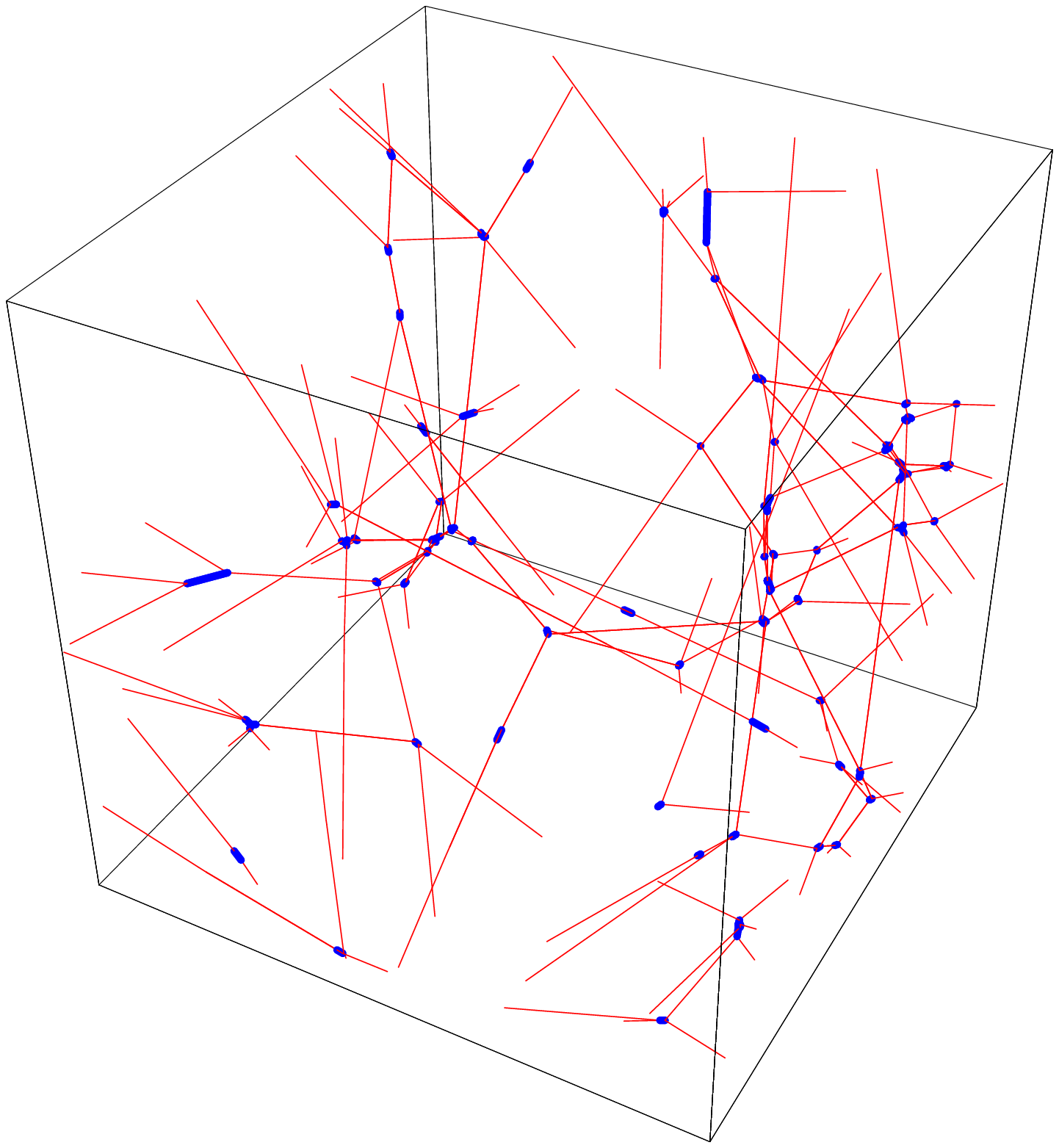}{\bcap{\heavySpics d}\/  Heavy $s$ strings, $t=6.0$}

\insertwiidefigpage{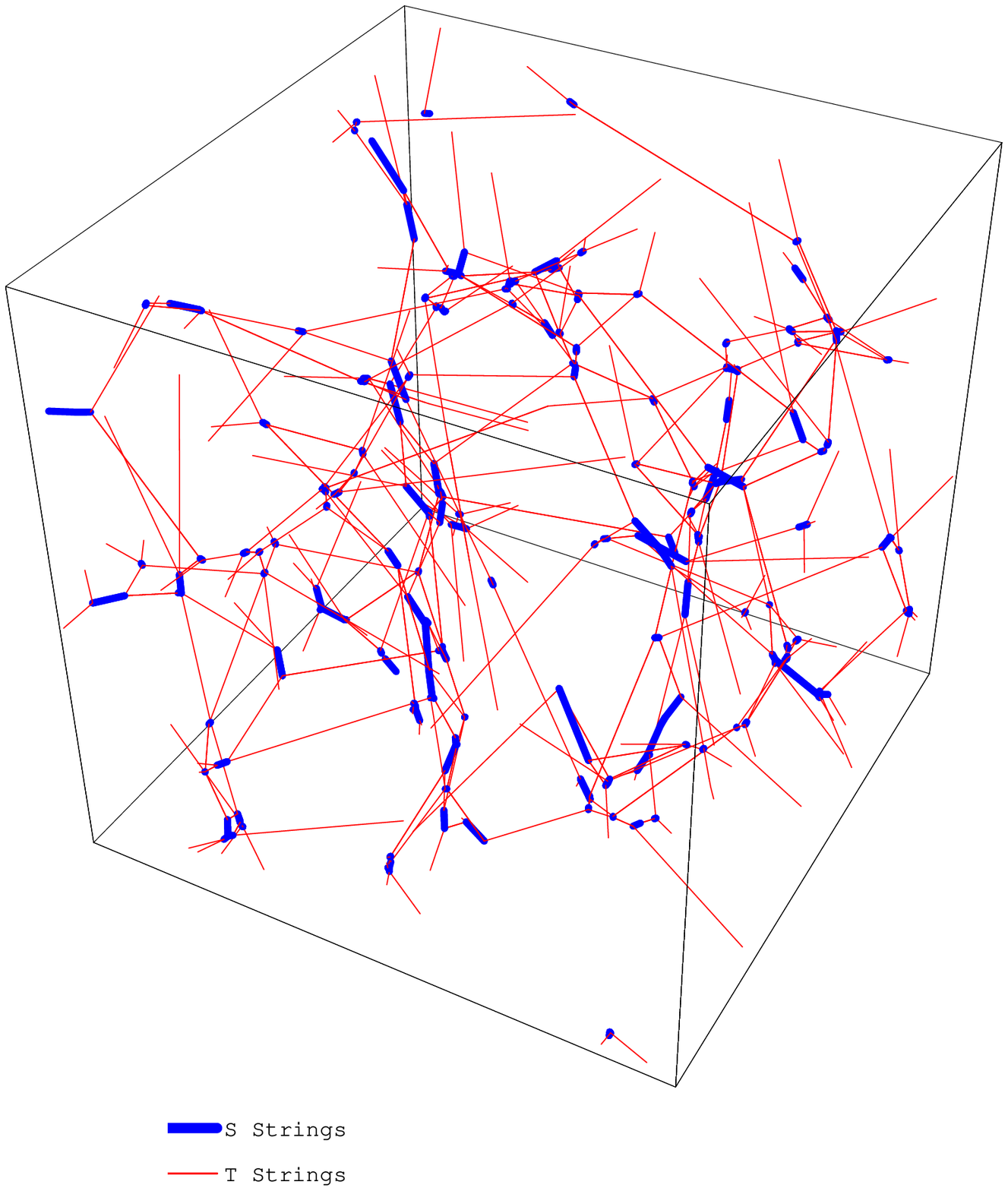}{\bcap{\heavyHSpics a}\/  Evolution of a 
network from Higgs initial conditions with heavy $S$ strings.  The qualitative
behavior is rather similar to that of the denser lattice gauge network, except
that the network disappears much more rapidly, with large voids opening up 
very quickly. This figure: $t=1.0$.}

\insertwiidefigpage{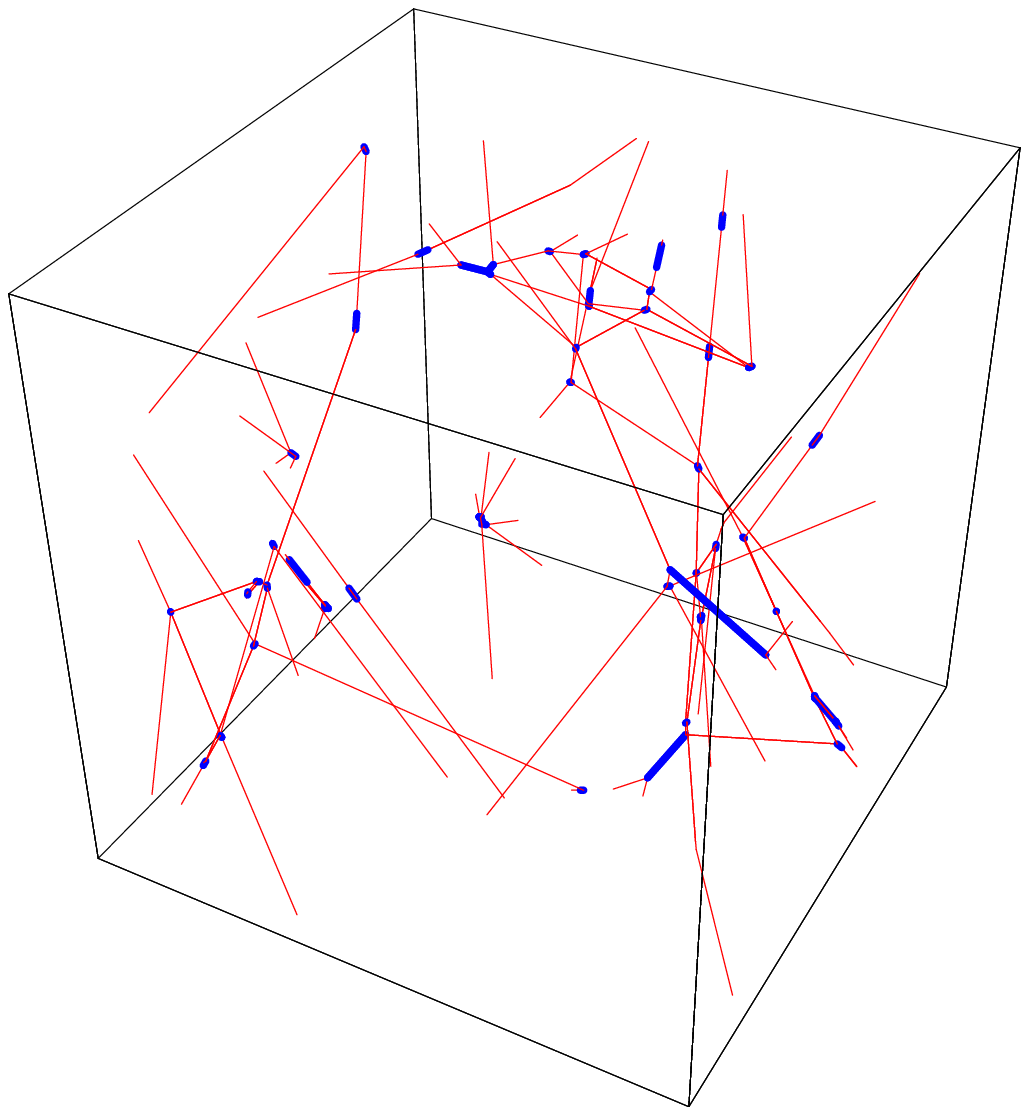}{\bcap{\heavyHSpics b}\/ Heavy $s$ strings, discrete Higgs initial conditions, $t=2.0$}

\insertwiidefigpage{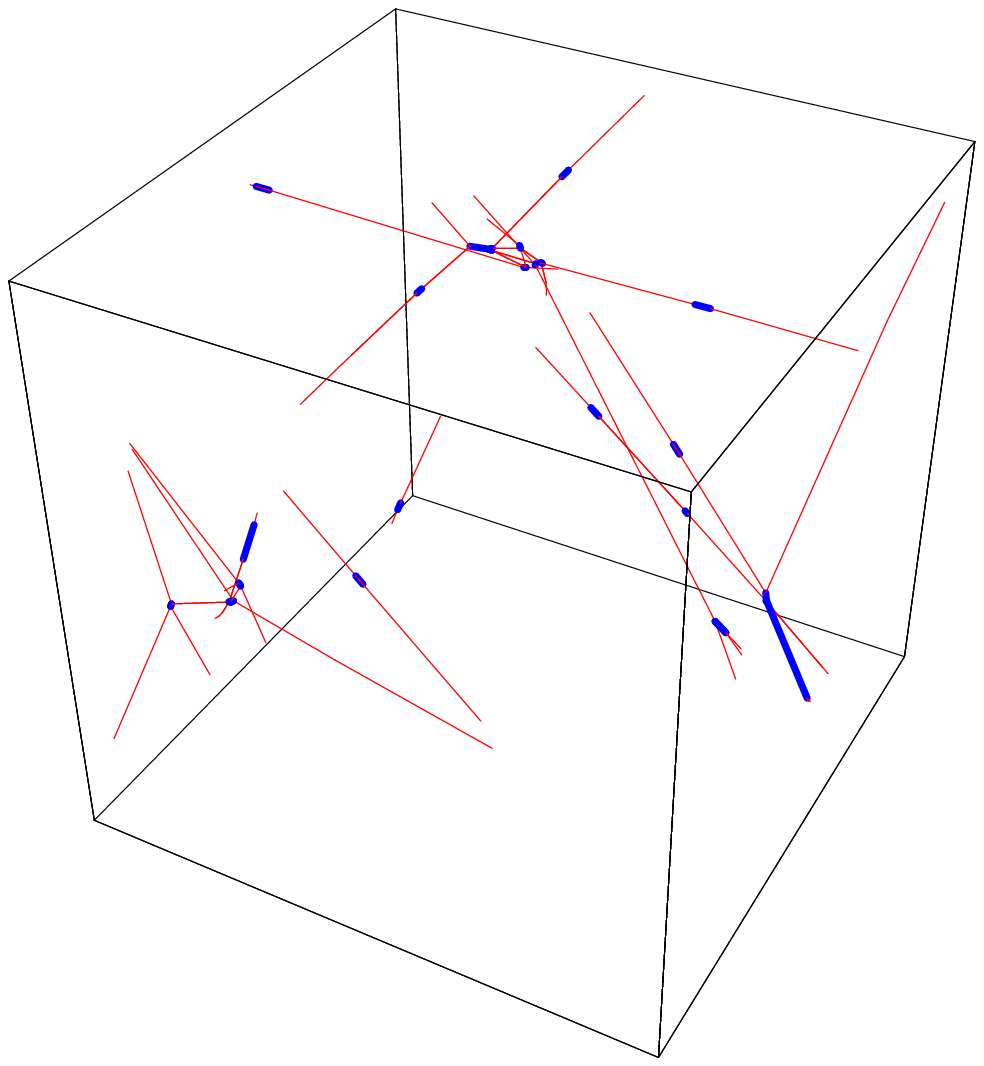}{\bcap{\heavyHSpics c}\/ Heavy $s$ strings, discrete Higgs initial conditions, $t=3.0$}

\insertwiidefigpage{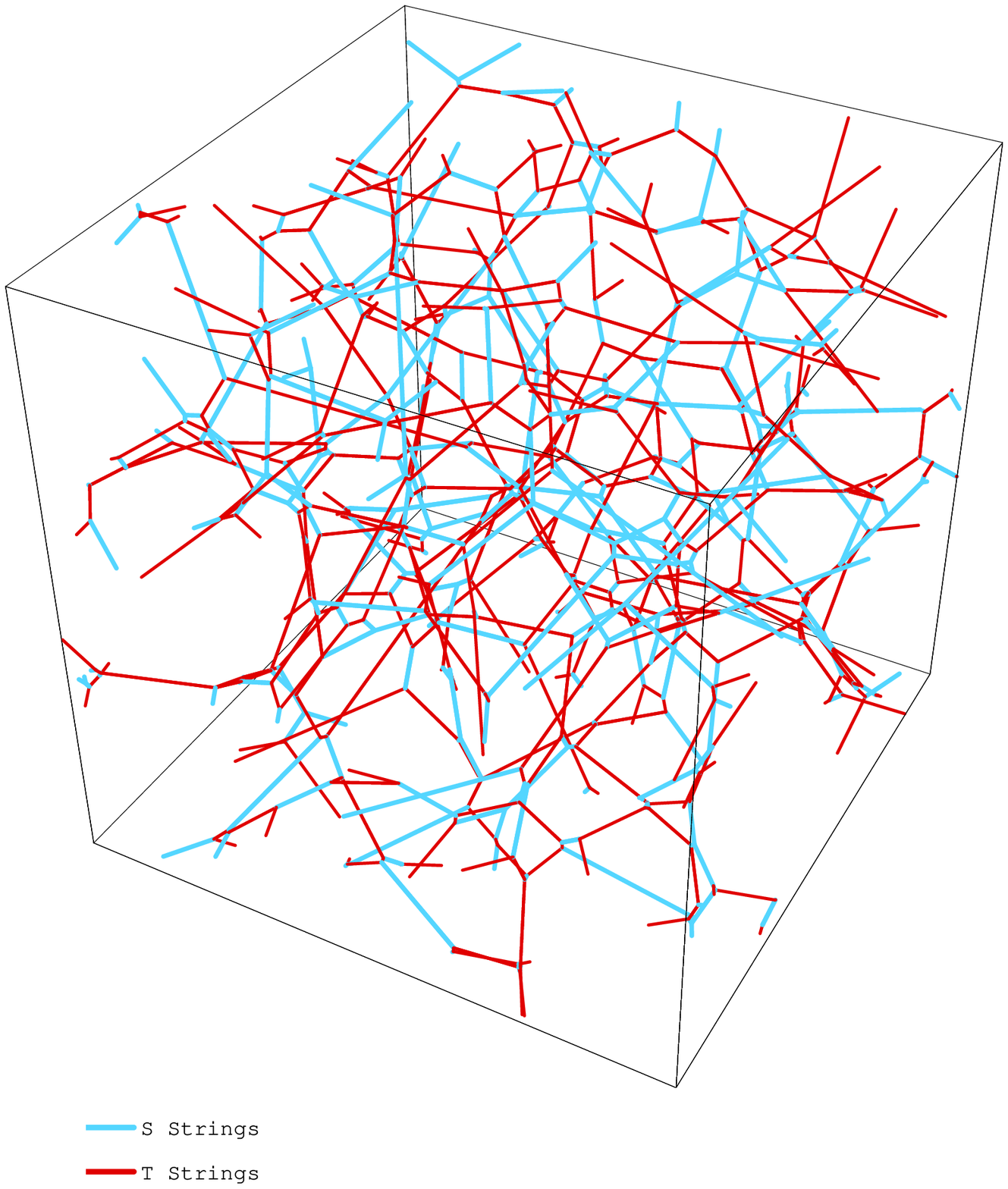}{\bcap{\Llpics a}\/  Evolution of a network
with equal string tensions, from lattice gauge initial conditions.  First
figure: $t=2.0$.}

\insertwiidefigpage{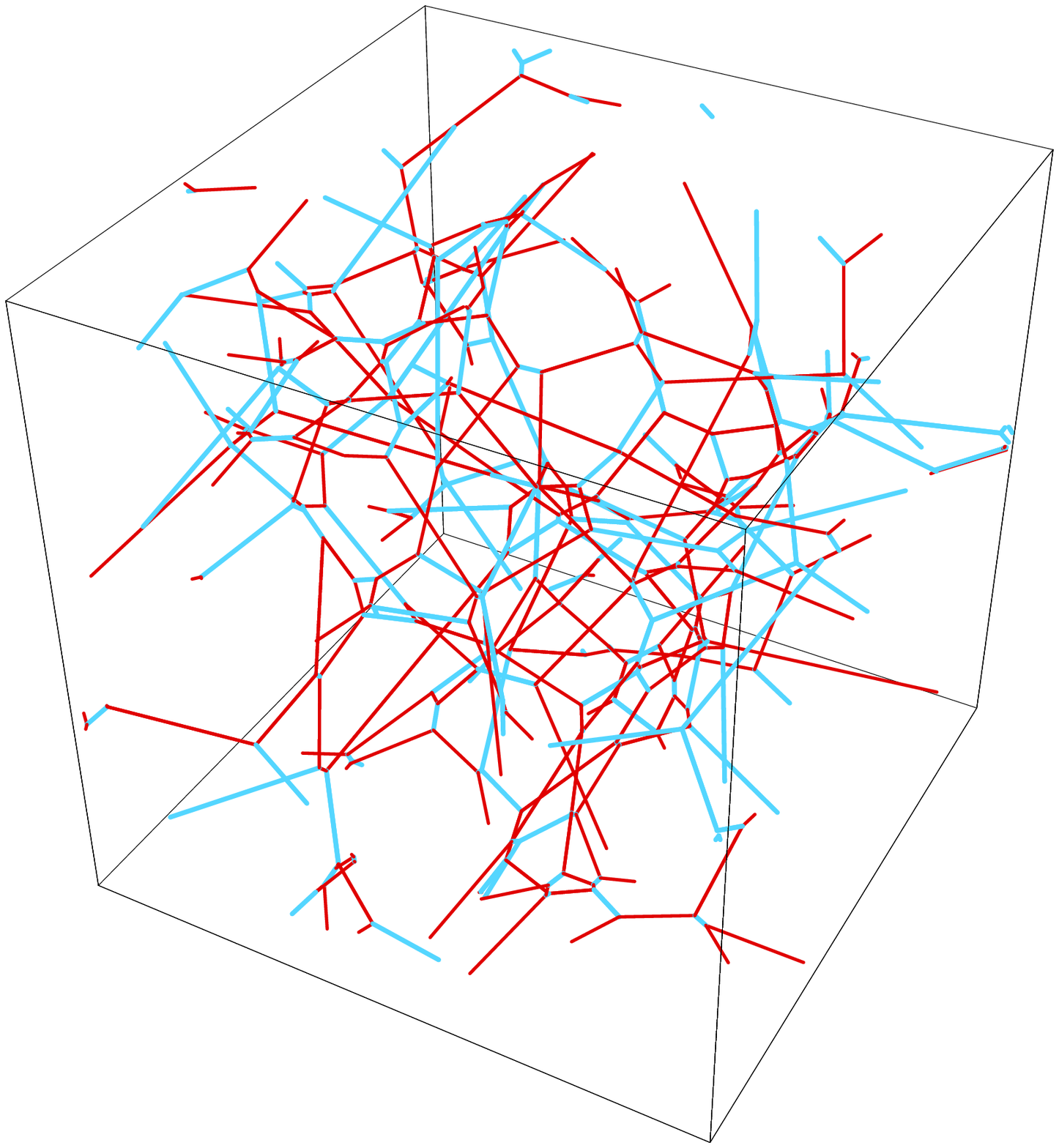}{\bcap{\Llpics b}\/ Equal tensions, $t=4.0$}

\insertwiidefigpage{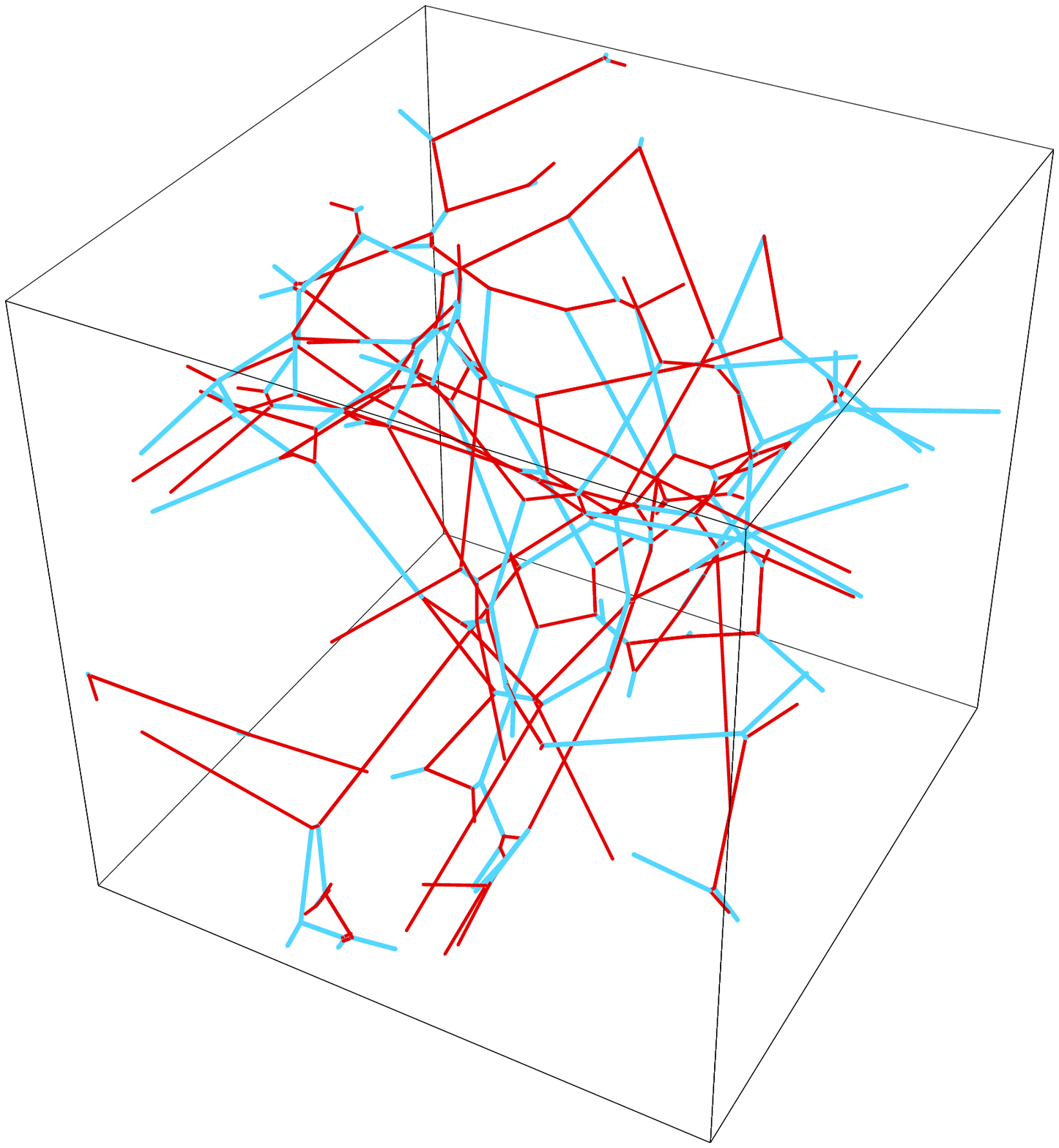}{\bcap{\Llpics c}\/ Equal tensions, $t=6.0$}

\insertwiidefigpage{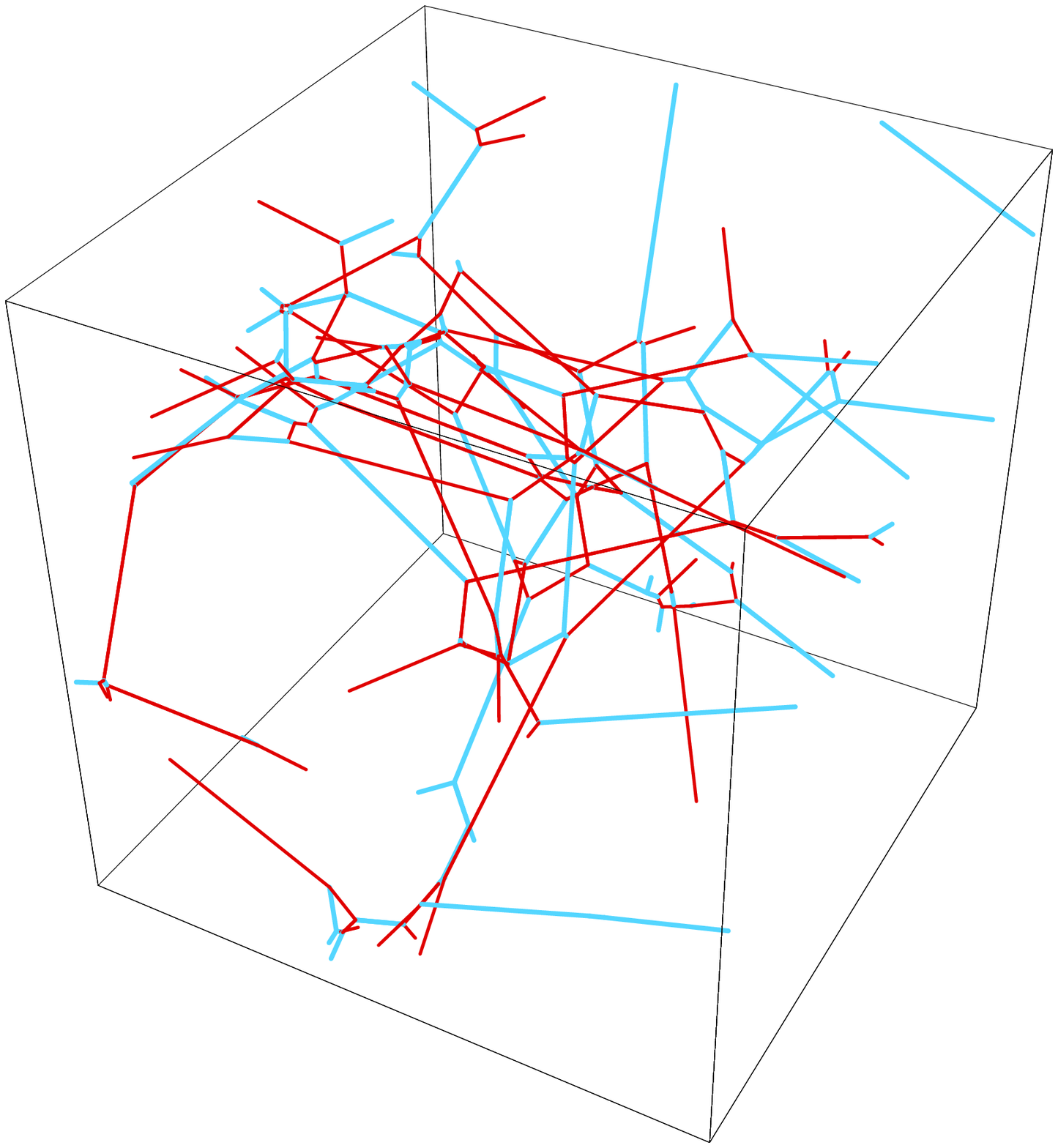}{\bcap{\Llpics d}\/  Equal tensions, $t=8.0$}

\insertwiidefigpage{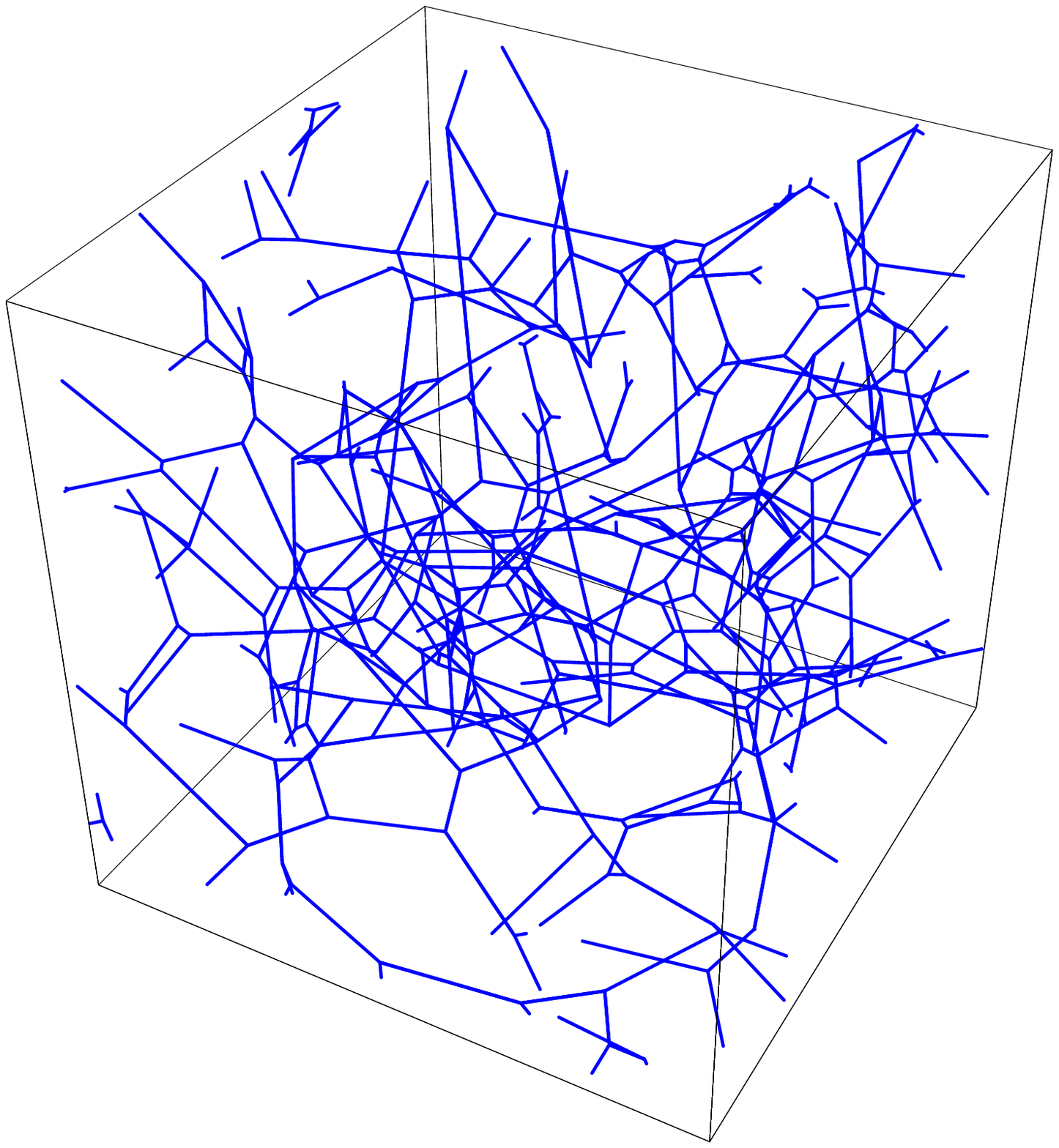}{\bcap{\Zpics a}\/ Snapshots from the evolution
of a pure $Z_3$ network, included for comparison with the non-Abelian network.
The first picture, shown here, is at $t=0$.}

\insertwiidefigpage{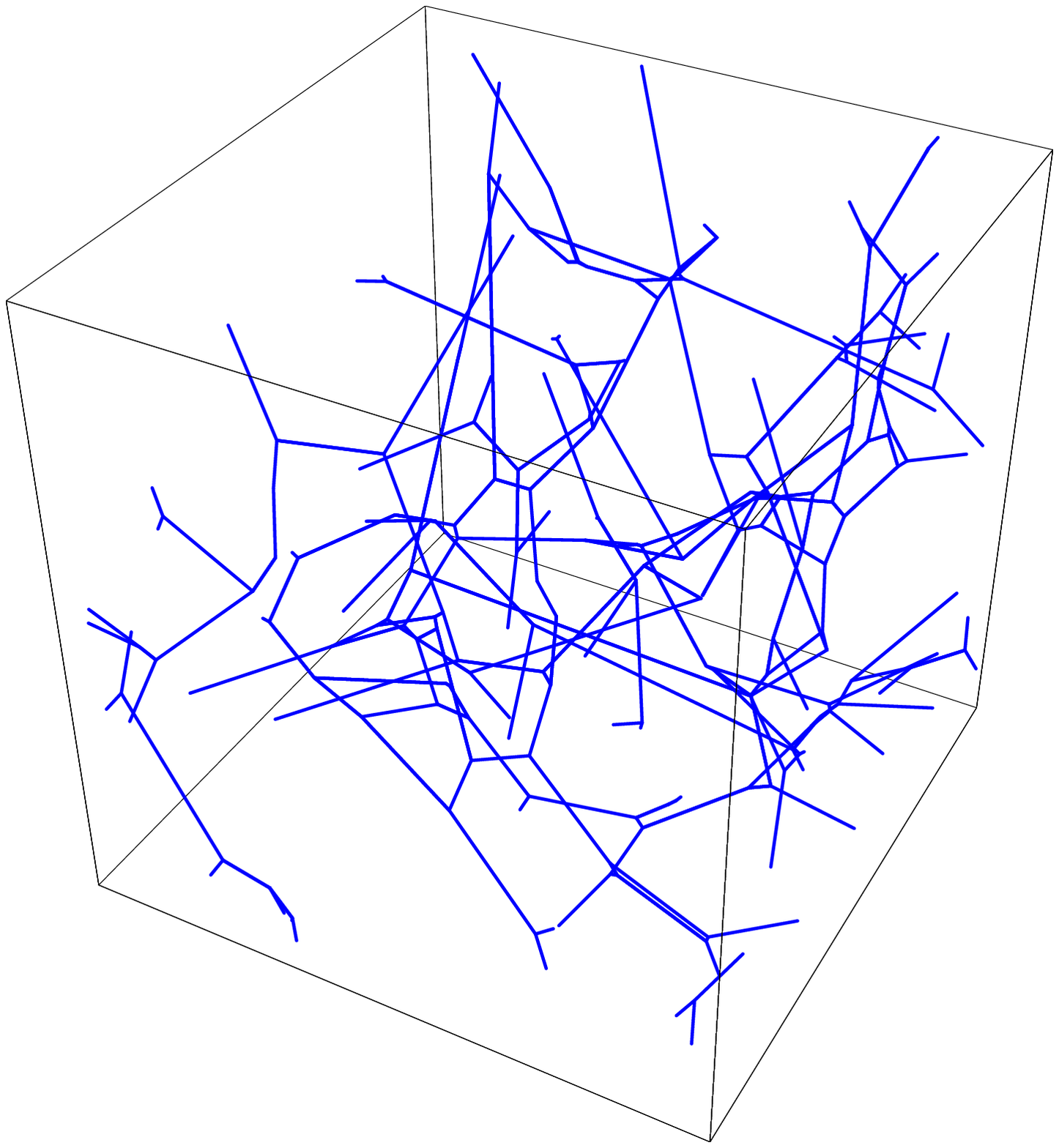}{\bcap{\Zpics b} $Z_3$ network, $t=2.0$}

\insertwiidefigpage{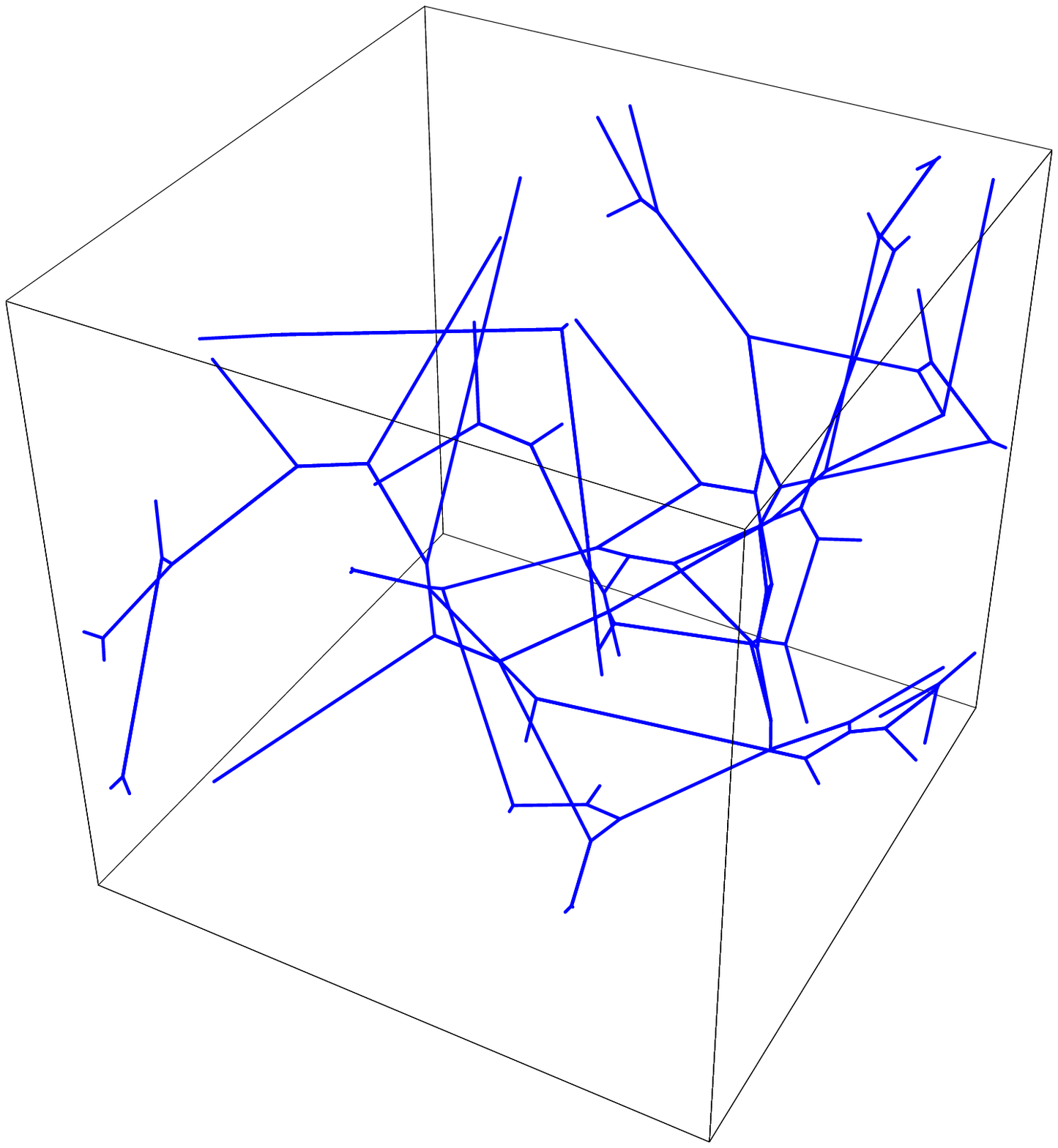}{\bcap{\Zpics b} $Z_3$ network, $t=4.0$}

\insertwiidefigpage{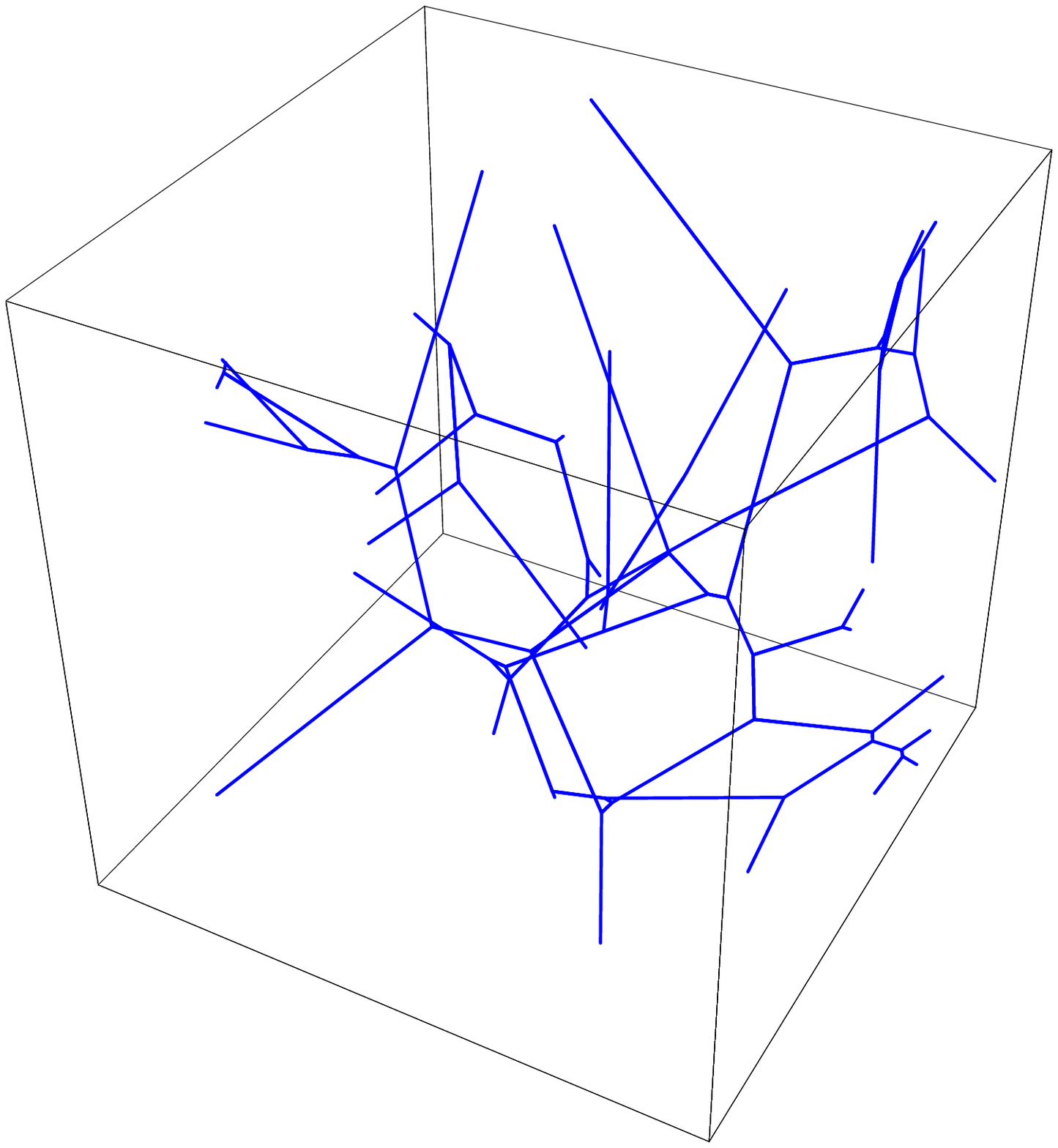}{\bcap{\Zpics b} $Z_3$ network, $t=6.0$}

\insertwiiderfigpage{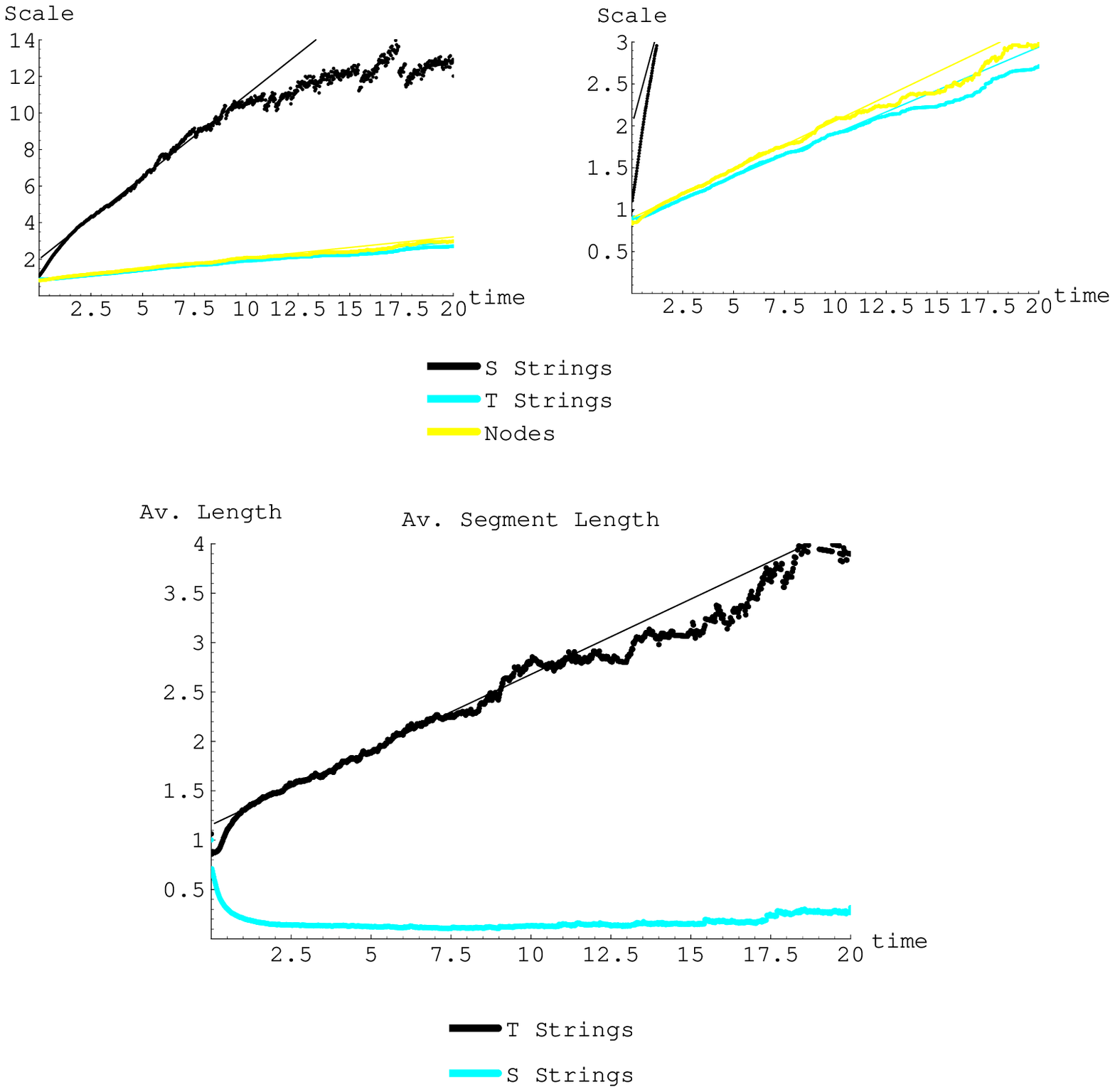}{\bcap\heavySplots\/ Scaling-law variables 
as a function of time for $S_3$ network with $T_s=2$, lattice gauge
initial conditions, bridge NCC and 
no rearrangement.} 

\insertlongfigpage{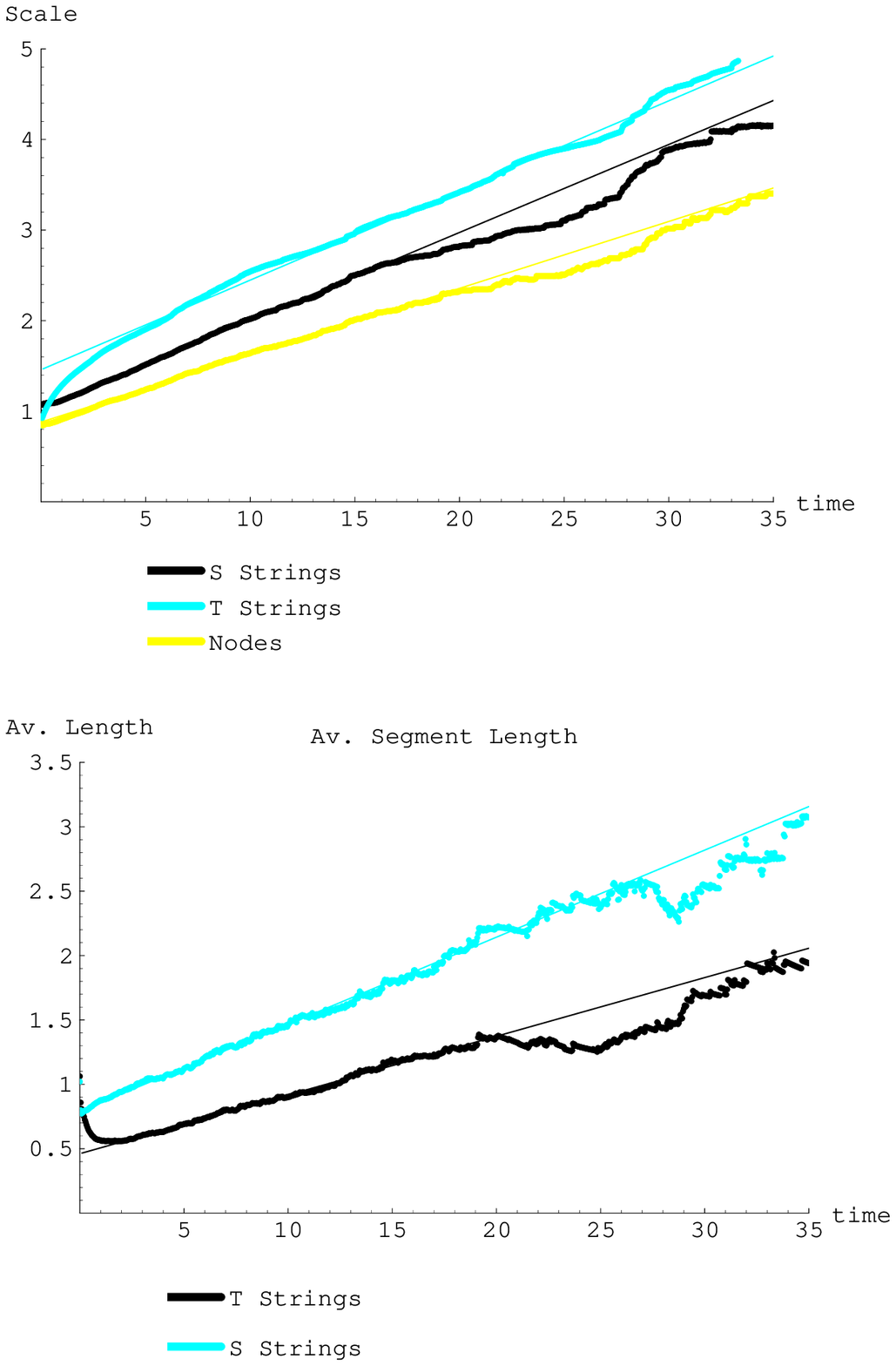}{\bcap\lightSPlots\/ Length scales for 
$S_3$ network with light $S$ strings. $T_s=0.5$, lattice gauge initial
conditions, bridge NCC, no rearrangements.}

\insertlongfigpage{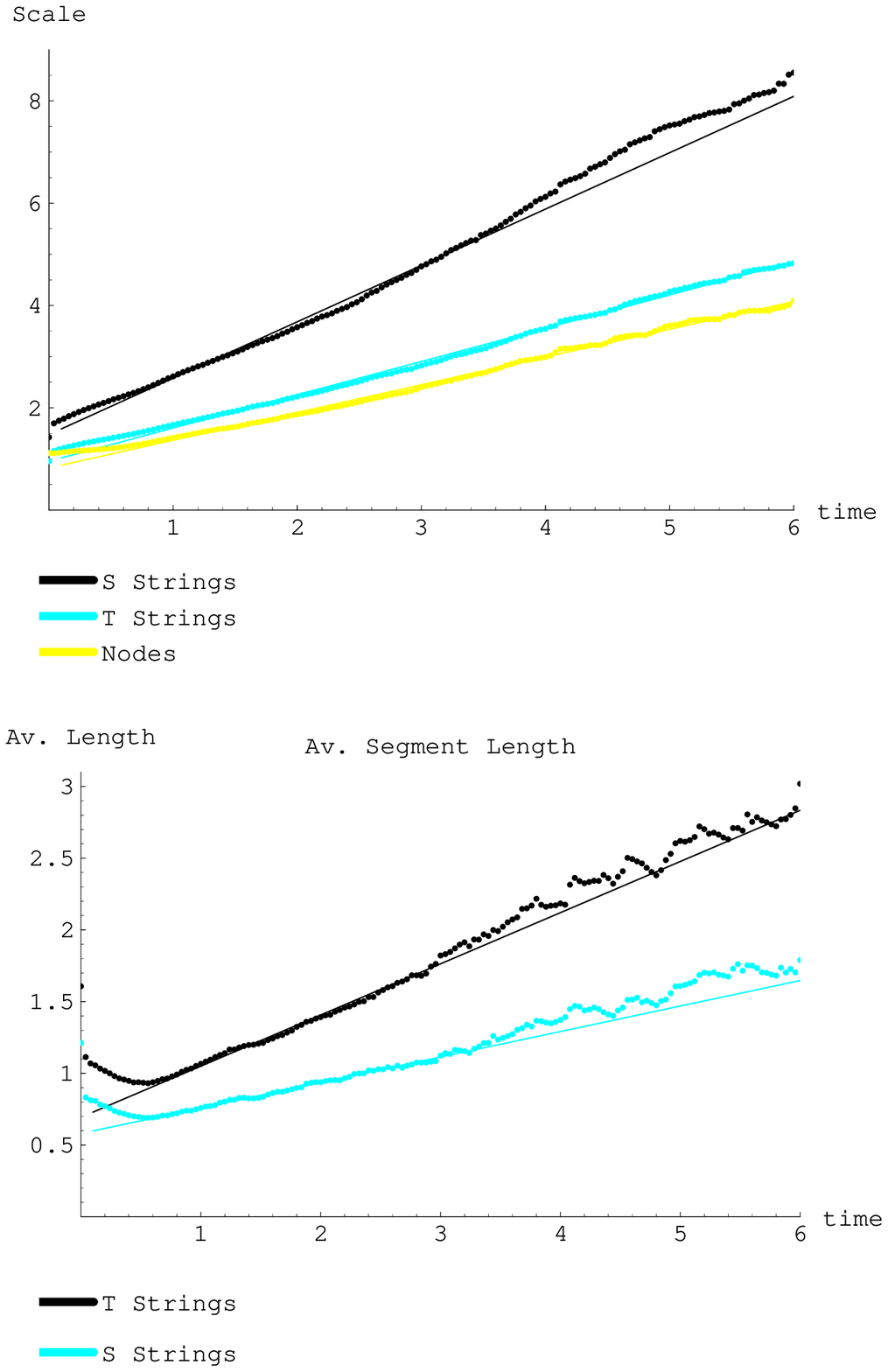}{\bcap\equalSPlots\/ Length scales for 
$S_3$ network with equal string tensions, Higgs initial conditions, bridge
NCC, rearrangement allowed.}

\insertlongfigpage{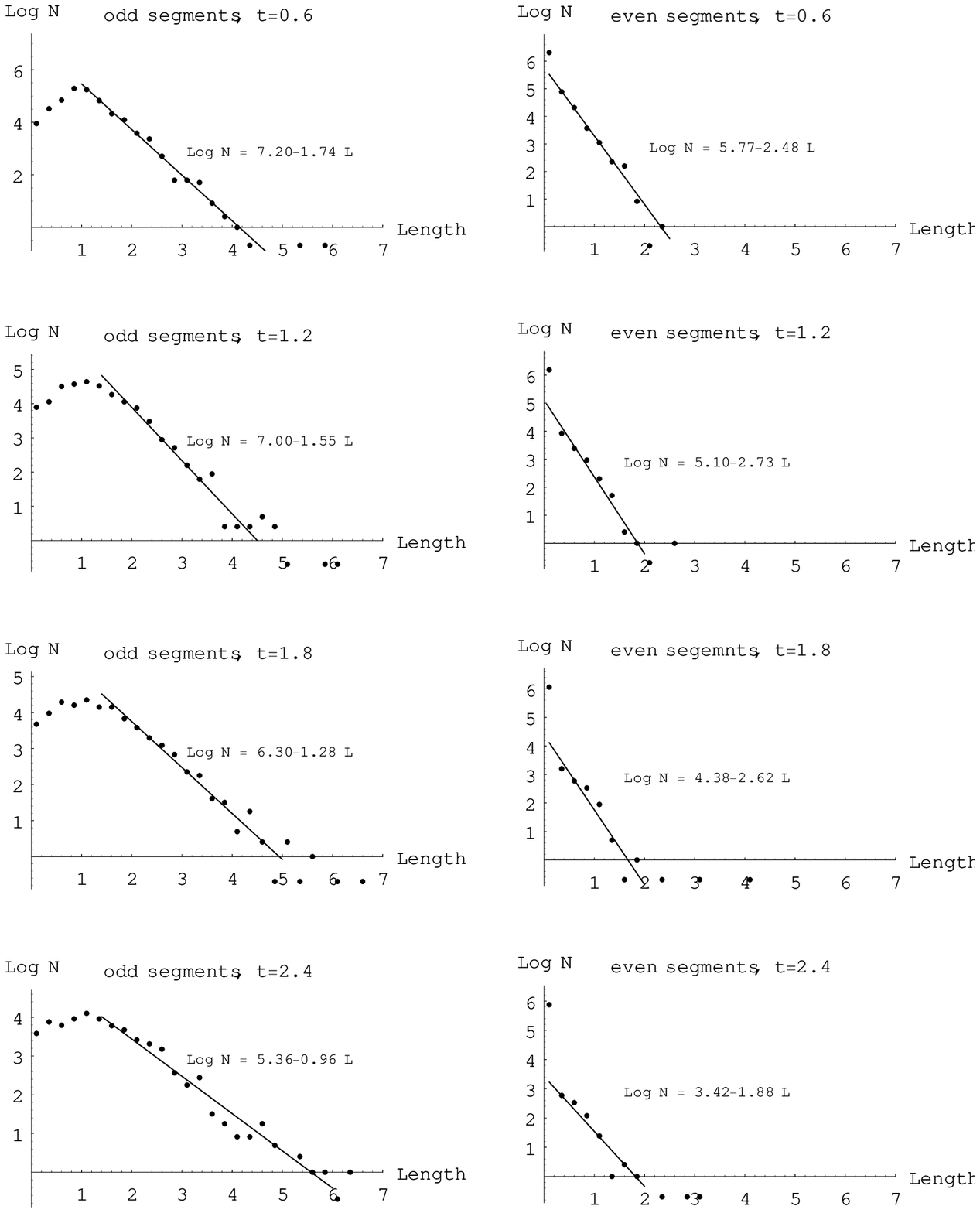}{\bcap\histograms\/ Evolution
of the distribution of length
of even ($S$) and odd ($T$) string segments between junctions, for a network
with heavy $S$ strings.}

\insertlongfigpage{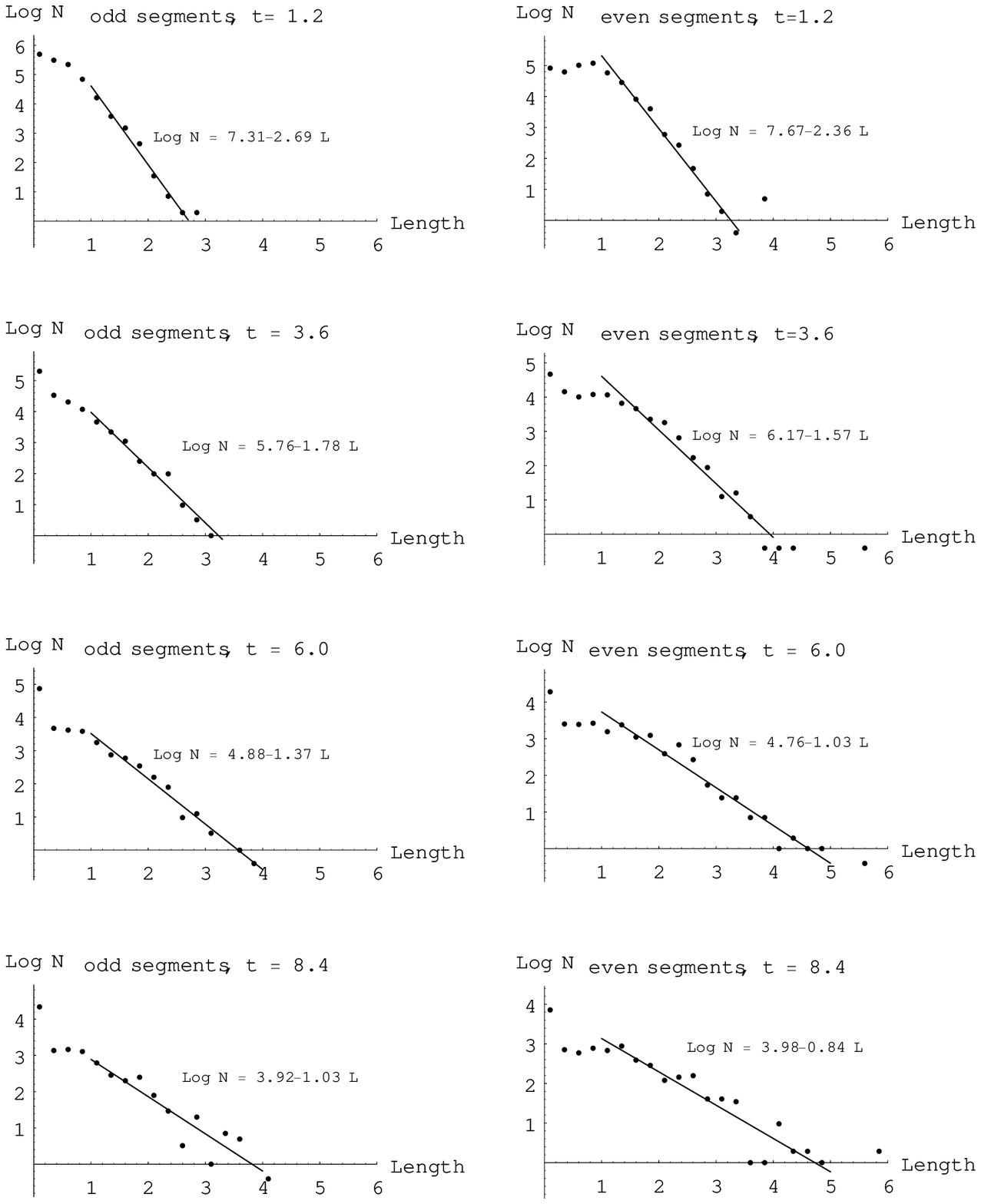}{\bcap\histogramt\/ Segment length
distributions for network with light $S$ strings.}

\insertlongfigpage{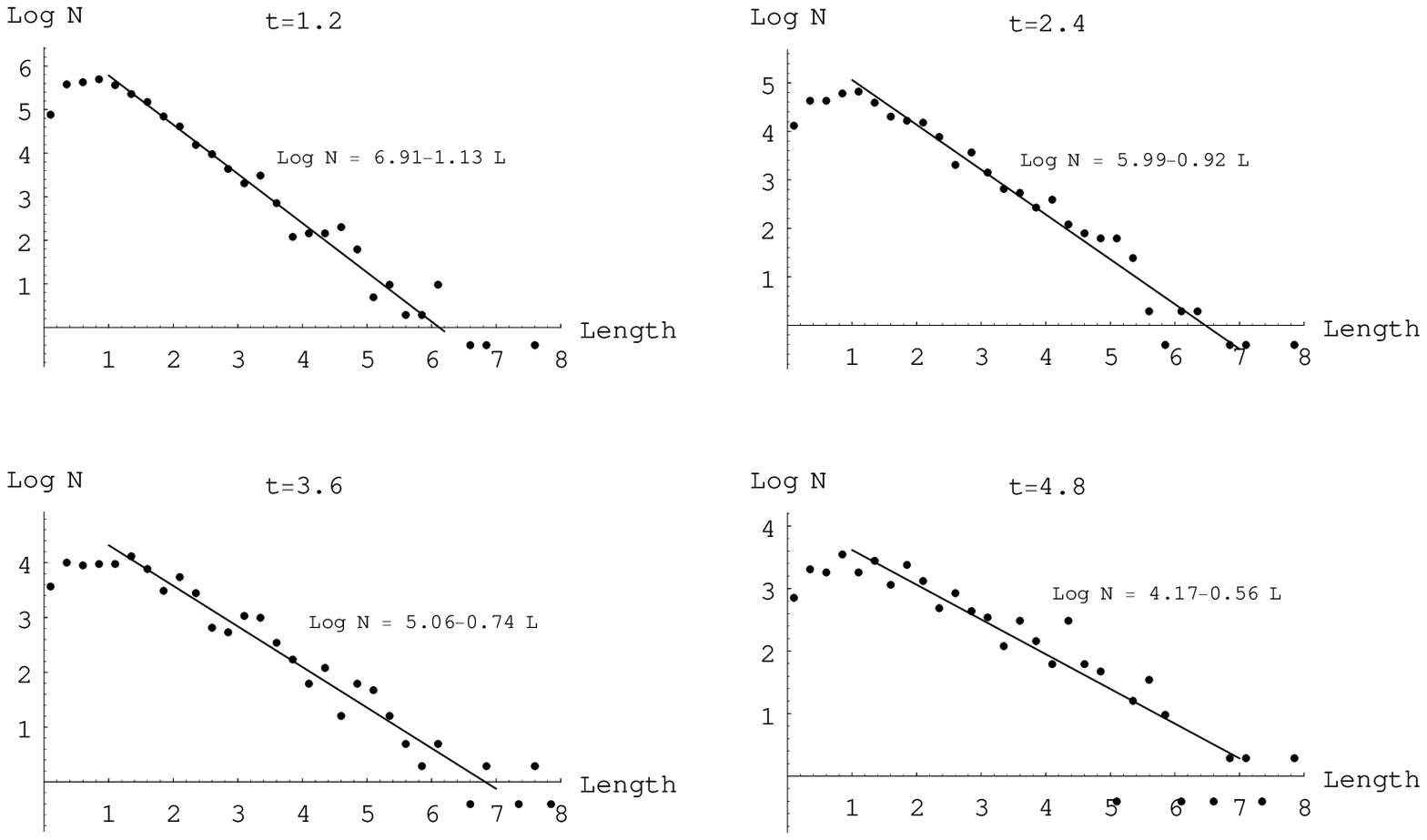}{\bcap\histogramz\/ Segment length
distributions for the Abelian $Z_3$ network, included here for comparison
with figures \histograms and \histogramt.}

Notice that the details of the behavior of closely approaching nodes 
makes some difference in the evolution.  The inability of some 
pairs of nodes to annihilate is apparently an important impediment to the
network's destruction.  Rearrangements (quasi-intercommutations) of the connections of neighboring
nodes evidently increase the mobility of flux, and increase the likelihood that
eventually some neighboring pairs of nodes will be able to annihilate.
As is apparent from Table 2, the inclusion of rearrangements almost invariably speeds up the decay of the network, often by roughly a factor of two.  The choice of either zipper or bridge configurations for
colliding non-commuting strings generally makes a smaller 
difference, if any. 

\subsection{Finite-Size Effects, Fluctuations and Uncertainties}

A few remarks are in order concerning the interpretation of 
results and the effects of simulating on a 
finite volume.  The size of our simulation volume is limited in practice
by computation time;  the flux computations (see Appendix) are computationally
intensive.
Typical plots of scale variables like those in figures 
\heavySplots\- - \equalSPlots\- exhibit some transient behavior at 
early times, followed by a period of linear increase. As the typical distance between nodes increases, however,
a point is eventually reached where only a few nodes remain within the simulation volume.  At this point, both systematic and random 
deviations from linear scale growth begin to occur.  With fewer nodes, curves
become bumpy due to lower statistics.  Many strings wind all the way around cycles of the periodic
boundary, so that the simulated network can no longer be expected to mimic
an infinite one.  A typical run meets one of two distinct fates at the end of its
period of self-similar evolution.  It may undergo a very sudden collapse as the 
last remaining nodes annihilate, leaving either no 
strings at all or a few strings stretched across the entire volume.  This sudden collapse is visible as a sharp upward turn in the
plots of length scale versus time.  Alternatively,  the network may reach
a stable or metastable configuration with a small number of nodes (typically
of order ten or fewer.)  The characteristic length scale ceases its linear
increase and reaches a plateau.  Needless to say, we cannot make reliable inferences about an infinite network once the scale of our simulation
volume becomes the important one, and we can only extract 
information on scaling behavior from the linear part of the curve. 
The conclusion that our data are consistent with scaling is based on the 
existence of a linear stretch which typically continues until at least
one of the length scale variables has grown to one half the width of the simulation volume.  We cannot rule out slow (e.g., logarithmic) deviations 
from scaling.

With respect to fluctuations, not all simulation runs are alike.  Runs
with lattice gauge initial conditions and $T_s=2$ seemed to undergo quite
a bumpy evolution characterized by periods of temporary freezing alternating
with cascades of annihilation.  The bumpiness was smoothed out only by averaging
over multiple runs. Runs with $T_s=2$ without rearrangement almost always 
end with a few nodes remaining rather than with complete collapse.
This is reflected in figure \heavySplots, where the curves of $D_t$ and 
$D_n$ versus time begin levelling off as the average segment length
approaches $4$ (which is half the width of the volume).  This levelling
is probably a finite-size effect and not indicative of the behavior 
of an infinite network for the following reasons:  It occurs
when only a few nodes are left in the simulation volume, and it occurs 
at later times as the simulation volume is increased.  In fact,
one run performed on a larger $10^3$ volume did not level off at all,
but collapsed entirely instead.  Higgs-
generated networks tend to evolve more smoothly and to be more likely to
undergo total collapse at the end of the simulation.

\section{Conclusions and Discussion}

One of the motivations for undertaking a simulation of non-Abelian string
dynamics was to test for deviations from the familiar power-law behavior of
the network's energy density as a function of time, and especially to
look for evidence for or against the conjecture that the tangling of strings would cause a non-Abelian network to
freeze into a static (fixed in comoving coordinates) equilibrium.  The results we have found suggest that such a scenario,
if it can occur at all, requires very special conditions.  Over a 
range of different regimes, the results
found here are consistent with some form of self-similar evolution,  and with
a density decaying at a rate commensurate with that of matter,  $n=C t^{-2}$.
The coefficient $C$, however, can vary in quite interesting ways due to
non-Abelian effects.  In a cosmological scenario, this coefficient controls
the fractional contribution of strings to the energy of the universe.

The interplay between
the different string species  causes the self-similar evolution to be realized
in some novel ways which can be quite different under different conditions.
Features such as the ratio of the populations of different string species 
may depend in rather complicated ways on various factors.   Particularly
striking and new is the strong influence of the initial distribution of 
strings
on the subsequent evolution.  Evidently,  different self-similar evolutionary trajectories
are possible, and initial conditions may be attracted to one trajectory or
another depending on some statistical feature other than the network's 
overall density.  This behavior seems almost paradoxical, going against the
notion that a scaling evolution is one which has no memory of its initial
state.  

There are interesting questions to investigate at both the ``microscopic''
and ``macroscopic'' level.  By microscopic questions, we mean those 
concerning the 
dynamics of individual strings and individual collisions.  Our results
indicate that the behavior of nodes in the network as they encounter each
other has a controlling influence on the network's evolution.  The 
inability of non-Abelian nodes to annihilate is an important impediment
to the removal of strings, and configurations with nodes close together
or coalescing into composites appear to be especially important in the
interesting case of heavy $s$ strings.  The process we have called ``rearrangement'' 
of two nearby nodes seems to increase the mobility of strings and
allow the network to decay more quickly.   This simulation was run with rather ad hoc
assumptions about node collisions.  A better microscopic understanding of
 nodes' 
behavior and that of multi-node tangles  will help provide input for improved simulations and understanding
of macroscopic questions.  

At the macroscopic level, it would be desirable to obtain
a better understanding of the principles governing
the interplay between different string types and  the
strong influence of initial conditions.  The types of network behavior
seen here may open up new possibilities for cosmological model-building, aside
from the string domination scenario.  Some of the types of structures seen here,  such as webs of light string stretched between 
heavier ones (as in the light-$s$-string scenario of this simulation) or
the tangled clusters seen the $T_s=2$ simulation might 
have interesting cosmological effects.  We have seen hints that
the strength of fluctuations in string density is different under different circumstances, being especially strong for $T_s=2$.  Might these fluctuations
be interesting sources of structure formation?    
Along with analytical study, 
more refined simulations of branched
and non-Abelian networks may well prove worthwhile.

\Appendix{A}

\centerline{\fourteenrm Simulation of non-Abelian Fluxes}

The subtle nature of non-Abelian magnetic flux in discrete gauge theories
has been the subject of quite interesting research,
\REF\ABOP{H.K. Lo, Phys. Rev. {\bf D52} (1995),7247;  preprints
hep-th/9502079, hep-th/9411133.}\REF\QHair{M. Alford, S. Coleman and
J. March-Russell, Nucl. Phys. {\bf B351} (1991), 735;  M. Alford, J. March-Russell and F. Wilczek, Nucl. Phys. {\bf B337} (1990),695.}  
\refmark{\Bais,\various,\Bucher,\ABOP,\QHair}  
and presented algorithmic and computational challenges for this
simulation.  Ambiguities in the definitions of
non-Abelian fluxes, mentioned in section 2,  require that we design a very careful
procedure to fix consistent definitions and maintain their consistency as
the strings move.  This must be done in order to allow us to make the appropriate comparisons of fluxes when two strings collide or two nodes attempt
to annihilate.  Additional subtleties occur as a result of our using 
periodic boundary conditions:  we are dealing with a discrete gauge theory
on a non-simply-connected space manifold,  and holonomies associated with non-trivial
cycles become important. 

This appendix describes our procedure for defining and comparing the 
fluxes of non-Abelian strings.  Further details may be found in [\mcgraw].
As in section 2, we use the formalism developed in [\various, \Bucher ].

\subsection{Gauge Fixing Conventions}
 
In our algorithm,  the strings and nodes exist inside a rectangular volume with opposite sides identified: a 3-torus.  The subtleties associated with the  periodic boundary conditions will be discussed later: for now we simply consider a network inside a rectangular volume with boundaries.  We choose a cubic volume
with one corner at $(0,0,0)$ and with side length $L$. As explained
in section 2, it is necessary to define fluxes using paths that begin and 
end at some basepoint.
We choose a basepoint $x_0$ at the center of our simulation 
volume, $x_0=(L/2,L/2,L/2)$.  Let each node be associated with a straight line segment (a ``tail'')
along the direction ${\vec{BN}}$ from the basepoint to the node's location.  Then let the flux
of each outgoing string be defined with respect to a path which runs
outward along this tail to a point which is taken to be vanishingly close
to the the node.  The path then encircles the string in a counterclockwise direction and returns to the basepoint along the node's tail. This is 
illustrated in figure \FIG\tail ~ \tail.  This will be our convention for defining the fluxes of the strings which join at a given junction.

\insertfigpage{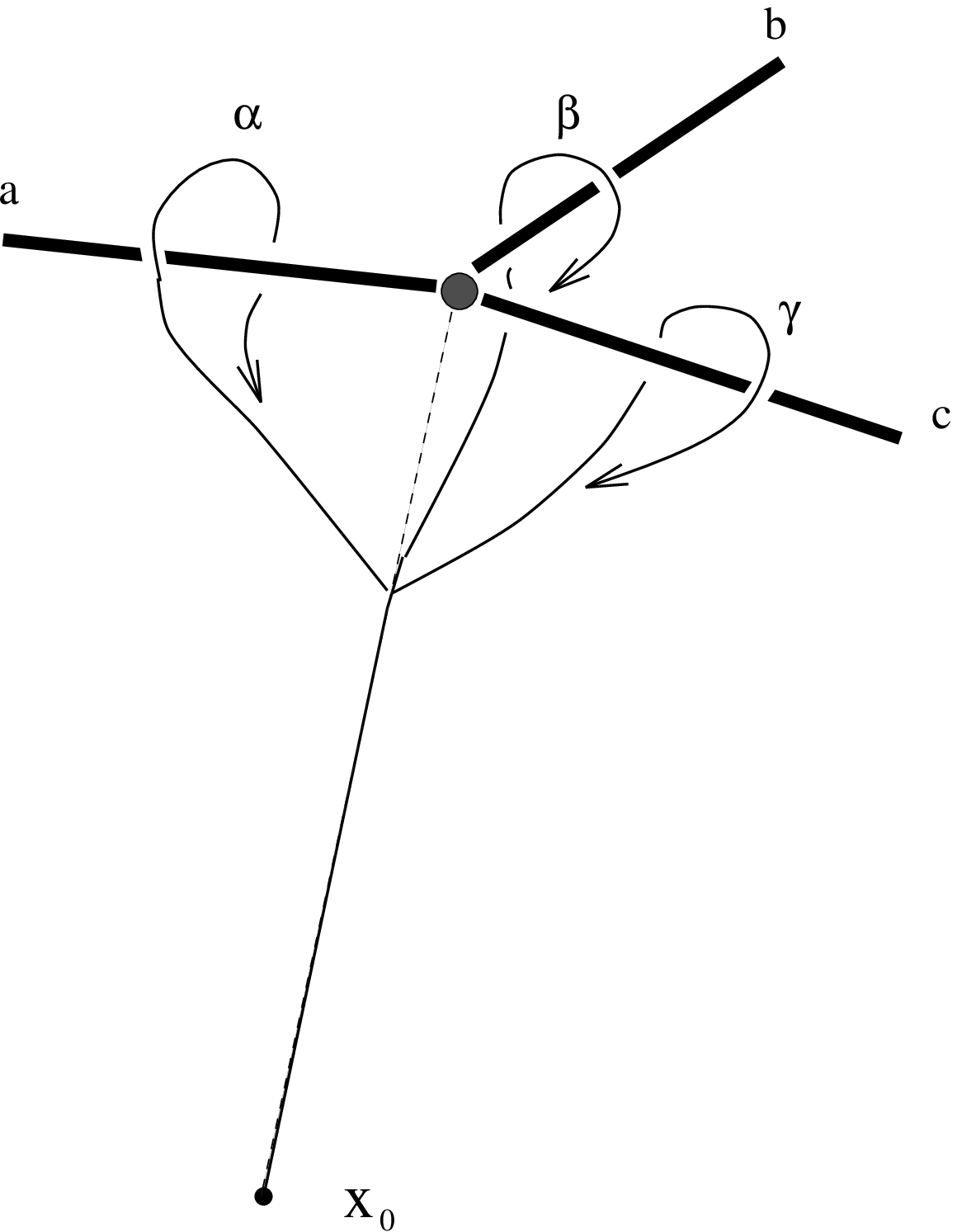}{\bcap\tail \/ Conventions for measuring the fluxes of the three strings emanating from a
node.  Each string's flux is defined as the flux through a path which leaves the basepoint $x_0$ along a straight line toward the node, then encircles the string in
a counterclockwise direction as seen from the far end of the string and returns
to the basepoint.
} 

 As illustrated in figure \FIG\product ~\product, flux conservation requires that the product of all three fluxes emanating from a node be trivial when the fluxes
are multiplied in a clockwise order with respect to the direction ${\vec{BN}}$.
i.e., if the strings in clockwise order are $a$, $b$, and $c$, then
$$cba = e.\eqn\fluxcons$$ 
  In our algorithm, a record is maintained of the geometry of each node:  the strings carry labels indicating the appropriate clockwise orientation.

In the case of a doubly linked pair of nodes (figure   ~\doubleone)   two  segments are collinear, and the order is therefore ill-defined.  In such a case we allow the order to be arbitrary, but the fluxes of the two strings must be defined in a way consistent with that order, such that the product of all three fluxes is as usual trivial.  The ordering must also be compatible between the two nodes which the segments join, so that the flux of a given segment is consistent at its two ends. (The consistency of segments from one end to another will be discussed below.)

\insertfigpage{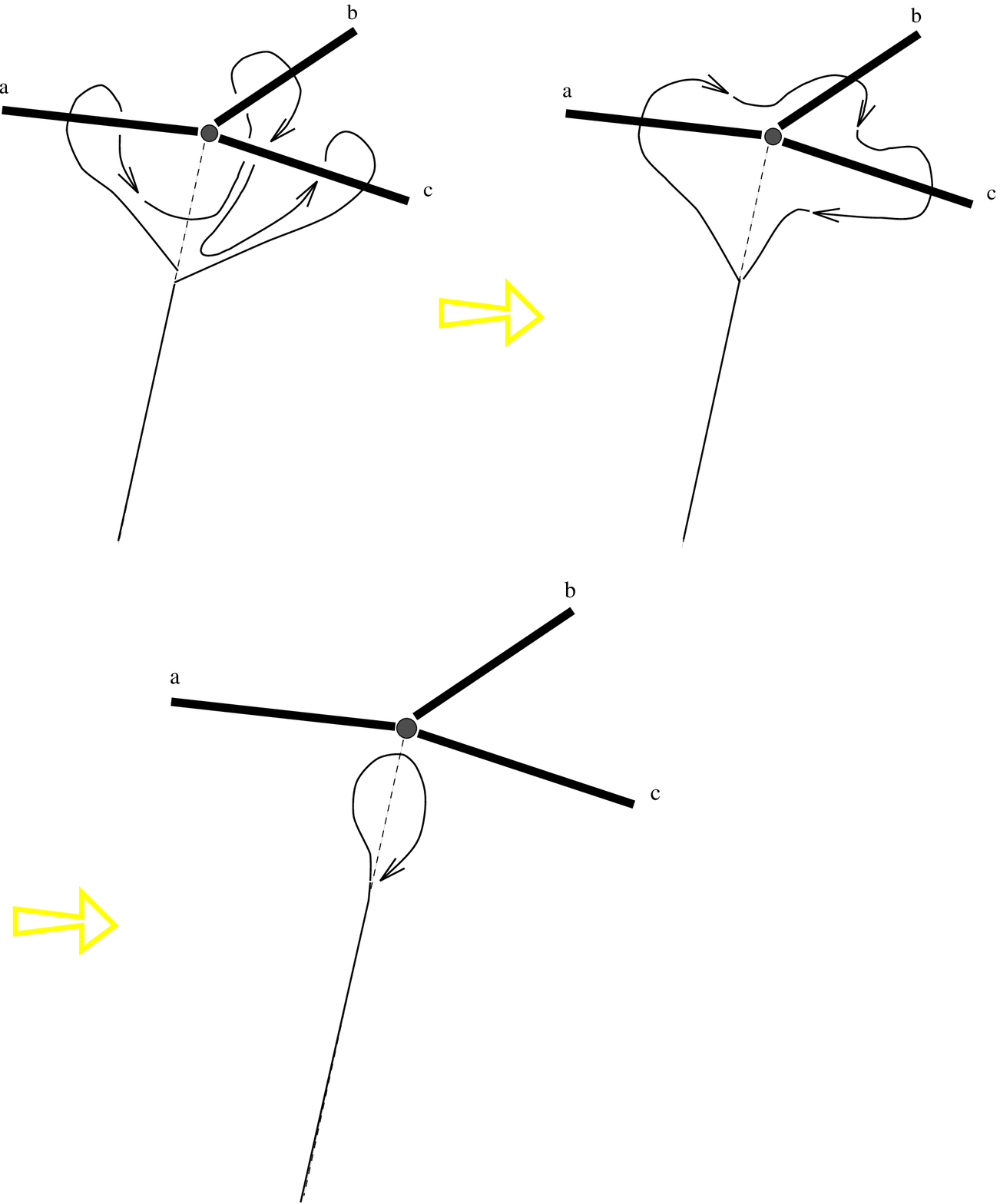}{{\bf Figure \product:}\/  The composition $\gamma\beta\alpha$  of all three paths can be continuously
deformed to a point.  Therefore $cba$, the product of all three fluxes taken in a
clockwise direction as seen from above the node, must be trivial.}

  The collection of standard paths defined above represents a set of generators for $\pi_1({\cal M}-\{D\})$.    The flux state of a network of strings is fully
specified when we know the fluxes enclosed by all of these standard paths. The condition \fluxcons \/ supplies one set of 
relations among these generators.  For each string segment, there is also a relation involving the fluxes defined at its two endpoints, as discussed below.

\subsection{``Sliding'' flux from the endpoint}

By the conventions above, the flux of each string is defined at its two
endpoints.  But for the purposes of this simulation it is necessary to make comparisons of the fluxes of strings at arbitrary points along their lengths.  For example, if two strings cross each other, their fluxes must be compared
at the crossing point in order to determine whether they commute.  A meaningful comparison of the fluxes of
nearby string segments can be obtained only if the paths used to define those two fluxes remain close to each other everywhere except in the immediate vicinity of the strings to be compared.   In particular, the ``tails'' of the paths must not pass on opposite sides of any string, because such paths would give different flux measurements for the same string.     It is possible to define the flux of a string at an arbitrary point
along its length by sliding the standard path to the one which encircles the 
string at the point we wish to measure, as illustrated in figure \FIG\basicslide ~\basicslide.  If another string with flux $b$ pierces the triangle
which is swept out by the sliding path, then the flux at the new position is conjugated by b.  If multiple strings occur, then the new flux $a'$ is given by $faf^{-1}$,  where $f$, the total flux inside the triangle, is defined as the product of the fluxes of all enclosed strings, taken in order of increasing angle from the initial ray $\vec{BP}$.  The flux of each other string at the point where it pierces
the triangle must in turn be defined by a similar sliding procedure from one of
its ends.  This procedure, applied recursively, can thus define the flux of any
string at an arbitrary point $P$ along its length, as measured by a path which follows a straight line from $x_0$ towards $P$ and encircles the string near 
$P$.  

If one slides the path all the way to the far end of the string, the resulting value of the flux must be consistent with the value
measured by the standard path at the other end.  This specifies an additional set of relations among the generators of  $\pi_1({\cal M}-\{D\})$ \/ and furnishes one way of testing for errors in the simulation, as well as being
necessary in order to define the fluxes of strings at the newly created nodes
that result from string collisions.

\insertfigpage{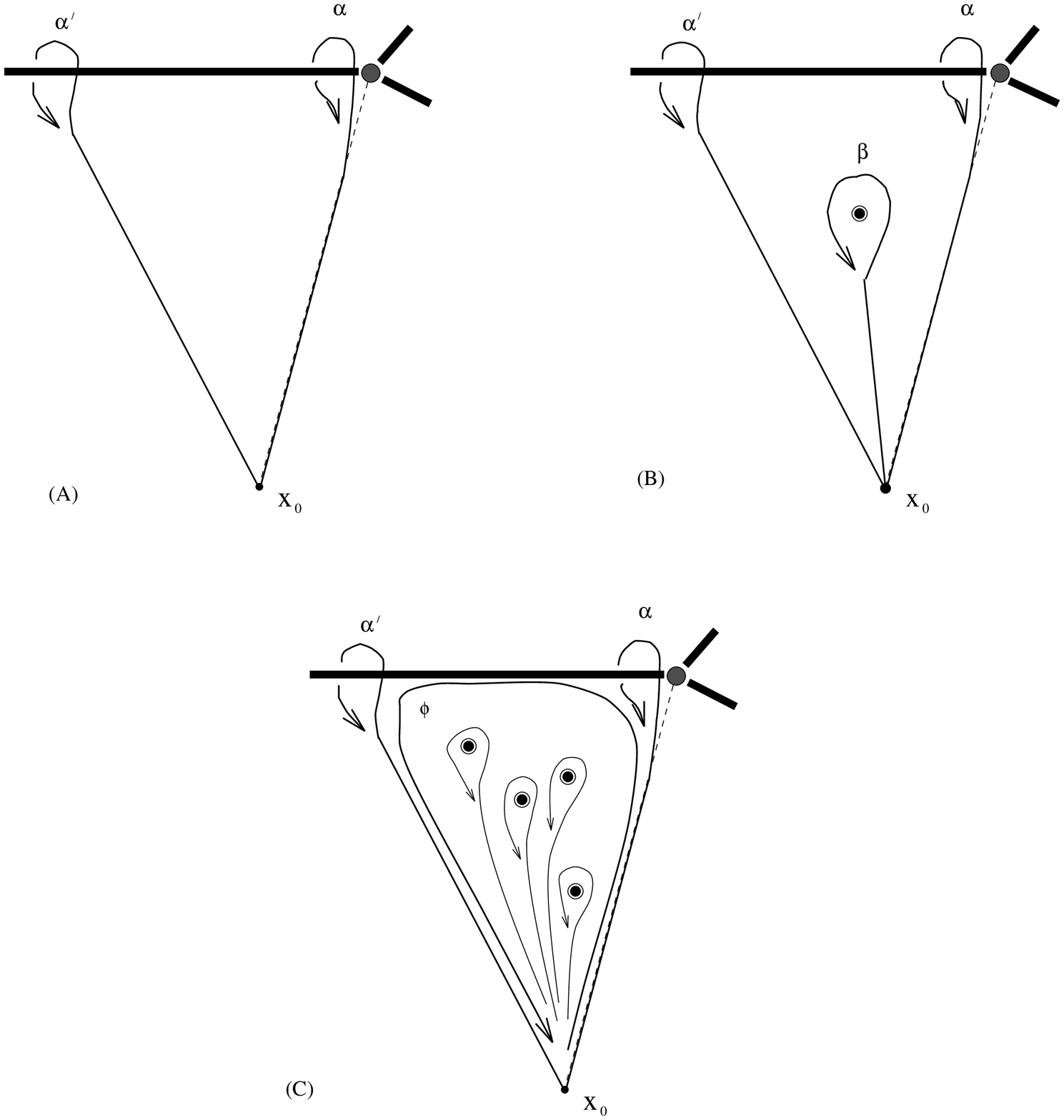}{\bcap\basicslide \/ 
When the flux $a$ of a string has been defined according to a path which encircles
it near one end,  the flux $a'$ of that string at another point along its length  can be defined by ``sliding'' the standard path $\alpha$ to $\alpha'$ as shown.
If no other strings pierce the triangle which is swept out, then this merely
represents a continuous deformation of $\alpha$, and thus $a'=a$\/ (fig. \basicslide A).  However, if the triangle is pierced by string with flux $b$ as measured by path $\beta$,
then the flux is conjugated by $b$:  $a'=bab^{-1}$\/ (\basicslide B).  More generally, if the 
triangle (or the oriented path $\phi$ shown in \basicslide C) encloses flux $f$, then $a$
is conjugated by the total flux $f$, i.e., $a'=faf^{-1}$. The total flux is given by the product of individual string fluxes, taken in order of increasing
angle from the initial tail. (This can be seen by deforming a product of loops
to a single loop enclosing all strings.)} 

A modified version of this sliding procedure is used to define all fluxes 
initially from the original lattice.  Paths composed of lattice links are
deformed by a series of steps to straight-line paths from the base point to
the location of each node.  

\subsection{Holonomy Interactions}

As the network evolves dynamically and nodes change their position, 
the fluxes defined by these conventions may change in several
different ways.  First, as a node moves, its tail may be dragged across
another string segment.  Conversely, a string segment may be dragged across
the node's tail by the motion of other nodes.  In both cases, the fluxes of
all strings at the node must be conjugated by the flux which is crossed, as
shown in figure \FIG\tailcrossing ~\tailcrossing.  In addition, the geometry of the strings
at a given junction may change, resulting in holonomy interactions among the 
three strings joined at that node.  Such a process is shown in figure
\FIG\geochange ~\geochange :  the motion of string $a$ causes its standard flux to change, 
and also changes the clockwise ordering of the strings $a$, $b$,and $c$. 
This requires both an adjustment of the flux definitions and of the order labels.

\FIG\wrapflux
\FIG\gammacrossing 
\insertfigpage{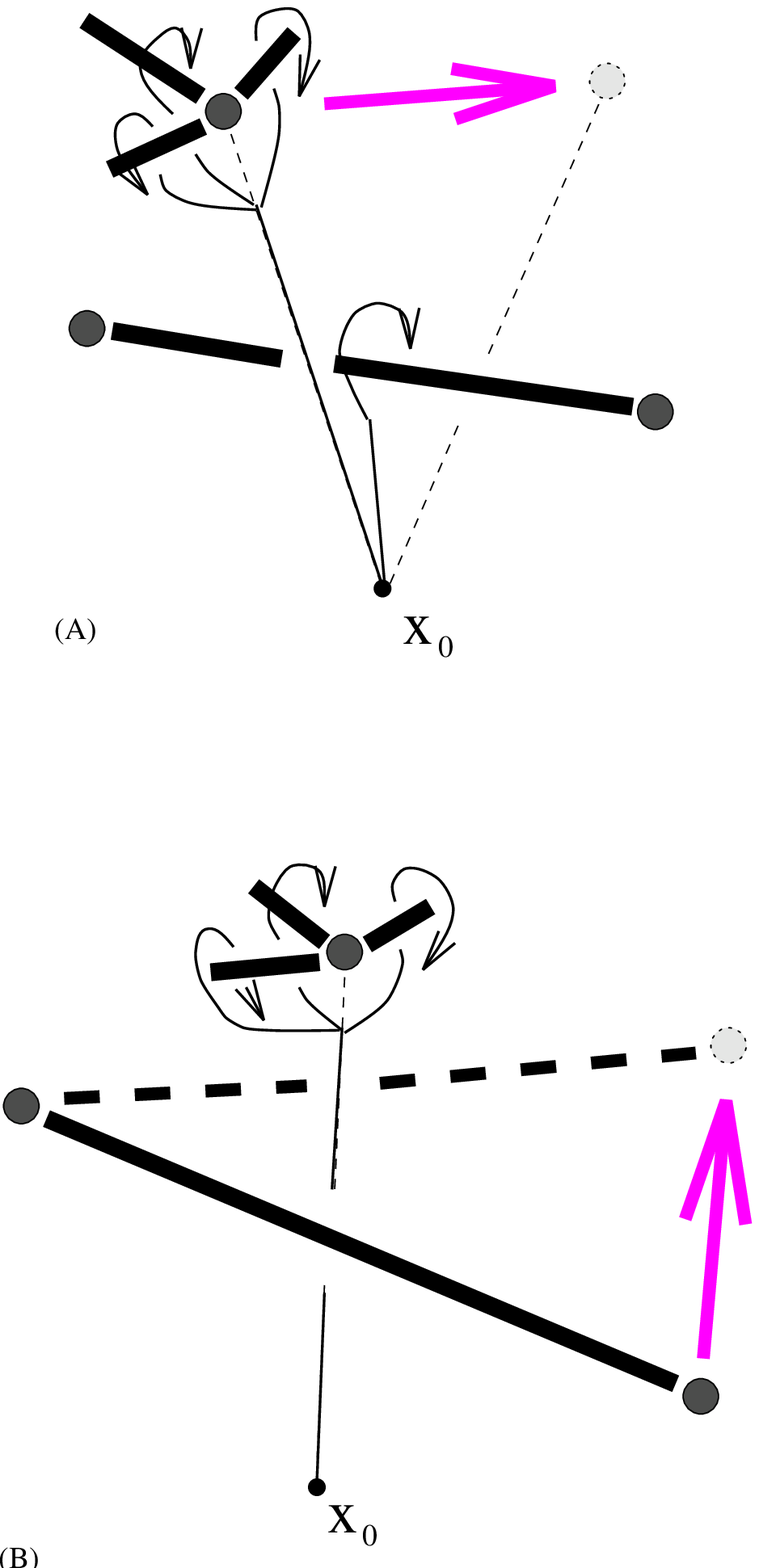}{\bcap\tailcrossing \/ Crossing of a node's tail
by a string.  This can happen either when the moving node drags its tail across
the string (A), or when the string is dragged across the tail due to the motion
of another node (B).  In both cases, the fluxes of all strings attached to the 
node whose tail is crossed must be conjugated by the flux of the crossing string.} 

\insertxfigpage{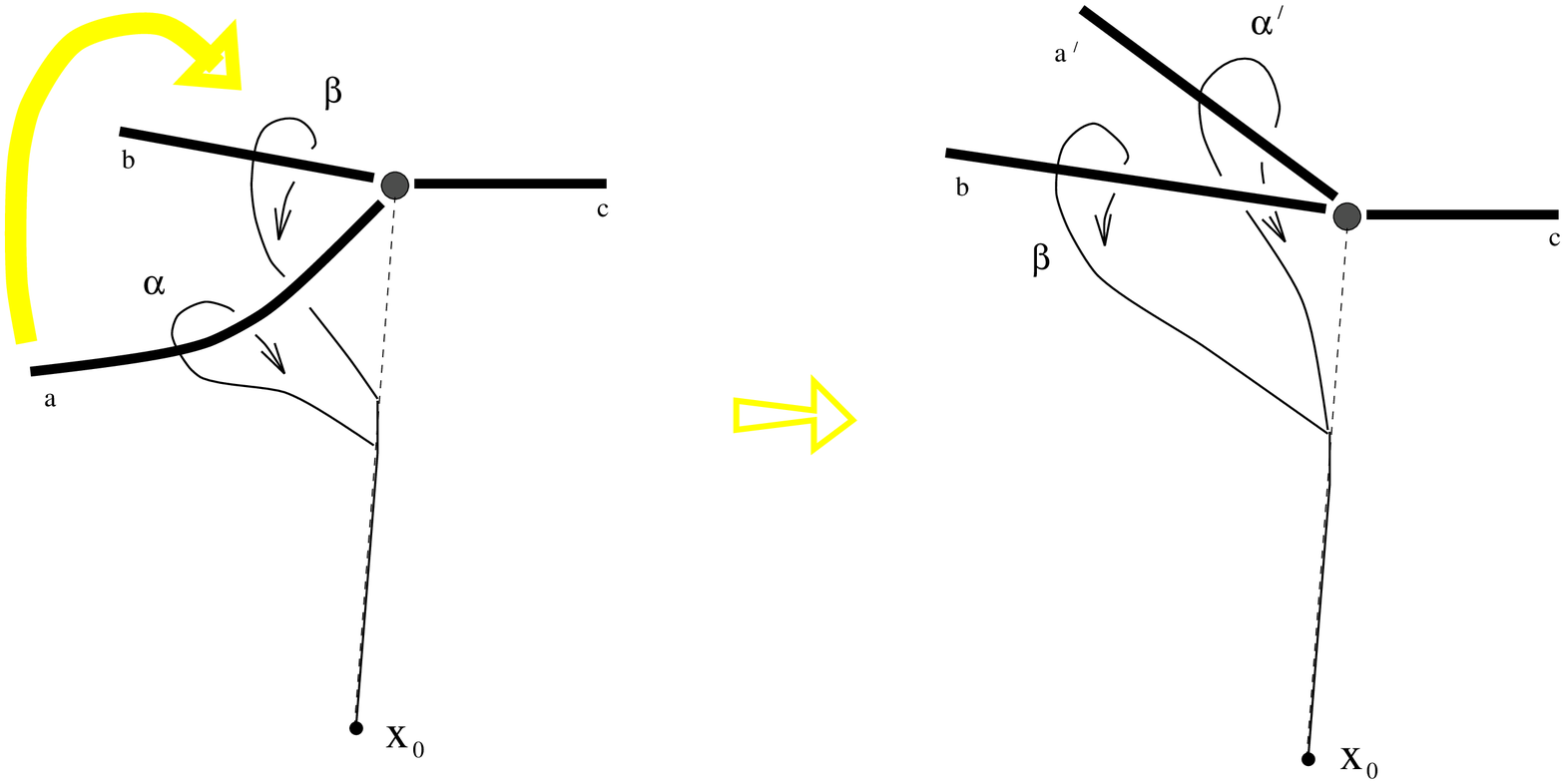}{\bcap\geochange \/ Example of a holonomy interaction between strings attached to the same node.  When the string carrying flux $a$ is lifted over the other string carrying flux $b$, its flux must be
redefined as $a'=bab^{-1}$, and the conventional clockwise order of the three strings changes, with $a$ and $b$ exchanging places.  The flux conservation condition is maintained:  if $cba=e$ originally, then also $ca'b=e$.}

\insertccomfigpage{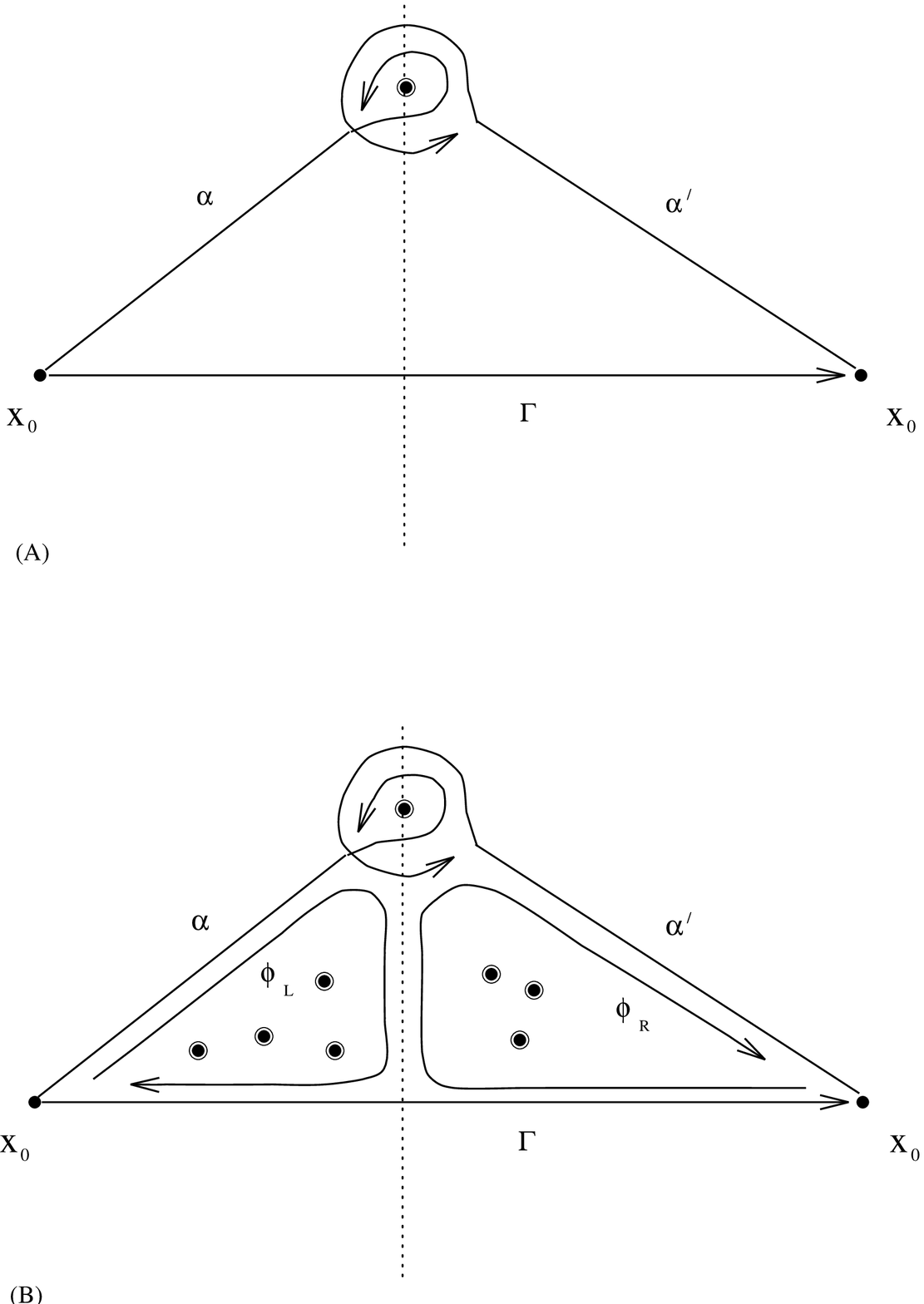}{\baselineskip=7pt \bcap\wrapflux \/ Transformation from one description of a flux to another at the boundary.  Here a string is shown intersecting the plane of the page precisely where it intersects the boundary of the cubic simulation volume (dotted line).  Under periodic boundary conditions, the two points labeled $x_0$ are identified. The flux of the string may be described in terms of a path whose tail extends to the right of $x_0$ ($\alpha$) or to the left ($\alpha'$).  If no other strings are present, then $\alpha$ is homotopically
equivalent to $\Gamma^{-1}\alpha'\Gamma$.  In the more general situation shown in (B), $\alpha\sim (\phi_L \Gamma \phi_R)^{-1}\alpha' (\phi_L \Gamma \phi_R)$, and so the two descriptions of the flux are related through conjugation by $f_L C f_R$,  where $C$ is the flux associated with the path $\Gamma$ and $f_L$ and $f_R$ are the overall fluxes enclosed by $\phi_L$ and $\phi_R$, respectively.  The latter can be defined in terms of paths lying entirely on one side or the other of the boundary.}

\insertcomfigpage{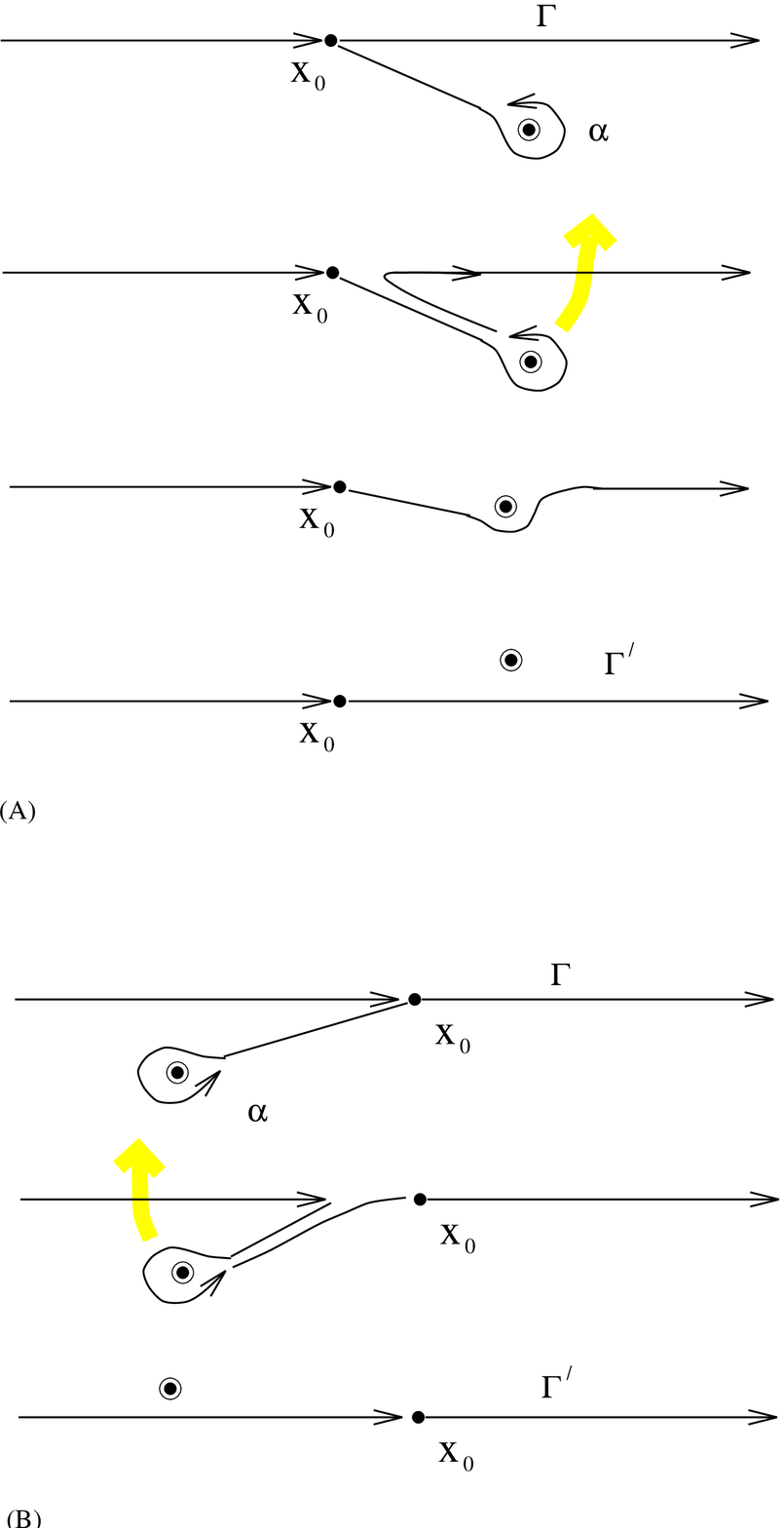}{\bcap\gammacrossing \/ Interaction between a string and one of the large loops of the 3-torus.  As a string with flux $a$ as defined by the path $\alpha$ crosses to the right of $x_0$ as shown in (A), the path $\alpha\Gamma$, where $\Gamma$ is a straight-line path which wraps around the 3-torus, can be continuously deformed to the new straight-line path 
$\Gamma'$.  Thus the flux $C$ associated with $\Gamma$ must be multiplied from the right by the string's flux, $C'=Ca$.  If the string crosses to the left as in (B), $\alpha\Gamma$ is deformed to $\Gamma'$ and so the multiplication is from the other side: $C'=aC$.}

\subsection{Periodic Boundary Conditions}
Due to the periodic boundary conditions, our simulation volume has the topology
of a 3-torus, $T_3$.  The non-trivial topology introduces three additional classes of noncontractible closed loops other than the ones associated 
with strings: namely those which wrap around one
of the boundaries.  These loops may be associated with nontrivial flux.  
As in the case of vortices on a Riemann surface,\Ref\Lee{K.M.~Lee, Phys. Rev. D 49, 2030 (1994).}
  the fluxes of these 
loops may have holonomy interactions with the fluxes of strings,  and therefore
a full description of string fluxes requires us to maintain a record of these 
large loops.

As representatives of the three ``wraparound'' classes, we
choose canonical straight-line paths  parallel to the coordinate 
axes, which we will refer to as $\Gamma_x,\/\Gamma_y$, and 
$\Gamma_z$.  $\Gamma_x$, for instance, leaves the basepoint along the $+\hat{x}$ direction, wraps around the boundary from $x=L$ to $x=0$, and then returns to the basepoint from
the $-\hat{x}$ side.  Along with values of the fluxes for all strings, our
algorithm maintains a record of the $S_3$ holonomies $C_x, C_y$, and $C_z$ associated with $\Gamma_x, \Gamma_y$, and $\Gamma_z$ respectively.  These
values must be known in order to make consistent comparisons of string fluxes
across the boundary. (The procedure for doing so is illustrated by figure  ~\wrapflux .)
The values of the $C_i$ may change if a string crosses 
one of the curves $\Gamma_i$  (see figure ~\gammacrossing ).

\vskip 0.4in
 
\ack

A portion of this work
was completed at Caltech, and the remaining portion at the University
of North Carolina, Chapel Hill.  This work was supported in 
part by U.S. Department of Energy Grant no.  DE-FG03-92-ER40701 and  
Grant no. DE-F605-85-ER-40219 Task A. The author thanks J. Preskill, T. Vachaspati, A. de Laix,
C. Thompson and P. Sikivie for helpful discussions and encouragement.

\par\penalty-400\vskip\chapterskip\spacecheck\referenceminspace
   \ifreferenceopen \Closeout\referencewrite \referenceopenfalse \fi
   \line{\fourteenrm\hfil REFERENCES\hfil}\vskip\headskip
   \input referenc.txa

\end